%% file: MultiL-KRIM.tex
\documentclass[journal]{IEEEtran}

\input{preamble.tex}

\title{Multilinear Kernel Regression and Imputation\\ via
  Manifold Learning}

\author{Duc Thien Nguyen and Konstantinos
  Slavakis\IEEEauthorrefmark{1}%
  \thanks{\IEEEauthorrefmark{1}D.~T.~Nguyen and K.~Slavakis
    are with the Department of Information and
    Communications Engineering, Tokyo Institute of
    Technology, Yokohama, 226-8502 Japan (e-mails:
    \{nguyen.t.au, slavakis.k.aa\}@m.titech.ac.jp).}%
}

\begin{document}
\maketitle

\begin{abstract}
  This paper introduces a novel nonparametric framework for
  data imputation, coined multilinear kernel regression and
  imputation via the manifold assumption
  (MultiL-KRIM). Motivated by manifold learning, MultiL-KRIM
  models data features as a point cloud located in or close
  to a user-unknown smooth manifold embedded in a
  reproducing kernel Hilbert space. Unlike typical
  manifold-learning routes, which seek low-dimensional
  patterns via regularizers based on graph-Laplacian
  matrices, MultiL-KRIM builds instead on the intuitive
  concept of tangent spaces to manifolds and incorporates
  collaboration among point-cloud neighbors (regressors)
  directly into the data-modeling term of the loss
  function. Multiple kernel functions are allowed to offer
  robustness and rich approximation properties, while
  multiple matrix factors offer low-rank modeling, integrate
  dimensionality reduction, and streamline computations with
  no need of training data. Two important application
  domains showcase the functionality of MultiL-KRIM:
  time-varying-graph-signal (TVGS) recovery, and
  reconstruction of highly accelerated
  dynamic-magnetic-resonance-imaging (dMRI) data. Extensive
  numerical tests on real and synthetic data demonstrate
  MultiL-KRIM's remarkable speedups over its predecessors,
  and outperformance over prevalent ``shallow''
  data-imputation techniques, with a more intuitive and
  explainable pipeline than deep-image-prior methods.
\end{abstract}


%%%%%%%%%%%%%%%%%%%%%%%%%%%%%%%%%%%%%%%%%%%%%%%%%%%%%%%%%%%%%%%%%%%%%%%%%%%%%%%%%%%%%%%%%%%%%%%

\section{Introduction}\label{sec:intro}

Missing data occur frequently across different fields, such
as recommender systems~\cite{ramlatchan2018survey}, remote
sensing~\cite{shen2015missing}, sensor
networks~\cite{du2020missing}, and accelerated medical
imaging~\cite{liang1994efficient, zhi2000principles},
inflicting bias and deficiency on downstream data-analysis
tasks. Amidst a wide range of strategies for data
imputation, \textit{imputation-by-regression}\/ is one of
the most popular frameworks, where measured data act as
regressors while missing data are imputed by the fitted
regression model and measured data~\cite{osman2018survey}.

Imputation of missing data is typically viewed as a
matrix/tensor completion problem, since (dynamic) images, or
spatio-temporal data of sensor networks are usually arranged
into matrices/tensors. Popular data-imputation approaches
are compressed sensing, \eg, \cite{candes2008exact}, and
low-rank methods which are realized via a decomposition of
the data matrix into low-rank matrix
factors~\cite{chen2004recovering, koren2009matrix}. On the
other hand, dictionary learning~\cite{mairal2008dictionary,
  tovsic2011dictionary, sun2016complete1, sun2016complete2}
factorizes the data matrix into an over-complete (``fat'')
matrix and a sparse one. Sequential multi-layer
factorization of the data matrix has been also
reported~\cite{cichocki2007multilayer}. Non-linear
activation function can be also applied to those layers,
resulting into deep matrix
factorizations~\cite{fan2018matrix, de2021survey}, similarly
to deep neural networks. Tensors can be reshaped
(``flattened'') into matrices to enjoy the tools of matrix
completion. However, there exist concepts of multilinear
rank inherent to tensors, such as the CP and Tucker rank,
yielding CP and Tucker tensor
decompositions~\cite{kolda2009tensor}. Recently, tensor
networks offer generalized tensor
decompositions~\cite{oseledets2011tensor, zhao2016tensor,
  zheng2021fctn}. All of the aforementioned methods fall
under the umbrella of parametric modeling, where the
characteristic is that model dimensions, that is, inner
dimensions of the matrix/tensor factorizations, do not scale
with the number of the observed data.

Another recently popular path of parametric modeling is deep
learning (DeepL), \eg, \cite{nguyen2019low,
  su2022survey}. Nevertheless, not only is DeepL usually
training-data hungry and computationally heavy, but concerns
were also raised in~\cite{antun2020instabilities} about
instabilities in medical-image reconstruction. Deep image
priors (DIP)~\cite{ulyanov2018deep} and untrained neural
networks offer user-defined priors to alleviate the need for
massive training data, and have been used in signal
recovery~\cite{heckel2020compressive, rey2022untrained} and
accelerated medical imaging~\cite{yoo2021time, Zou:TMI:21}.

In contrast to the previous parametric methods,
nonparametric regression operates without any priors and
statistical assumptions on the data in an effort to reduce
as much as possible the bias inflicted on data modeling by
the user~\cite{Gyorfi:DistrFree:10}. The price to be paid
for this distribution-free approach is that the dimensions
of the nonparametric-regression model scales with the number
of observed data. Within nonparametric-regression methods,
the kernel-based ones stand out thanks to their highly
intuitive and explainable functional forms to approximate
involved non-linear
functions~\cite{bazerque2013nonparametric, fan2018non,
  stock2018comparative, gimenez2019matrix,
  chen2021kernel}. The crux is to use a non-linear feature
mapping $\varphi(\cdot)$ to map data from their input space
to a high-, even infinite-dimensional feature space, where
linear operations are used to reflect non-linear operations
in the original input space. If the feature space is chosen
to be a reproducing kernel Hilbert space (RKHS)
$\mathscr{H}$, with well-known and versatile properties in
approximation theory~\cite{aronszajn1950theory}, inner
products in $\mathscr{H}$ can be calculated via simple
functional evaluations in the input space (``kernel
trick'')~\cite{scholkopf2002learning}.

Aiming at learning methods for high-dimensional data,
manifold learning (ManL) offers a framework where
data/features are considered to be points in or close to a
low-dimensional (non-linear) smooth manifold $\mathscr{M}$
embedded in a high-dimensional ambient
space~\cite{lin2008riemannian}. ManL affords many degrees of
freedom in data modeling because it adopts a minimal number
of assumptions; no specifics of $\mathscr{M}$ are presumed
other than its smoothness. ManL identifies low-dimensional
data patterns to effect dimensionality reduction. More
specifically, ``local approaches,'' including local linear
embedding (LLE)~\cite{roweis2000nonlinear} and Laplacian
eigenmaps~\cite{belkin2001laplacian}, effect dimensionality
reduction by preserving local proximity among data, while
``global approaches,'' such as
ISOMAP~\cite{tenenbaum2000global}, preserve geodesic
distances between all pairs of points. While local
approaches may fail to capture long-range dependencies in
the point-cloud, global ones are usually computationally
costly. The standard ManL route for data imputation is to
use a graph Laplacian matrix to model the interdependencies
among the rows/columns of the data
matrix~\cite{kalofolias2014matrix, poddar2016dynamic}. This
approach has been widely used in recommender
systems~\cite{ma2011recommender, rao2015collaborative,
  mongia2019matrix} to construct users/items
interaction. % From a ``collaborative-filtering''
% standpoint, each group of users may share similar ratings
% of some groups of items, which is manifested by
% neighborhoods on a manifold.
Due to the focus on ``locality'' by the graph Laplacian
matrix, the computed interpolation functions appear to be
non-continuous at sampled points, especially in cases of
severe under-sampling~\cite{shi2017weighted}.

Different from parametric modeling and motivated by ManL,
this paper develops a generalization, abbreviated hereafter
\textit{MultiLinear Kernel Regression and Imputation via the
  Manifold assumption (MultiL-KRIM),} of the nonparametric
kernel-based approach of~\cite{shetty2020bilmdm,
  slavakis2022krim}. Features are assumed to form a
point-cloud which lies in or close to a smooth manifold
$\mathscr{M}$ embedded in an RKHS $\mathscr{H}$. Unlike
popular decomposition-based methods, such as
low-rank-~\cite{davenport2016overview, chi2019nonconvex},
tensor-based~\cite{zhou2017tensor, cai2019nonconvex} and
dictionary learning \cite{mairal2008dictionary,
  tovsic2011dictionary, sun2016complete1, sun2016complete2},
which promote a ``blind decomposition'' of the data
matrix/tensor, and unlike standard ManL routes which are
based on Laplacian-matrix-based regularization of a
regression loss~\cite{kalofolias2014matrix,
  poddar2016dynamic}, MultiL-KRIM explores
collaborative-filtering and tangent-space modeling ideas to
extract and at the same time incorporate latent feature
geometry \textit{directly}\/ into data-matrix
decompositions.
% Owing to its minimal assumptions and flexible matrix
% factorization, the framework of MultiL-KRIM can
% accommodate several state-of-the-art kernel-based
% schemes~\cite{bazerque2013nonparametric,
% venkitaraman2019predicting, pu2021kernel,
% cichocki2007multilayer}.
MultiL-KRIM needs no training data to operate, belongs to
the family of nonparametric regression to reduce its
dependence on the probability distribution of the data as
much as possible~\cite{Gyorfi:DistrFree:10}, and offers an
explainable and simple geometric learning paradigm, unlike
the majority of DeepL schemes which are based on perplexed
and cascading non-linear function layers.

Unlike its bilinear predecessors
BiLMDM~\cite{shetty2020bilmdm} and
KRIM~\cite{slavakis2022krim}, where dimensionality-reduction
\textit{pre-steps}\/ are detached from the regression task,
MultiL-KRIM affords multiple matrix factors to allow
dimensionality reduction be employed directly into the
regression task by letting its inverse-problem automatically
identify the ``optimal low-dimensional'' rendition of a
kernel matrix. In addition, MultiL-KRIM leverages its
flexible matrix factors for higher computational efficiency
in its inverse-problem algorithmic solution.

To showcase the functionality of MultiL-KRIM, two imputation
tasks, corresponding to different application domains, are
considered:
\begin{enumerate*}[label = \textbf{(\roman*)}]
\item the recently popular problem of time-varying
  graph-signal (TVGS) recovery (\cref{sec:graph}), and
\item signal reconstruction in dynamic magnetic resonance
  imaging (dMRI) under severe under-sampling
  (\cref{sec:dmri}). This manuscript extends the short
  paper~\cite{Thien:ICASSP24}, which considers the dMRI
  case.
\end{enumerate*}
The extensive numerical tests in \Cref{sec:graph,sec:dmri}
illustrate that MultiL-KRIM outperforms various
state-of-the-art ``shallow'' parametric and nonparametric
methods.

The rest of the paper is organized as
follows. \cref{sec:problem.formulation} describes rigorously
the missing-data imputation problem and briefly introduces
state-of-the-art techniques. \cref{sec:modeling} presents
the MultiL-KRIM data modeling, its inverse problem and an
algorithmic solution. Finally, \cref{sec:conclusion}
summarizes the paper and provides hints about future
research directions. To abide by the Journal's page limits
for a first paper submission, several of the discussions are
moved into the accompanying supplementary file.

%%%%%%%%%%%%%%%%%%%%%%%%%%%%%%%%%%%%%%%%%%%%%%%%%%%%%%%%%%%%%%%%%%%%%%%%%%%%%%%%%%%%%%%%%%%%%%%

\section{The Imputation-by-Regression Problem and Prior Art}\label{sec:problem.formulation}

Data are arranged in an $N$-mode tensor $\mathbfscr{Y}$ of
size $I_1 \times I_2 \times \cdots \times I_N$, where
$\Set{I_n}_{n=1}^N \subset \IntegerPP$ ($\IntegerPP$ is the
set of all positive integers). The observed data take values
from the set of all complex numbers $\Complex$, while
entries of $\mathbfscr{Y}$ which correspond to missing data
are set to be equal to $+\infty$. In other words,
$\mathbfscr{Y}\in (\Complex \cup \{ +\infty \})^{I_1 \times
  I_2 \times \cdots \times I_N}$.

Tensor $\mathbfscr{Y}$ can be
\textit{unfolded/matricized/flattened}\/ along mode $k$ into
a matrix $\vect{Y}_{(k)}$~\cite{kolda2009tensor}. For
simplicity, assume that $\mathbfscr{Y}$ is unfolded along
mode $N$, and
$\vect{Y} \coloneqq \vect{Y}_{(N)} \in (\Complex \cup \{
+\infty \})^{I_0 \times I_N}$, where
$I_0 \coloneqq I_1I_2\cdots I_{N-1}$. Let the $(i,t)$th
entry of $\vect{Y}$ be denoted by
$y_{it} \coloneqq [\vect{Y}]_{it}$. Denote also the rows and
columns of $\vect{Y}$ as
$\mathcal{Z}\coloneqq\Set{\vect{z}_i}_{i=1}^{I_0}$ and
$\mathcal{Y}\coloneqq\Set{\vect{y}_t}_{t=1}^{I_N}$,
respectively; that is,
$\vect{Y} = [ \vect{z}_1, \ldots, \vect{z}_{I_0}
]^{\intercal} = [ \vect{y}_1, \ldots, \vect{y}_{I_N} ]$,
where $\intercal$ denotes vector/matrix transposition. For
rigorous discussions, let the index set of observed entries
$\Omega \coloneqq \Set{ (i,t) \in \Set{1, \ldots, I_0}
  \times \Set{1, \ldots, I_N} \given y_{it} \neq +\infty }$,
and define the linear \textit{sampling mapping}\/
$\mathscr{S}_{\Omega} \colon (\Complex \cup \{ +\infty
\})^{I_0 \times I_N} \to \Complex^{I_0 \times I_N} \colon
\vect{Y} \mapsto \mathscr{S}_{\Omega}(\vect{Y})$, which
operates entry-wisely as follows:
$[\mathscr{S}_{\Omega}(\vect{Y})]_{it} \coloneqq
[\vect{Y}]_{it}$, if $(i,t) \in \Omega$, while
$[\mathscr{S}_{\Omega}(\vect{Y})]_{it} \coloneqq 0$, if
$(i,t) \notin \Omega$.

To cover a wide range of application domains, it is assumed
that data $\vect{Y}$, observed in a input-data domain
$\mathscr{D}_y \subset (\Complex \cup \{ +\infty \})^{I_0
  \times I_N}$, possess also a representation over another
domain $\mathscr{D}_x \subset \Complex^{I_0 \times I_N}$,
and that the ``link'' between $\mathscr{D}_y$ and
$\mathscr{D}_x$ is a transform mapping
$\mathscr{T} \colon \mathscr{D}_x \to \mathscr{D}_y$, which
is usually considered to be invertible. In other words, it
is assumed that there exists an
$\vect{X}\in \mathscr{D}_x \subset \Complex^{I_0 \times
  I_N}$ such that (s.t.) $\mathscr{T}(\vect{X})$ agrees with
$\vect{Y}\in \mathscr{D}_y$ on its observed entries at
positions $\Omega$. For example, in the context of
\cref{sec:dmri}, $\mathscr{D}_y$ is the k-space domain and
$\mathscr{D}_x$ is the image one, with $\mathscr{T}$ being
the two-dimensional discrete Fourier transform (DFT)
$\mathscr{F}$~\cite{zhi2000principles}. In cases where there
is no need for the extra domain $\mathscr{D}_x$ (see
\cref{sec:graph}), then
$\mathscr{D}_x \coloneqq \mathscr{D}_y$, and $\mathscr{T}$
is set to be the identity mapping $\Id$.

Typically, imputation-by-regression is achieved by solving the inverse problem
\begin{subequations}\label{recovery.generic}
  \begin{align}
    \min\nolimits_{ \vect{X}\in \Complex^{I_0 \times I_N} }
    {} & {} \mathcal{L}( \vect{X} ) + \mathcal{R}( \vect{X} ) {} \label{IbyR.loss}\\
    \text{s.to}\ {}
       & {} \mathscr{S}_{\Omega}(\vect{Y}) = \mathscr{S}_{\Omega} \mathscr{T}
         (\vect{X}) \label{IbyR.consistency} \\
       & \text{and other constraints,} \notag
  \end{align}
\end{subequations}
where $\mathcal{L}(\cdot)$ is a loss that incorporates the
user-defined data model into the design, and
$\mathcal{R}(\cdot)$ is the regularizer used to impose
structural priors to the data model. Moreover,
\eqref{IbyR.consistency} is the ``data-consistency''
constraint. A popular strategy for low-rank models is to
have data consistency appear as the ``soft'' loss
$\mathcal{L}( \vect{X} ) \coloneqq \norm{
  \mathscr{S}_{\Omega}(\vect{Y}) - \mathscr{S}_{\Omega}
  \mathscr{T} (\vect{X}) }^2_{\textnormal{F}}$
in~\eqref{IbyR.loss}, where $\norm{\cdot}_{\textnormal{F}}$
stands for the Frobenius norm of a matrix, together with the
nuclear-norm regularizer
$\mathcal{R}(\vect{X}) \coloneqq
\norm{\vect{X}}_*$~\cite{candes2008exact}, instead of using
the ``hard'' constraint
in~\eqref{IbyR.consistency}. Although such a path can be
also addressed by the proposed MultiL-KRIM, the present
manuscript keeps the ``hard'' constraint
in~\eqref{IbyR.consistency} to allow for
$\mathcal{L}(\cdot)$ obtain simpler, ``cleaner,'' and more
familiar expressions than those after involving the operator
$\mathscr{S}_{\Omega}$ in $\mathcal{L}( \cdot)$.

A popular low-rank data model is expressed by the loss
$\mathcal{L}( \vect{X}, \vect{U}, \vect{V} ) = \norm{
  \vect{X} - \vect{U} \vect{V} }^2_{\textnormal{F}}$, where
$\vect{U}$ and $\vect{V}$ are added as auxiliary variables
to the inverse problem~\eqref{recovery.generic}, with
$\vect{U}$ having more rows than
columns~\cite{chen2004recovering, koren2009matrix}. In
contrast, if $\vect{U}$ is over-complete, \ie, more columns
than rows, and $\vect{V}$ is sparse, the previous loss
yields dictionary learning~\cite{tovsic2011dictionary}. More
generally, the loss
$\norm{\vect{X} - \vect{U}_1 \vect{U}_2 \cdots \vect{U}_Q
  \vect{V} }_{\textnormal{F}}^2$, where all of $\vect{X}$,
$\Set{ \vect{U}_q }_{q=1}^Q$ and $\vect{V}$ are variables to
be optimized, leads to multi-layer matrix factorization
(MMF)~\cite{cichocki2007multilayer}, which serves as a
multilinear representation of the data. With regards to
multilinear algebra, inverse problems can also be formulated
directly on the tensor $\mathbfscr{Y}$, where
tensor-decomposition frameworks such as Tucker and CP
decomposition~\cite{kolda2009tensor}, or the more recent
tensor networks~\cite{oseledets2011tensor, zhao2016tensor,
  zheng2021fctn}, replace the standard matrix ones. All of
the aforementioned parametric factorization schemes are
``blind,'' in a sense that all matrix/tensor factors are
variables to be computed, and no data/feature geometry is
incorporated in the data model. Typically, any prior
knowledge and additional model assumptions are realized via
the regularizer $\mathcal{R}(\cdot)$.

The nonparametric basis pursuit
(NBP)~\cite{bazerque2013nonparametric} introduces the
following data-modeling loss
\begin{align}
  \mathcal{L}(\vect{X}, \vect{B}, \vect{C}) = \norm{ \vect{X} - \vect{K}_{\mathcal{Z}} \vect{B}
  \vect{C} \vect{K}_{\mathcal{Y}} }^2_{\textnormal{F}} \,, \label{eq:NBP.approx}
\end{align}
where
$[\vect{K}_{\mathcal{Z}}]_{ i^{\prime}i^{\prime\prime} } =
\kappa_{\mathcal{Z}} (\vect{z}_{i^{\prime}},
\vect{z}_{i^{\prime\prime}})$ and
$[\vect{K}_{\mathcal{Y}}]_{ t^{\prime}t^{\prime\prime} } =
\kappa_{\mathcal{Y}} (\vect{y}_{t^{\prime}},
\vect{y}_{t^{\prime\prime}})$ are kernel matrices, generated
by the rows $\mathcal{Z}$ and columns $\mathcal{Y}$, with
reproducing kernel functions
$\kappa_{\mathcal{Z}}(\cdot, \cdot)$ and
$\kappa_{\mathcal{Y}}(\cdot, \cdot)$ corresponding to some
user-defined RKHSs $\mathcal{H}_{\mathcal{Z}}$ and
$\mathcal{H}_{\mathcal{Y}}$, respectively. A very short
primer on RKHSs can be found in \cref{app:RKHS}
(supplementary file). The motivation
behind~\eqref{eq:NBP.approx} is model
$x_{it} \approx \sum_{k=1}^d f_k(\vect{z}_i)
g_k(\vect{y}_t)$, where
$f_k \coloneqq \sum_{i^{\prime} = 1}^{I_0} b_{i^{\prime}k}
\varphi_{\mathcal{Z}}(\vect{z}_{i'}) \in
\mathcal{H}_{\mathcal{Z}}$,
$g_k \coloneqq \sum_{t^{\prime} = 1}^{I_N} c_{kt^{\prime}}
\varphi_{\mathcal{Y}}(\vect{y}_{t'}) \in
\mathcal{H}_{\mathcal{Y}}$, with $\varphi_{\mathcal{Z}}$ and
$\varphi_{\mathcal{Y}}$ being feature maps (see
\cref{app:RKHS}), and $d \leq \min\{ I_0, I_N\}$ is a
user-defined upper bound on $\vect{X}$'s rank. Matrices
$\vect{B}$ and $\vect{C}$ in~\eqref{eq:NBP.approx} gather
all coefficients $\{ b_{i^{\prime}k} \}$ and
$\{ c_{kt^{\prime}} \}$, respectively. Similarly to
NBP~\cite{bazerque2013nonparametric}, kernel graph learning
(KGL)~\cite{pu2021kernel} assumes the loss
$\mathcal{L}(\vect{X}, \vect{G}) = \norm{ \vect{X} -
  \vect{K}_{\mathcal{Z}} \vect{G} \vect{K}_{\mathcal{Y}}
}^2_{\textnormal{F}}$, which yields~\eqref{eq:NBP.approx} by
decomposing $\vect{G}$ as the product $\vect{BC}$ of
low-rank matrices. Moreover, kernel regression over graphs
(KRG)~\cite{venkitaraman2019predicting} adopts the loss
$\mathcal{L}(\vect{X}, \vect{H}) = \norm{ \vect{X} -
  \vect{H} \vect{K}_{\mathcal{Y}} }^2_{\textnormal{F}}$,
which yields NBP and KGL by decomposing $\vect{H}$ into
$\vect{K}_{\mathcal{Z}} \vect{BC}$ and
$\vect{K}_{\mathcal{Z}} \vect{G}$, respectively.  To avoid
alternating minimization for solving for the bilinear
\eqref{eq:NBP.approx}, \cite{gimenez2019matrix} proposed the
loss
$\mathcal{L}(\vect{X}, \bm{\gamma}) = \norm{
  \tovec(\vect{X}) - \vect{K}_\otimes \bm{\gamma}
}^2_{\textnormal{F}}$, where $\vect{K}_\otimes$ is the
Kronecker product
$\vect{K}_\mathcal{Y}\otimes \vect{K}_\mathcal{Z}$ and
$\tovec(\vect{X})$ is the vectorization of
$\vect{X}$. Although~\cite{gimenez2019matrix} offers an
efficient closed-form solution, its data modeling lacks
low-rankness and bilinearity. Along the lines of
KRG~\cite{venkitaraman2019predicting} and motivated by
LLE~\cite{roweis2000nonlinear}, \cite{carreira2011manifold}
introduces
$\mathcal{L}(\vect{X}, \vect{H}, \vect{F}, \vect{Q}) =
\norm{\vect{X} - \vect{H} \vect{K}_{F}^2}_{\textnormal{F}} +
\norm{ \vect{F} - \vect{Q} \vect{K}_{\mathcal{Y}}
}^2_{\textnormal{F}}$, where the kernel matrix
$\vect{K}_{F}$ is defined by the low-dimensional
representation $\vect{F}$ of $\vect{X}$. This method
requires two optimization steps per iteration, first on
$(\vect{H}, \vect{Q})$, and then on $(\vect{X}, \vect{F})$.

Notice that kernel matrices
$\vect{K}_{\mathcal{Z}}, \vect{K}_{\mathcal{Y}}$ are square,
with their dimensions scaling with the dimensions of
$\vect{X}$. This causes the scaling of the dimensions of the
parameter matrices $\vect{B}, \vect{C}, \vect{G}$, and
$\vect{H}$ with the dimensions of the data, which is
characteristic of nonparametric methods, but at the same
time raises significant computational obstacles in cases of
high-dimensional data. In fact, the CPU setting, used in
this study for numerical validations, was not able to carry
through the NBP computations for the high-dimensional data
of \cref{sec:dmri}. Moreover, all approximating functions
$\{ f_k\}$ and $\{ g_k \}$ in NBP, KGL, and KRG are only
required to belong to some RKHSs, without any consideration
of more elaborate geometric and ManL arguments.

\section{Data Modeling by MultiL-KRIM}\label{sec:modeling}

\begin{figure*}[!ht]
  \centering
  \begin{tikzpicture}
    \small
    % Create nodes for boxes and add labels directly
    % \node [block] (step1) {
    %   Step 1 \\
    %   Define matrix $\vect{Y}$ and sampling operator $\mathscr{S}_{\Omega}$
    % };
    \node [block] (step2) {
      \textbf{Step~1} \\
      Identify navigator data $\check{\vect{Y}}_{\textnormal{nav}}$
    };
    \node [block, right=.3cm of step2] (step3) {
      \textbf{Step~2} \\
      Select landmark points $\check{\vect{L}}$
    };
    \node [block, right=.3cm of step3] (step4) {
      \textbf{Step~3} \\
      Compute kernel matrix $\vectcal{K}$ in~\eqref{app:eq.kernel.supermatrix} (supplementary file)
    };
    \node [block, right=.3cm of step4] (step5) {
      \textbf{Step~4} \\
      Design and solve the inverse problem~\eqref{eq:multil.manifold.right}
    };

    % Create arrows
    % \draw[->] (step1) -- (step2);
    \draw[->] (step2) -- (step3);
    \draw[->] (step3) -- (step4);
    \draw[->] (step4) -- (step5);
  \end{tikzpicture}
  \caption{Pipeline of MultiL-KRIM}\label{fig:pipeline}
\end{figure*}
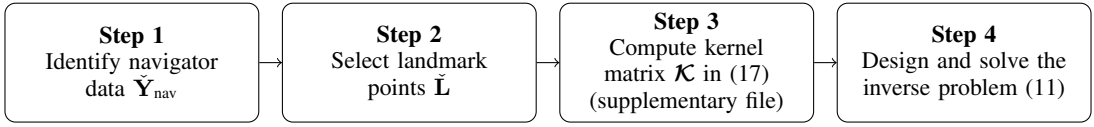

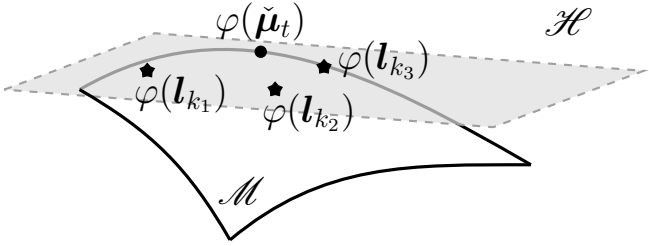
\begin{figure}[t!]
  \centering
  \begin{tikzpicture}

    \draw [very thick, black] (-3,1) to [out=30, in=180-30]
    coordinate[pos=.4] (CenterPoint) (3,0);
    \path (CenterPoint);
    \pgfgetlastxy{\Cx}{\Cy};

    \draw[thick, dashed, color=gray!75, fill=gray!25, fill
    opacity=0.7] (-4,1) -- (-2,1.75) -- (4.5,0.75 + 0.5) --
    (2.5,0.5) -- cycle;

    \node (LandA) at (-1.9,1) {};
    \node[draw,fill,star,star points=5,scale=0.4] at (-2.1,1.25) {};
    \node[text opacity=1, font = \Large] at
    ([shift={(0.25,-0.05)}] LandA) {$\varphi (\mathbfit{l}_{k_1})$};

    \node (LandB) at (-0.2,0.7) {};
    \node[draw,fill,star,star points=5,scale=0.4] at (-0.4,1) {};
    \node[text opacity=1, font = \Large] at
    ([shift={(0.25,0)}] LandB) {$\varphi (\mathbfit{l}_{k_2})$};

    \node (LandC) at (0.55,1.4) {};
    \node[draw,fill,star,star points=5,scale=0.4] at (0.25,1.3) {};
    \node[text opacity=1, font = \Large] at
    ([shift={(0.5,0)}] LandC) {$\varphi (\mathbfit{l}_{k_3})$};

    \draw[very thick, black] (-3,1) to [out=-30, in=90+30] (-1,-1);
    \draw[very thick, black] (-1,-1) to [out=40, in=180] (3,0);

    \node [font = \Large] at ([shift={(0,0.4)}] CenterPoint)
    {$\varphi ( \check{\bm{\mu}}_t )$}; \draw [fill] (\Cx,\Cy) circle(.075);

    \node [font = \Large] at (-0.85,-.35) {$\mathscr{M}$};
    \node (v) at (.4,.9) {};

    \node [font = \Large] at (3.5,1.9) {$\mathscr{H}$};

  \end{tikzpicture}
  \caption{The manifold-learning (ManL) modeling assumption:
    points
    $\Set{ \varphi (\mathbfit{l}_{k})
    }_{k=1}^{N_{\mathit{l}}}$ lie into or close to an
    unknown-to-the-user smooth manifold $\mathscr{M}$ which
    is embedded into an ambient RKHS $\mathscr{H}$. The
    ``collaborative-filtering'' modeling assumption: only a
    few points,
    $\Set{ \varphi (\mathbfit{l}_{k_n}) }_{n=1}^3$ here,
    collaborate \textit{affinely}\/ to approximate
    $\varphi (\check{\bm{\mu}}_t)$. The small number of
    ``collaborating neighbors'' implies a low-dimensional
    structure. All affine combinations of
    $\Set{ \varphi (\mathbfit{l}_{k_n})}_{n=1}^3$ define the
    approximating ``linear patch'' (affine hull,
    gray-colored plane), which mimics the concept of a
    tangent space to
    $\mathscr{M}$.} \label{fig:manifold.kernel.space}
\end{figure}

As a first step (see \cref{fig:pipeline}), MultiL-KRIM
extracts geometric information from a subset of the observed
data, coined ``navigator data'' and denoted hereafter as
$\check{\vect{Y}}_{\textnormal{nav}} \coloneqq [
\check{\vect{y}}_1^{\textnormal{nav}}, \ldots,
\check{\vect{y}}_{N_\text{nav}}^{\textnormal{nav}}] \in
\Complex^{\nu \times N_\text{nav}}$, for some
$\nu, N_\text{nav}\in \IntegerPP$. Their specific definition
depends on the application domain. For example, navigator
data can be vectorized patches from the heavily sampled
region of an image, or, in a network with time-series data,
they can be samples of time-series collected from a subset
of the network nodes. This approach of extracting geometric
information from the data themselves, without imposing any
user-defined statistical priors, and incorporating the
extracted information into MultiL-KRIM's inverse problem
differs from ``blind'' decomposition methods,
\eg~\cite{davenport2016overview, chi2019nonconvex,
  zhou2017tensor, cai2019nonconvex, mairal2008dictionary,
  tovsic2011dictionary, sun2016complete1, sun2016complete2}
(see also \cref{sec:intro}), where no geometric information
from the data is infused into their inverse problems.

As the number of navigator data points $N_\text{nav}$ grows
for large datasets, to avoid computational complexity
issues, a subset
$\{ \mathbfit{l}_k \}_{k = 1}^{N_{\mathit{l}}}$, coined
\textit{landmark/representative}\/ points, with
$N_{\mathit{l}} \leq N_{\text{nav}}$, is selected from
$\{ \check{\vect{y}}^{\textnormal{nav}}_k \}_{k =
  1}^{N_{\text{nav}}}$ (see \cref{fig:pipeline}). To avoid
interrupting the flow of the discussion, several strategies
to form navigator data $\check{\vect{Y}}_{\textnormal{nav}}$
and to select landmark points
$\{ \mathbfit{l}_k \}_{k=1}^{N_{\mathit{l}}}$ are deferred
to \cref{sec:graph.exp}. Let
$\check{\vect{L}} \coloneqq [ \mathbfit{l}_1,
\mathbfit{l}_2, \ldots, \mathbfit{l}_{N_{\mathit{l}}}] \in
\Complex^{\nu \times N_{\mathit{l}}}$. Motivated by ManL,
this study assumes that there is some latent geometry hidden
within
$\{ \check{\vect{y}}^{\textnormal{nav}}_k \}_{k =
  1}^{N_{\text{nav}}}$ and consequently
$\{ \mathbfit{l}_k \}_{k=1}^{N_{\mathit{l}}}$, which
manifests itself in an appropriately user-defined feature
space $\mathscr{H}$. Here, $\mathscr{H}$ is taken to be an
RKHS, associated with a reproducing kernel
$\kappa(\cdot, \cdot) \colon \Complex^{\nu} \times
\Complex^{\nu} \to \Real$, and a feature mapping
$\varphi\colon \Complex^{\nu} \to \mathscr{H}$ (see
\cref{app:RKHS} in the supplementary file). By a slight
abuse of notation, let
$\bm{\Phi}( \check{\vect{L}} ) \coloneqq [
\varphi(\mathbfit{l}_1), \ldots, \varphi(
\mathbfit{l}_{N_{\mathit{l}}}) ]$, where
$\varphi(\mathbfit{l}_k) \in \mathscr{H}$,
$\forall k\in \Set{1, \ldots, N_{\mathit{l}}}$ (see
\cref{fig:manifold.kernel.space}).

The $(i,t)$th entry $x_{it}$ of $\vect{X}$ is modeled as
\begin{alignat}{2}
  [\vect{X}]_{it} = x_{it}
  & {} \approx {} && f_i( \check{\bm{\mu}}_{t}
                     ) = \innerp{f_i}{ \varphi(
                     \check{\bm{\mu}}_{t})}_{\mathscr{H}}
                     \,, \label{xij.reproducing.property}
\end{alignat}
where $f_i(\cdot): \Complex^{\nu} \to \Complex$ is an
unknown non-linear function in the RKHS functional space
$\mathscr{H}$, $\check{\bm{\mu}}_{t}$ is an unknown vector
in $\Complex^{\nu}$, and the latter part
of~\eqref{xij.reproducing.property} is because of the
reproducing property of
$\mathscr{H}$~\cite{aronszajn1950theory}. It is assumed that
$f_i$ and $\varphi( \check{\bm{\mu}}_{t})$ belong to the
linear span of
$\Set{ \varphi( \mathbfit{l}_k) }_{k=1}^{ N_{\mathit{l}} }$,
\ie, there exist complex-valued $N_{\mathit{l}}\times 1$
vectors
$\vect{d}_i \coloneqq [ d_{i1}, \ldots,
d_{iN_{\mathit{l}}}]^{\intercal}$ and
$\vect{b}_t \coloneqq [ b_{1t}, \ldots,
b_{N_{\mathit{l}}t}]^{\intercal}$ s.t.\
\begin{subequations}\label{subspace.assumptions}
  \begin{align}
    f_i & = \sum\nolimits_{k=1}^{N_{\mathit{l}}} d_{ik}
          \varphi( \mathbfit{l}_k ) = \bm{\Phi}(
          \check{\vect{L}} ) \vect{d}_i \,, \\
    \varphi( \check{\bm{\mu}}_{t})
        & = \sum\nolimits_{k=1}^{N_{\mathit{l}}} b_{kt} \varphi( \mathbfit{l}_k )
          = \bm{\Phi}(\check{\vect{L}}) \vect{b}_t \,. \label{eq:model.basis}
  \end{align}
\end{subequations}
Altogether, by back-substitutions
into~\eqref{xij.reproducing.property},
\begin{align}
    x_{it} \approx \innerp{\bm{\Phi}( \check{\vect{L}} ) \vect{d}_i} {\bm{\Phi}(\check{\vect{L}})
  \vect{b}_t}_{\mathscr{H}} = \vect{d}_i^{\hermconj} \vect{K} \vect{b}_t \,,
\end{align}
where $\vect{K}$ is the complex-valued
$N_{\mathit{l}} \times N_{\mathit{l}}$ matrix whose
$(k, k^{\prime})$-th entry is equal to
$\innerp{\varphi(\mathbfit{l}_k)}
{\varphi(\mathbfit{l}_{k^{\prime}})}_{\mathscr{H}} =
\kappa(\mathbfit{l}_k, \mathbfit{l}_{k^{\prime}})$, and
superscript $\hermconj$ stands for complex conjugate
transposition. To offer a simple data model and compact
notations, if
$\vect{D} \coloneqq [ \vect{d}_1, \ldots, \vect{d}_{I_0}
]^{{\hermconj}} \in \Complex^{ I_0 \times N_{\mathit{l}}}$
and
$\vect{B} \coloneqq [ \vect{b}_1, \ldots, \vect{b}_{I_N} ]
\in \Complex^{ N_{\mathit{l}} \times I_N }$ are the
parameters that need to be identified, then data are modeled
as the bilinear form
\begin{align}
  \vect{X} \approx \vect{D} \vect{K} \vect{B}
  \,. \label{KRIM.step1}
\end{align}

So far, \eqref{subspace.assumptions} follow the standard
subspace-modeling assumption of kernel methods, met also in
the prior nonparametric designs
NBP~\cite{bazerque2013nonparametric},
KRG~\cite{venkitaraman2019predicting} and
KGL~\cite{pu2021kernel} (see also the discussion after
\eqref{eq:NBP.approx}): both $f_i$ and
$\varphi( \check{\bm{\mu}}_{t} )$ are assumed to belong to a
linear subspace spanned by
$\Set{ \varphi (\mathbfit{l}_{k})
}_{k=1}^{N_{\mathit{l}}}$. However, inspired by ManL
and~\cite{shetty2020bilmdm, slavakis2022krim}, and in quest
for low-dimensional structures, MultiL-KRIM takes the
standard kernel modeling assumptions a few steps further:
points
$\Set{ \varphi (\mathbfit{l}_{k}) }_{k=1}^{N_{\mathit{l}}}$
are assumed to lie into or close to an unknown-to-the-user
smooth manifold $\mathscr{M}$ which is embedded into an
ambient RKHS $\mathscr{H}$ (see
\cref{fig:manifold.kernel.space}). Furthermore, a
``collaborative-filtering'' modeling flavor is added to the
design: only a few points within
$\Set{ \varphi (\mathbfit{l}_{k}) }_{k=1}^{N_{\mathit{l}}}$
($\Set{ \varphi (\mathbfit{l}_{k_n}) }_{n=1}^3$ in
\cref{fig:manifold.kernel.space}) collaborate
\textit{affinely}\/ to approximate
$\varphi (\check{\bm{\mu}}_t)$. Affine combinations,
expressed via
$\vect{1}_{N_{\mathit{l}}}^{{\hermconj}} \vect{b}_t = 1$,
where $\vect{1}_{N_{\mathit{l}}}$ is the
$N_{\mathit{l}} \times 1$ all-one vector, are motivated by
the fundamental concept of tangent spaces (``linear
patches'', simply put) to smooth
manifolds~\cite{RobbinSalamon:22}. Moreover, the assumption
that only a small number of ``collaborating neighbors'' is
sufficient to approximate $\varphi (\check{\bm{\mu}}_t)$,
expressed explicitly by considering a sparse vector
$\vect{b}_t$, is motivated by the desire to identify
low-dimensional structures within
$\Set{ \varphi (\mathbfit{l}_{k}) }_{k=1}^{N_{\mathit{l}}}$
and by the success of sparse approximations in
low-dimensional data modeling.

Choosing $\kappa(\cdot, \cdot)$ to define $\vect{K}$
requires the intensive tasks of cross validation and fine
tuning. A popular and straightforward way to overcome such
tasks and to generalize~\eqref{KRIM.step1}, implemented also
in~\cite{slavakis2022krim}, is via multiple kernels:
\begin{align}
  \vect{X} \approx \sum\nolimits_{m=1}^M \vect{D}_m
  \vect{K}_m \vect{B}_m \,, \label{KRIM.step2}
\end{align}
with user-defined reproducing kernels
$\Set{ \kappa_m(\cdot, \cdot) }_{m=1}^M$, and thus kernel
matrices $\Set{ \vect{K}_m }_{m=1}^M$, complex-valued
$I_0 \times N_{\mathit{l}}$ matrices
$\Set{ \vect{D}_m }_{m=1}^M$, and
$N_{\mathit{l}} \times I_N$ matrices
$\Set{ \vect{B}_m }_{m=1}^M$.

The larger the cardinality $N_\text{nav}$ of the navigator
data, the larger the number $N_{\mathit{l}}$ of landmark
points may become, justifying the nonparametric character of
the present design. Likewise, the larger the dimensionality
$I_0$ of the data $\vect{Y}$, the larger the number of rows
of $\vect{D}_m$s in~\eqref{KRIM.step2}. In other words, for
large and high-dimensional datasets (see \cref{sec:dmri}),
the size $I_0 \times N_{\mathit{l}}$ of $\vect{D}_m$s may
become unbearably large. To reduce computational complexity
and effect low-rank structure on the approximation,
KRIM~\cite{slavakis2022krim} compressed each
$N_{\mathit{l}} \times N_{\mathit{l}}$ matrix $\vect{K}_m$
into a low-dimensional $d \times N_{\mathit{l}}$
($d\ll N_{\mathit{l}}$) rendition $\check{\vect{K}}_m$ to
yield the model
$\vect{X} \approx \sum_{m=1}^M \check{\vect{D}}_m
\check{\vect{K}}_m \vect{B}_m$, where
$\Set{ \check{\vect{D}}_m }_{m=1}^M$ are now low-rank
$I_0 \times d$ matrices. Matrix $\check{\vect{K}}_m$ was
computed from $\vect{K}_m$ via a dimensionality-reduction
module (pre-step) in~\cite{slavakis2022krim}, inspired by
LLE~\cite{roweis2000nonlinear}. However, such a pre-step
inflicts the following drawbacks on KRIM's design:
\begin{enumerate*}[label=\textbf{(\roman*)}]

\item with large numbers $M$ of kernels and $N_{\mathit{l}}$
  of landmark points, the LLE-inspired pre-step raises
  computational obstacles, while fine-tuning its
  hyperparameters consumes extra effort; and

\item if not carefully fine-tuned, the error from
  compressing $\vect{K}_m$ into $\check{\vect{K}}_m$ may
  propagate to the next stage of KRIM's pipeline.
\end{enumerate*}

% \begin{figure*}[!t]
%   \begin{tikzpicture}
%     \node[text width=\textwidth] (compact) {
%       \begin{align}
%         \vect{X} \approx
%         \underbrace{
%         \left[\begin{smallmatrix}
%           \vect{D}_1^{(1)} & \vect{D}_2^{(1)} & \ldots & \vect{D}_M^{(1)}
%         \end{smallmatrix}\right]
%         }_{\vectcal{D}_1}
%         \underbrace{\left[\begin{smallmatrix}
%           \vect{D}_1^{(2)} & 0 & \cdots & 0 \\
%           0 & \vect{D}_2^{(2)} & \cdots & 0 \\
%           \vdots & \vdots & \ddots & \vdots \\
%           0 & 0 & \cdots & \vect{D}_M^{(2)} \\
%         \end{smallmatrix}\right]}_{\vectcal{D}_2}
%         \ldots
%         \underbrace{\left[\begin{smallmatrix}
%           \vect{D}_1^{(Q)} & 0 & \cdots & 0 \\
%           0 & \vect{D}_2^{(Q)} & \cdots & 0 \\
%           \vdots & \vdots & \ddots & \vdots \\
%           0 & 0 & \cdots & \vect{D}_M^{(Q)} \\
%         \end{smallmatrix}\right]}_{\vectcal{D}_Q}
%         \underbrace{\left[\begin{smallmatrix}
%           \vect{K}_1 & 0 & \cdots & 0 \\
%           0 & \vect{K}_2 & \cdots & 0 \\
%           \vdots & \vdots & \ddots & \vdots \\
%           0 & 0 & \cdots & \vect{K}_M \\
%         \end{smallmatrix}\right]}_{\vectcal{K}}
%         \underbrace{\left[\begin{smallmatrix}
%           \vect{B}_1 \\
%           \vect{B}_2 \\
%           \vdots \\
%           \vect{B}_M
%         \end{smallmatrix}\right]}_{\vectcal{B}} \,.
%         \label{eq:fig.multil.compact}
%       \end{align}
%     };
%   \end{tikzpicture}
% \end{figure*}

The present MultiL-KRIM avoids the previous KRIM's drawbacks
by the following multilinear factorization,
\renewcommand\matscale{.5}
\begin{align}
  \vect{X}
  & \approx \sum\nolimits_{m=1}^M \vect{D}_m^{(1)}
    \vect{D}_m^{(2)} \cdots \vect{D}_m^{(Q)} \vect{K}_m
    \vect{B}_m \,, \label{eq:multi.kernel.nodimred} \\[15pt]
  \matbox{9}{6}{I_0}{I_N}{\vect{X}}{tableSingle}
  & \approx \sum_{m=1}^{M} {\matbox{9}{3}{I_0}{d_1}{\vect{D}_m^{(1)}}{tableSingle}} \
    \cdots
    \raiserows{3.25}{\matbox{2.5}{4}{d_{Q-1}}
    {N_{\mathit{l}}}{\vect{D}_m^{(Q)}}{tableSingle}}
    \raiserows{2.5}{\matbox{4}{4}{N_{\mathit{l}}}
    {N_{\mathit{l}}}{\vect{K}_m}{white}}
    \raiserows{2.5}{\matbox{4}{6}{N_{\mathit{l}}}
    {I_N}{\vect{B}_m}{tableSingle}} \notag
\end{align}
where $\vect{D}_m^{(q)} \in \Complex^{d_{q-1}\times d_q}$,
$d_0 \coloneqq I_0$, $d_Q \coloneqq N_{\mathit{l}}$, and the
inner matrix dimensions $\Set{ d_q }_{q=1}^{Q-1}$ are user
defined. Notice that for $Q=2$, the term
$\vect{D}_m^{(2)} \vect{K}_m$ may be seen as the
low-dimensional $\check{\vect{K}}_m$
in~\cite{slavakis2022krim}. Nonetheless,
$\Set{\vect{D}_m^{(q)}}_{(q, m)}$ are identified during a
single-stage learning task, avoiding any KRIM-like pre-steps
with their hyperparameter tuning and errors. It is worth
stressing here that in the special case where $M=1$,
$\vect{K}_1 \coloneqq \vect{I}_{N_{\mathit{l}}}$, with
$\vect{I}_{N_{\mathit{l}}}$ denoting the identity matrix,
and with no affine and sparsity constraints on $\vect{B}_1$,
\eqref{eq:multi.kernel.nodimred} yields the
latent-geometry-agnostic
MMF~\cite{cichocki2007multilayer}. See \cref{sec:tests.tvgs}
for numerical tests of MultiL-KRIM vs.\ MMF on real-world
data.

It can be verified that $Q>1$ in
\eqref{eq:multi.kernel.nodimred} may offer substantial
computational savings compared to the $Q = 1$ case.  Indeed,
the number of unknowns to be identified in
\eqref{eq:multi.kernel.nodimred} for $Q > 1$ is
\begin{align}
    N_{Q>1} = M \left( \sum\nolimits_{q=1}^Q d_{q-1}d_q +
  I_N N_{\mathit{l}} \right) \,, \label{eq:N.Q>1}
\end{align}
as opposed to
\begin{align}
    N_{Q=1} = M ( I_0N_{\mathit{l}} + I_N N_{\mathit{l}} ) \label{eq:N.Q=1}
\end{align}
in the $Q = 1$ case. If $I_0$ and $N_{\mathit{l}}$ are
large, then $\Set{ d_q }_{q=1}^{Q-1}$ can be chosen so that
$N_{Q>1} \ll N_{Q=1}$ (see \cref{app:tab.N.Q} (supplementary
file)).
% can be chosen so that $N_{Q>1} \ll N_{Q=1}$ (see \cref{tab:N.Q}).

By constructing the ``supermatrices''
$\Set{\vectcal{D}_q}_{q=1}^Q\,, \vectcal{K}$, and
$\vectcal{B}$, which are described in
\cref{app:modeling.compact},
% \begin{alignat}{2}
%   \vectcal{D}_1
%   & {} \coloneqq {} [\vect{D}_1^{(1)}, \vect{D}_2^{(1)}, \ldots, \vect{D}_M^{(1)}]
%   && \in \Complex^{I_0 \times d_1M} \,, \notag\\
%   \vectcal{D}_q
%   & \coloneqq \bdiag{(\vect{D}_1^{(q)}, \vect{D}_2^{(q)}, \ldots, \vect{D}_M^{(q)})}
%   && \in \Complex^{d_{q-1}M \times d_{q}M} \,, \notag\\
%   & \hphantom{{} \coloneqq {}} \forall q\in \{ 2, \ldots, Q\}\,, && \notag\\
%   \vectcal{K}
%   & \coloneqq \bdiag{(\vect{K}_1, \vect{K}_2, \ldots, \vect{K}_M)}
%   && \in \Complex^{MN_{\mathit{l}} \times MN_{\mathit{l}}} \,, \label{kernel.supermatrix}\\
%   \vectcal{B}
%   & \coloneqq [\vect{B}_1^{\hermconj}, \vect{B}_2^{\hermconj}, \ldots,
%     \vect{B}_{M}^{\hermconj}]^{\hermconj}
%   && \in \Complex^{MN_{\mathit{l}} \times I_N} \,, \notag
% \end{alignat}
% where $\bdiag$ denotes a block-diagonal structure, 
the previous considerations yield the compact data
% model~\eqref{eq:fig.multil.compact}.
model~\eqref{app:eq.fig.multil.compact} (supplementary
file).  Under the previous definitions, the generic
MultiL-KRIM inverse problem becomes%
\begin{subequations}\label{eq:multil.manifold.right}%
  \begin{align}
    \min_{ (\vect{X}, \Set{\vectcal{D}_q}_{q=1}^Q, \vectcal{B}) }
    {}\ & {} \tfrac{1}{2} \norm{ \vect{X} - \vectcal{D}_1
          \vectcal{D}_2 \cdots \vectcal{D}_Q \vectcal{K}
          \vectcal{B} }^2_{\textnormal{F}} \notag \\
        & + \underbrace{\lambda_1 \norm{\vectcal{B}}_1 +
          \mathcal{R}_2(\vect{X},
          \Set{\vectcal{D}_q}_{q=1}^Q)}_{ \mathcal{R}
          (\vect{X}, \Set{\vectcal{D}_q}_{q=1}^Q,
          \vectcal{B}) } \label{generic.inv.problem.loss} \\
    \text{s.to} {}\
        & {} \mathscr{S}_{\Omega}(\vect{Y}) =
          \mathscr{S}_{\Omega} \mathscr{T} (\vect{X})
          \,, \label{generic.inv.problem.consistency} \\
        & {} \vect{1}_{N_{\mathit{l}}}^{{\hermconj}}
          \vect{B}_m = \vect{1}_{I_N}^{{\hermconj}}\,,
          \forall m \in \{1, \ldots,
          M\}\,, \label{generic.inv.problem.affine} \\
        & \text{and other
          constraints,} \label{generic.inv.problem.other}
  \end{align}%
\end{subequations}%
where the $\ell_1$-norm
$\mathcal{R}_1(\cdot) \coloneqq \lambda_1 \norm{\cdot}_1$,
used here to impose sparsity on $\vectcal{B}$, and
$\mathcal{R}_2(\cdot)$, used to enforce prior information on
the model, constitute the generic regularizer
$\mathcal{R} \coloneqq \mathcal{R}_1 + \mathcal{R}_2$
of~\eqref{IbyR.loss}. Constraint~\eqref{generic.inv.problem.consistency}
states the requirement for data consistency over the index
set $\Omega$ of observed data values,
\eqref{generic.inv.problem.affine} stands for the
aforementioned ManL-inspired affine combinations,
and~\eqref{generic.inv.problem.other} is used to
conveniently include any potential application-domain
specific
constraint.% Furthermore, for factorization schemes such
% as MMF, NBP, KGL and KRG, which are described in \cref{sec:problem.formulation}, their
% inverse problems may follow the generic inverse problem of MultiL-KRIM in
% \eqref{eq:multil.manifold.right}.

The pipeline of MultiL-KRIM framework is illustrated in
\cref{fig:pipeline}. To validate the proposed framework, the
following sections will demonstrate two important
applications, will provide solutions to specific instances
of the generic inverse
problem~\eqref{eq:multil.manifold.right}, and will leverage
extensive numerical tests to take a thorough look at the
pipeline of \cref{fig:pipeline}.

%%%%%%%%%%%%%%%%%%%%%%%%%%%%%%%%%%%%%%%%%%%%%%%%%%%%%%%%%%%%%%%%%%%%%%%%%%%%%%%%%%%%%%%%%%%%%%%

\section{Application: Time-Varying Graph-Signal Recovery}\label{sec:graph}

This section begins with an overview of the problem of
time-varying graph-signal (TVGS) recovery, followed by a
short review of several state-of-the-art prior works in
\cref{sec:graph.context}. \cref{sec:graph.inv} formulates
and provides solutions to a special instance of the
MultiL-KRIM's inverse
problem~\eqref{eq:multil.manifold.right}. Finally,
\cref{sec:graph.exp} compares empirically MultiL-KRIM with
several state-of-the-art methods, and investigates several
variations of MultiL-KRIM.

\subsection{TVGS recovery}\label{sec:graph.context}

Graph signal processing (GSP) is an emerging field at the
intersection of graph theory and signal
processing~\cite{ortega2018graph, hu2021graph}. A graph is
an abstract mathematical object equipped with nodes/vertices
and edges, where nodes correspond to entities such as
sensors of a physical network or data points, and edges
represent the connections/relationships between these
entities. In GSP, each node is annotated with a
(time-varying) signal, so that all signals gathered across
all nodes constitute the so-called graph signal.

To be more specific, a graph is denoted by
$G = (\mathcal{V}, \mathcal{E})$, where
$\mathcal{V} \coloneqq \Set{1, 2, \ldots, I_0}$ represents
the set of $I_0$ nodes, and
$\mathcal{E} \subseteq \mathcal{V} \times \mathcal{V}$ is
the set of edges. The topology of the graph is described by
the adjacency matrix $\vect{W}\in \Real^{I_0 \times I_0}$,
whose $(i,j)$th entry $w_{ij}\in \RealP$ denotes the edge
weight between nodes $i$ and $j$. In this study, $G$ is
considered to be undirected, thus, $\vect{W}$ is
symmetric. The graph Laplacian $I_0 \times I_0$ matrix is
defined as
$\vect{L} \coloneqq \diag (\vect{W} \vect{1}_{I_0}) -
\vect{W}$, where $\vect{1}_{I_0}$ denotes the $I_0 \times 1$
all-one vector, and $\diag(\vect{W} \vect{1}_{I_0})$ stands
for the $I_0 \times I_0$ diagonal matrix whose diagonal
entries are the corresponding entries of the $I_0\times 1$
vector $\vect{W} \vect{1}_{I_0}$. The graph signal is
denoted by
$\vect{Y} = [\vect{y}_1, \vect{y}_2, \ldots, \vect{y}_{I_N}]
\in \Real^{I_0\times I_N}$, where the columns
$\mathcal{Y} \coloneqq \Set{\vect{y}_t}_{t=1}^{I_N}$ are the
graph-signal ``snapshots'' at $I_N$ time points, while rows
$\mathcal{Z}\coloneqq\Set{\vect{z}_i}_{i=1}^{I_0}$ of
$\vect{Y}$ are the time profiles of the $I_0$ signals which
annotate the nodes of the graph.

The analysis of graph signals is pivotal in fields such as
recommender systems~\cite{huang2018rating}, power
grids~\cite{ramakrishna2021grid},
sensor~\cite{jablonski2017graph},
urban~\cite{laharotte2014spatiotemporal}, and vehicular
networks~\cite{placzek2012selective}. Due to reasons like
users' privacy, sensors fault, or resources conservation,
such application domains often encounter the problem of
missing data. Naturally, there have been numerous studies to
address the problem, most of which leverage low-rank and
smoothness assumptions on $\vect{Y}$. Typically, it is
assumed that $\vect{Y}$ is low rank due to dependencies
among time points and/or nodes~\cite{cheng2012stcdg}. On the
other hand, there are studies which adopt the graph
Laplacian operator $\vect{L}$~\cite{ortega2018graph} to
impose graph-wide (a.k.a.\ spatial) smoothness, \ie,
neighboring nodes share similar signal
attributes~\cite{chen2015signal, chen2016signal}. For
example, the regularization function $\mathcal{R}(\cdot)$ of
both KGL and KRG (see \cref{sec:problem.formulation})
comprises a classical Tikhonov regularizer and an
$\vect{L}$-based quadratic form. Instead, NBP uses only
Tikhonov regularization for $\mathcal{R}(\cdot)$. Although
KGL and KRG jointly recover $\vect{Y}$ and the underlying
graph topology $\vect{W}$ or $\vect{L}$, since this study
emphasizes on TVGS recovery, $\vect{L}$ is assumed to be
given and fixed in the sequel. Another kernel-based
regression method can be also found
in~\cite{romero2017kernel}, where kernel ridge regression is
used to develop a kernel Kalman-filtering (KKF) algorithm
for TVGS recovery, with an $\vect{L}$-based regularizer
$\mathcal{R}(\cdot)$. Furthermore, \cite{qiu2017time,
  mao2018spatio} introduce differential smoothness and
impose low-rank structure by nuclear-norm
regularizers. Low-rank-cognizant matrix factorizations have
been also studied extensively~\cite{roughan2011spatio,
  kong2013data, xie2016recover, song2020novel,
  lei2022bayesian}.

By using the one-step difference operator
$\mathbf{\Delta}$~\cite{roughan2011spatio},
% \begin{align*}
%   \mathbf{\Delta} \coloneqq \left[\begin{smallmatrix}
%     -1 & & \\
%     1 & \ddots & \\
%        & \ddots & -1 \\
%        & & 1 \\
%   \end{smallmatrix} \right]\,,
% \end{align*}
study~\cite{qiu2017time} introduces the
spatio-temporal-smoothness regularizer
$\tr(\mathbf{\Delta}^\intercal \vect{X}^\intercal \vect{L}
\vect{X} \mathbf{\Delta})$ in the place of
$\mathcal{R}(\cdot)$ in~\eqref{recovery.generic}, where
$\tr(\cdot)$ stands for the trace operator.  Specifically,
$\mathbf{\Delta}$ is an $I_N \times (I_N-1)$ matrix, s.t.\
$[\mathbf{\Delta}]_{i,i} = -1, [\mathbf{\Delta}]_{i+1,i} =
1$ for all $i=1, \ldots, I_N-1$, while other entries are
zeros.  Subsequently, \cite{mao2018spatio} considered an
additional low-rank constraint in the form of a nuclear-norm
regularizer, which was shown to outperform other popular
regularizers such as spatial
$\tr(\vect{X}^\intercal \vect{LX})$ and temporal
$\norm{\vect{X} \mathbf{\Delta} }_{\textnormal{F}}^2$
smoothness~\cite{piao2014correlated}. On the other hand,
\cite{giraldo2022reconstruction} introduces the following
Sobolev-smoothness regularizer
$\tr(\mathbf{\Delta}^{\intercal} \vect{X}^\intercal
(\vect{L} + \epsilon \vect{I}_{I_0})^{\beta} \vect{X}
\mathbf{\Delta})$, where $\epsilon, \beta\in \RealPP$ are
user-defined parameters. For convenience, denote
$\vect{L}_{\epsilon}^{\beta} \coloneqq (\vect{L} + \epsilon
\vect{I}_{I_0})^{\beta}$. Works~\cite{qiu2017time,
  mao2018spatio, giraldo2022reconstruction} use a ``soft''
data-consistency loss, \ie, the quadratic loss
$(1/2) \norm{\mathscr{S}_{\Omega} (\vect{X}) -
  \mathscr{S}_{\Omega} (\vect{Y})}_{\textnormal{F}}^2$ is
used instead of the ``hard'' constraint
$\mathscr{S}_{\Omega}(\vect{X}) =
\mathscr{S}_{\Omega}(\vect{Y})$.

The extensive numerical tests of the present study have
shown that the Sobolev-smoothness regularizer is currently
offering state-of-the-art performance in TVGS
recovery. Consequently, to ensure fairness during numerical
tests, \textit{all}\/ competing methods, as well as the
proposed MultiL-KRIM, will be equipped hereafter with the
Sobolev-smoothness regularizer.

\subsection{The inverse problem}\label{sec:graph.inv}

\begin{algorithm}[!t]
  \caption{Solving MultiL-KRIM's inverse problem}\label{alg:multil.general}
  \begin{algorithmic}[1]

    % \REQUIRE Under-sampled data $\mathscr{S}_{\Omega}(\vect{Y})$ and faithful data
    % $\check{\vect{Y}}_{\textnormal{nav}}$; learning rate $\gamma_0\in (0,1]$, decay rate $\zeta\in
    % (0,1)$ and relevant parameters;
    % % $\lambda_1, \lambda_2, \lambda_3, C_{A}, C_{G}, \tau_A, \tau_G, \tau_B \in
    % % \Real_{>0}$, $\zeta\in (0,1)$ and $\gamma_0\in (0,1]$;
    % $N_l$ landmark points as well as $M$ kernels and parameters for constructing kernel
    % matrices.

    \ENSURE Limit point $\hat{\vect{X}}^{(*)}$ of sequence
    $(\hat{\vect{X}}^{(n)})_{n\in\IntegerP}$.

    % $\Set{\hat{\tilde{\vect{A}}}_i^{*}}_{i=1}^Q$, and $\hat{\tilde{\vect{B}}}^*$.
    % of sequences $(\hat{\tilde{\vect{X}}}_n)_n$, $(\hat{\tilde{\vect{A}}}_n)_n$,
    % $(\hat{\tilde{\vect{G}}}_n)_n$ and $(\hat{\tilde{\vect{B}}}_n)_n$, respectively.

    % \STATE Identify landmark points
    % \STATE Compute $\tilde{\vect{K}}$

    \STATE Fix $\hat{\mathbfcal{O}}^{(0)}$, $\gamma_0\in
    (0,1]$, and $\zeta\in (0,1)$.

    \WHILE{$n\geq 0$} \label{alg.step:resume.k}

    \STATE {Available are
      % $( \Set{\hat{\tilde{\vect{A}}}_i^{(n)}}_{i=1}^Q, \hat{\tilde{\vect{B}}}^{(n)},
      % \hat{\vect{X}}^{(n)}, \hat{\vect{Z}}^{(n)}
      % )$
      $\hat{\vectcal{O}}^{(n)}$ (\eqref{Oh.tuple.tvgs} for
      TVGS or \eqref{Oh.tuple.dmri} for dMRI) and
      $\gamma_n$.}

    \STATE {$\gamma_{n+1} \coloneqq \gamma_n (1 - \zeta
      \gamma_n)$.}

    \STATE Solve sub-tasks \eqref{eq:graph.subtasks} for
    TVGS or \eqref{eq:dmri.subtasks} for dMRI.

    % \STATE {Obtain  $\hat{\tilde{\vect{A}}}_{n+1/2}$,
    % $\hat{\tilde{\vect{G}}}_{n+1/2}$ and $\hat{\tilde{\vect{B}}}_{n+1/2}$ by
    % \eqref{eq:task.min.A} to \eqref{eq:task.min.B}, respectively,
    % $\hat{\vect{X}}_{n+1/2}$ by \eqref{eq:task.min.X} and
    % \eqref{eq:stationary.X}, and $\hat{\vect{Z}}_{n+1/2}$ by \eqref{eq:task.Z}.
    % }\label{alg.step:convex.tasks.k}

    \STATE
    {$\hat{\vectcal{O}}^{(n+1)} \coloneqq \gamma_{n+1}
      \hat{\vectcal{O}}^{(n+1/2)} + (1-\gamma_{n+1})
      \hat{\vectcal{O}}^{(n)}$.}

    % \STATE {Update $\begin{aligned}[t] & (
    %   \Set{\hat{\tilde{\vect{A}}}_i^{(n+1)}}_{i=1}^Q, \hat{\tilde{\vect{B}}}^{(n+1)},
    %   \hat{\vect{X}}^{(n+1)}, \hat{\vect{Z}}^{(n+1)}
    %   )\\
    %   & \coloneqq \gamma_{n+1} ( \Set{\hat{\tilde{\vect{A}}}_i^{(n+1/2)}}_{i=1}^Q,
    %   \hat{\tilde{\vect{B}}}^{(n+1/2)}, \hat{\vect{X}}^{(n+1/2)},
    %   \hat{\vect{Z}}^{(n+1/2)}
    %   ) \\
    %   & \hphantom{\coloneqq\ } + (1-\gamma_{n+1}) (
    %   \Set{\hat{\tilde{\vect{A}}}_i^{(n)}}_{i=1}^Q, \hat{\tilde{\vect{B}}}^{(n)},
    %   \hat{\vect{X}}^{(n)}, \hat{\vect{Z}}^{(n)}
    %   ) \,.
    % \end{aligned}$}

    \STATE {Set $n \leftarrow n+1$ and go to step~\ref{alg.step:resume.k}.}

    \ENDWHILE
  \end{algorithmic}
\end{algorithm}

The inverse problem for TVGS recovery becomes a special
instance of the generic~\eqref{eq:multil.manifold.right}
one. More specifically,
\begin{subequations}\label{eq:graph.task.general}
  \begin{align}
    \min_{ (\vect{X}, \Set{\vectcal{D}_q}_{q=1}^Q, \vectcal{B}) }
    {}\ & {} \tfrac{1}{2} \norm{ \vect{X} - \vectcal{D}_1
          \vectcal{D}_2 \cdots \vectcal{D}_Q \vectcal{K}
          \vectcal{B} }^2_{\textnormal{F}} + \underbrace{
          \lambda_1 \norm{\vectcal{B}}_1 }_{
          \mathcal{R}_1(\vectcal{B}) } \notag \\
        & + \underbrace{\tfrac{\lambda_L}{2}
          \tr(\mathbf{\Delta}^\intercal \vect{X}^\intercal
          \vect{L}_{\epsilon}^{\beta} \vect{X}
          \mathbf{\Delta}) + \tfrac{\lambda_2}{2}
          \sum_{q=1}^Q
          \norm{\vectcal{D}_q}_{\textnormal{F}}^2}_{
          \mathcal{R}_2( \vect{X},
          \Set{\vectcal{D}_q}_{q=1}^Q )
          } \label{eq:graph.task.loss} \\
    \text{s.to} {}\
        & \mathscr{S}_{\Omega}(\vect{Y}) = \mathscr{S}_{\Omega}
          (\vect{X})\,, \label{graph.task.consistency} \\
        & \vect{1}_{N_{\mathit{l}}}^{{\hermconj}} \vect{B}_m
          = \vect{1}_{I_N}^{{\hermconj}}\,,
          \forall m \in \{1, \ldots,
          M\}\,, \label{graph.task.right}
  \end{align}
\end{subequations}
where the generic regularizer
in~\eqref{eq:multil.manifold.right} becomes here
$\mathcal{R}( \vect{X}, \Set{\vectcal{D}_q}_{q=1}^Q,
\vectcal{B} ) \coloneqq \mathcal{R}_1( \vectcal{B} ) +
\mathcal{R}_2 ( \vect{X}, \Set{\vectcal{D}_q}_{q=1}^Q )$. In
the current TVGS setting, the transform mapping
$\mathscr{T}$ in~\eqref{generic.inv.problem.consistency}
takes the form of the identity operator $\Id$. The loss
function is non-convex, and to guarantee convergence to a
critical point, the successive-convex-approximation
framework of~\cite{facchinei2015parallel} is utilized. The
algorithm is summarized in \cref{alg:multil.general}, where
the following tuple of estimates
$\forall n\in\IntegerP, \forall k\in\Set{0,1}$,
\begin{align}
    \hat{\mathbfcal{O}}^{(n+k/2)} \coloneqq (
      {} & {} \hat{\vect{X}}^{(n+k/2)},
           \hat{\vectcal{D}}_1^{(n+k/2)}, \notag \\
         & \ldots, \hat{\vectcal{D}}_Q^{(n+k/2)}, \hat{\vectcal{B}}^{(n+k/2)}
           )\,, \label{Oh.tuple.tvgs}
\end{align}
is recursively updated via the following convex sub-tasks
which need to be solved at every iteration $n$:
\begin{subequations}\label{eq:graph.subtasks}
  \begin{alignat}{3}
    && \hat{\vect{X}}^{(n + 1/2) }
    && \coloneqq \arg\min_{ {\vect{X}} } {}
    & {}\ \tfrac{1}{2} \norm{{\vect{X}} -
      \hat{\vectcal{D}}_1^{(n)} \cdots
      \hat{\vectcal{D}}_Q^{(n)} \vectcal{K}
      \hat{\vectcal{B}}^{(n)} }_{\textnormal{F}}^2
      \notag \\
    &&&&& {}\ + \tfrac{\lambda_L}{2}
          \tr(\mathbf{\Delta}^\intercal \vect{X}^\intercal
          \vect{L}_{\epsilon}^{\beta} \vect{X}
          \mathbf{\Delta}) \notag \\
    &&&&& {}\ + \tfrac{\tau_X}{2} \norm{
          \vect{X} - \hat{\vect{X}}^{(n)
          }}_{\textnormal{F}}^2 \notag \\
    &&&& \text{s.to} {} & {}\ \mathscr{S}_{\Omega}(\vect{Y})
                          = \mathscr{S}_{\Omega} (
                          {\vect{X}} )
                          \,, \label{eq:graph.min.X} \\
    && \hat{\vectcal{D}}_q^{(n + 1/2)} {}
    && {} \coloneqq \arg\min_{\vectcal{D}_q} {}
    & {}\ \tfrac{1}{2} \norm{ \hat{\vect{X}}^{(n)} - \hat{\vectcal{D}}_1^{(n)} \cdots
      \vectcal{D}_q \cdots \hat{\vectcal{D}}_Q^{(n)} \vectcal{K}
      \hat{\vectcal{B}}^{(n)} }_{\textnormal{F}}^2 \notag \\
    &&&& {} & {}\ + \tfrac{\lambda_2}{2}
              \norm{\vectcal{D}_q}_{\textnormal{F}}^2 +
              \tfrac{\tau_D}{2} \norm{ \vectcal{D}_q -
              \hat{\vectcal{D}}_q^{(n)}
              }_{\textnormal{F}}^2  \notag \\
    &&&& \text{s.to} {} & {}\ \vectcal{D}_q\ \textnormal{is
                          block diagonal}\ \forall q\in
                          \Set{2, \ldots, Q}
                          \,, \label{eq:graph.min.D} \\
    && \hat{\vectcal{B}}^{(n + 1/2)}
    && \coloneqq \arg \min_{\vectcal{B}} {}
    & {}\ \tfrac{1}{2} \norm{ \hat{\vect{X}}^{(n)}-
      \hat{\vectcal{D}}_1^{(n)} \cdots
      \hat{\vectcal{D}}_Q^{(n)} \vectcal{K} \vectcal{B} }_{\textnormal{F}}^2 \notag \\
    &&&&& {}\ + \lambda_1 \norm{\vectcal{B}}_1 +
          \tfrac{\tau_B}{2} \norm{ \vectcal{B} -
          \hat{\vectcal{B}}^{(n)} }_{\textnormal{F}}^2 \notag \\
    &&&& \text{s.to} {}
    & {}\ \vect{1}_{N_{\mathit{l}}}^{{\hermconj}} \vect{B}_m
      = \vect{1}_{I_N}^{{\hermconj}}\,,
          \forall m \in \{1, \ldots, M\}
      \,, \label{eq:graph.min.B}
  \end{alignat}
\end{subequations}
where the user-defined $\tau_X, \tau_D, \tau_B \in
\RealPP$. Sub-task~\eqref{eq:graph.min.B} is a composite
convex minimization task under affine constraints, and can
be thus solved iteratively by~\cite{slavakis2018fejer},
while~\eqref{eq:graph.min.X} and~\eqref{eq:graph.min.D} have
closed-form solutions (see \cref{app:TVGS.solve}
(supplementary file)).

\subsection{Experimental setting}\label{sec:graph.exp}

\begin{figure*}[ht]
  \centering
  \subfloat[(D1, P1) \label{fig:plot.d1p1}]
  {\includegraphics[height=135pt]{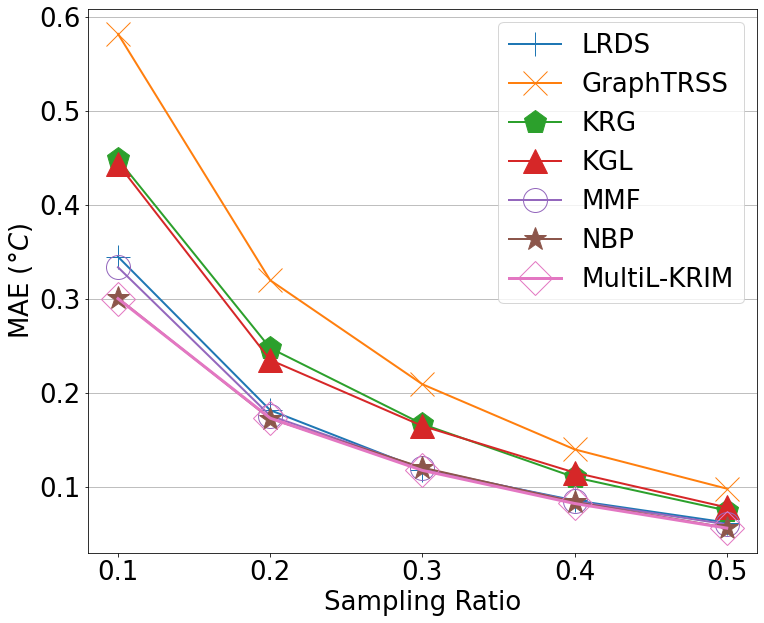}}\hspace{2pt}
  \subfloat[(D2, P1) \label{fig:plot.d2p1}]
  {\includegraphics[height=135pt]{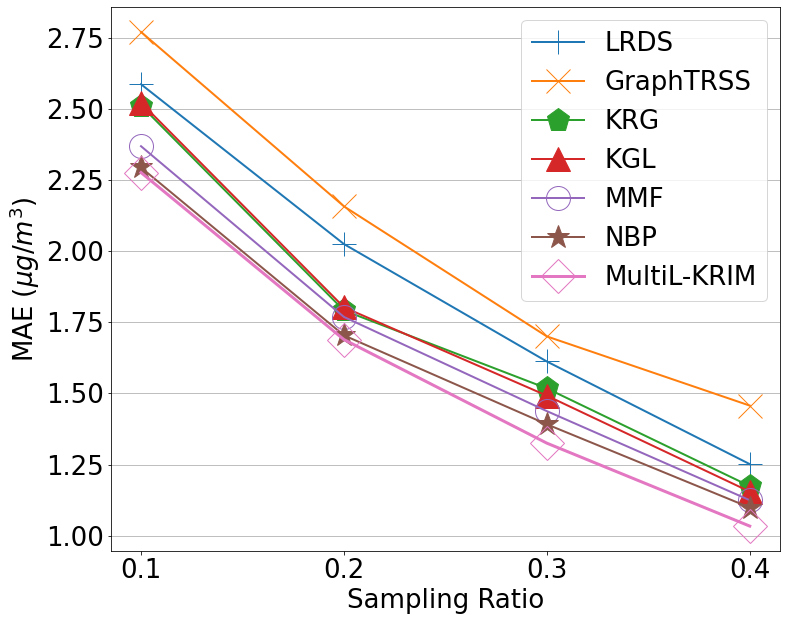}}\hspace{2pt}
  \subfloat[(D3, P1) \label{fig:plot.d3p1}]
  {\includegraphics[height=135pt]{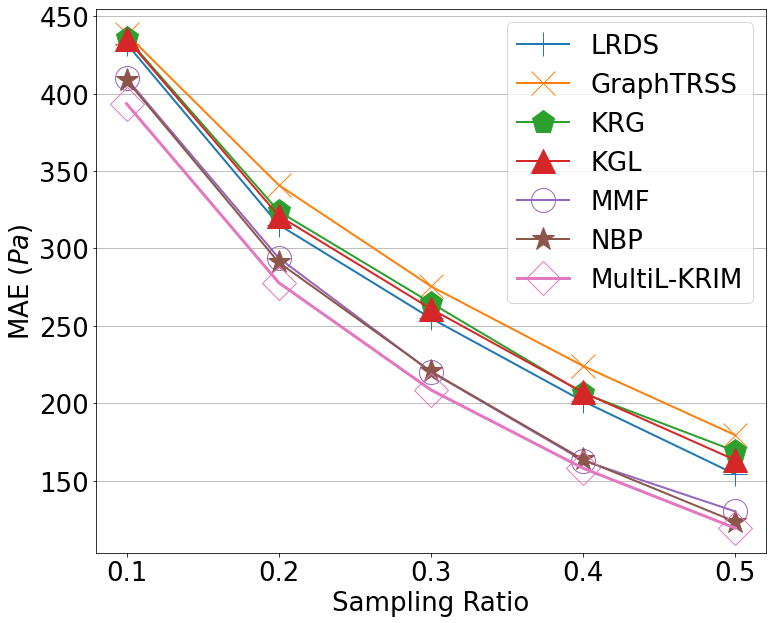}} \\
  \subfloat[(D1, P2) \label{fig:plot.d1p2}]
  {\includegraphics[height=135pt]{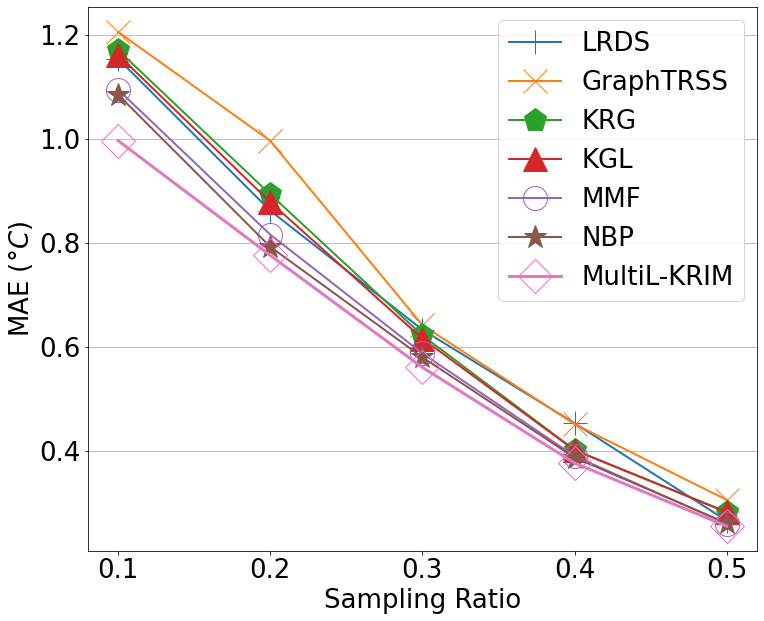}}\hspace{2pt}
  \subfloat[(D2, P2) \label{fig:plot.d2p2}]
  {\includegraphics[height=135pt]{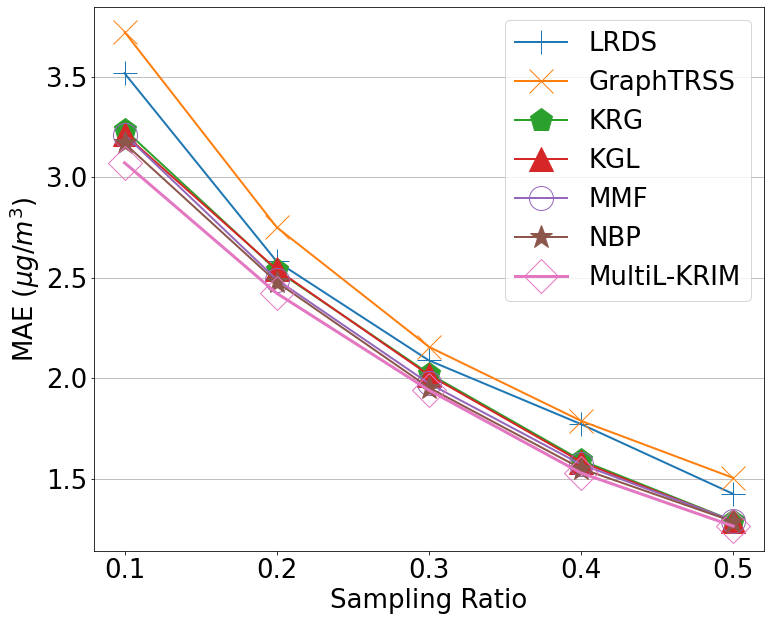}}\hspace{2pt}
  \subfloat[(D3, P2) \label{fig:plot.d3p2}]
  {\includegraphics[height=135pt]{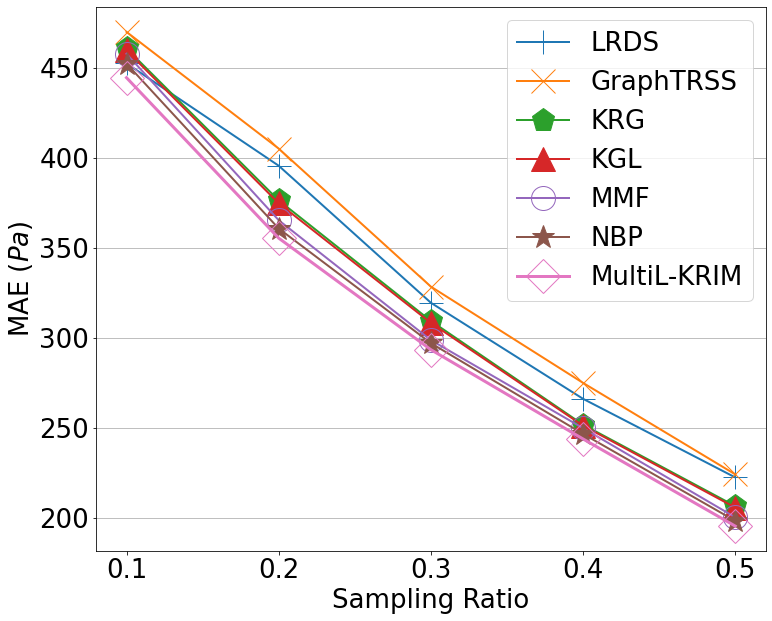}}
  \caption{Performance of MultiL-KRIM vs. other methods on
    different datasets (D1, D2, D3) and sampling patterns
    (P1, P2).}
  \label{fig:plot.p1}
\end{figure*}
\begin{table*}[htb!]
\renewcommand*{\arraystretch}{1.0}
\caption{Average error metrics over sampling ratios in all
  datasets for sampling schemes P1 and P2. The best and
  second-best performing methods on each category are shown
  in red and blue, respectively.}
    \centering
    \resizebox{.8\textwidth}{!}{%
    \begin{tabular}{c|ccc|ccc|ccc} 
    \hline 
 \multirow{3}{*}{Method}& \multicolumn{9}{c}{Random Sampling (P1)}\\
    \cline{2-10}
         &  \multicolumn{3}{c|}{Sea Surface Temperature}&  \multicolumn{3}{c|}{PM 2.5 Concentration}&  \multicolumn{3}{c}{Sea Level Pressure}\\
         &  MAE &  RMSE &  MAPE &  MAE &  RMSE &  MAPE &  MAE &  RMSE & MAPE \\ \hline  
         GraphTRSS&  0.2699&  0.4404&  0.0250&  2.0212&  3.5152&  0.3449&  291.7323&  494.7112& 0.0029\\ 
         KRG&  0.2097&  0.3301&  0.0195&  1.7488&  3.4578&  0.3042&  280.1288&  472.0558& 0.0027\\ 
         KGL&  0.2075&  0.3287&  0.0188&  1.7430&  3.4657&  0.3025&  277.5922&  461.9560& 0.0025\\ 
         LRDS&  0.1586&  0.2559&  0.0162&  1.8688&  3.4736&  0.3438&  271.6200&  444.6673& 0.0027\\ 
         MMF&  0.1549&  0.2417&  \nbest{0.0147}&  1.6751&  3.2526&  0.2809&  243.4078&  413.8005& \nbest{0.0024}\\ 
         NBP&  \nbest{0.1469}&  \best{0.2242}&  \best{0.0145}&  \nbest{1.6222}&  \nbest{3.1857}&  \best{0.2726}&  \nbest{241.5242}&  \nbest{412.2467}& \nbest{0.0024}\\ \hline  
         MultiL-KRIM&  \best{0.1460}&  \nbest{0.2315}&  0.0148&  \best{1.5812}&  \best{3.0745}&  \nbest{0.2743}&   \best{231.4998}&   \best{395.8148}& \best{0.0023}\\ 

         \hline \hline
         \multirow{3}{*}{Method}& \multicolumn{9}{c}{Entire Snapshots Sampling (P2)}\\
    \cline{2-10}
         &  \multicolumn{3}{c|}{Sea Surface Temperature}&  \multicolumn{3}{c|}{PM 2.5 Concentration}&  \multicolumn{3}{c}{Sea Level Pressure}\\
         &  MAE &  RMSE &  MAPE &  MAE &  RMSE &  MAPE &  MAE &  RMSE & MAPE \\ \hline  
         GraphTRSS&  0.7204&  1.1669&  0.0751&  2.3844&  4.4152&  0.4072&  340.3902&  550.2937& 0.0034\\ 
         KRG&  0.6744&  1.0759&  0.0695&  2.1357&  4.0605&  0.3862&  320.8218&  563.9427& \nbest{0.0032}\\ 
         KGL&  0.6682&  1.0727&  0.0691&  2.1284&  4.0418&  \best{0.3631}&  319.7752&  553.7329& \nbest{0.0032}\\ 
         LRDS&  0.6739&  1.0754&  0.0702&  2.2765&  4.1946&  0.3724&  331.0909&  540.7169& 0.0034\\ 
         MMF&  0.6299&  0.9938&  0.0682&  2.1077&  4.0294&  \nbest{0.3668}&  314.2487&  540.5447& \nbest{0.0032}\\ 
         NBP&  \nbest{0.6214}&  \nbest{0.9875}&  \nbest{0.0665}&  \nbest{2.0875}&  \nbest{3.9952}&  0.3756&  \nbest{310.8744}&  \best{528.2204}& \best{0.0031}\\ \hline  
         MultiL-KRIM&  \best{0.5934}&  \best{0.9566}&  \best{0.0651}&  \best{2.0456}& \best{3.8761}&  0.3738&  \best{306.3696}&  \nbest{528.7843}& \best{0.0031}\\ \hline 
         
    \end{tabular}}
    \label{tab:results.avg}
\end{table*}

MultiL-KRIM is compared against several state-of-the-art
methods on TVGS recovery, such as the
low-rank-and-differential-smoothness (LRDS)
method~\cite{mao2018spatio}, the Sobolev-smoothness
(GraphTRSS) one~\cite{giraldo2022reconstruction}, the
nonparametric kernel-based
KRG~\cite{venkitaraman2019predicting},
KGL~\cite{pu2021kernel}, and
NBP~\cite{bazerque2013nonparametric}, as well as the
multi-layer matrix factorization
(MMF)~\cite{cichocki2007multilayer}. Among these, LRDS, MMF,
NBP, and the proposed MultiL-KRIM promote low-rank models,
while the others do not.  In another view, MultiL-KRIM, NBP,
MMF, KGL, and KRG are factorization/structured methods,
while LRDS and GraphTRSS are unstructured.  More readable
categorization of methods can be found at
\cref{app:tab.TVGS.category} (supplementary file).  Notice
that numerical tests of other methods on the same datasets
can be found in~\cite{qiu2017time, mao2018spatio,
  venkitaraman2019predicting,
  giraldo2022reconstruction}. MultiL-KRIM, NBP, KGL, KRG,
and MMF are all solved by the
successive-convex-approximation framework
of~\cite{facchinei2015parallel}, and realized in software by
the Julia programming language~\cite{bezanson2017julia}. The
software code of LRDS and GraphTRSS can be found available
online. All tests were run on an 8-core Intel(R) i7-11700
2.50GHz CPU with 32GB RAM. The following sub-sections
explore all steps of the MultiL-KRIM pipeline, portraited in
\cref{fig:pipeline}.

\noindent\textbf{Performance metrics.} Root mean square
error (RMSE), mean absolute error (MAE), and mean absolute
percentage error (MAPE) are defined in
\cref{app:exp.TVGS.metrics} (supplementary file).  Detailed
error values of all methods across all datasets and sampling
patterns are displayed in
\cref{app:tab.result_random_sampling,app:tab.result_entire_snapshots}
(supplementary file).  For all error metrics RMSE, MAE, and
MAPE, lower values are better.  All methods are finely tuned
to achieve the lowest MAE. The reported metric values are
the uniform averages of 20 independent runs.

\noindent\textbf{Data (Step 1 of \cref{fig:pipeline}).} Three real-world datasets are
considered~\cite{qiu2017time, mao2018spatio,
  giraldo2022reconstruction}.
\begin{enumerate}

\item D1 (sea surface temperature): Monthly data from 1870
  to 2014 are collected with a spatial resolution of
  $1^\circ~\textnormal{latitude} \times
  1^\circ~\textnormal{longitude}$ global grid. 500-month
  data of 100 random locations on the Pacific ocean are
  selected. The dimensions of data $\vect{Y}$ are
  $(I_0, I_N) = (100, 500)$.

\item D2 (particulate matter 2.5): The California daily mean
  PM2.5 concentration data are collected from 93 observation
  sites over 200 days starting from January 1, 2015. The
  dimensions of data $\vect{Y}$ are
  $(I_0, I_N) = (93, 200)$.

\item D3 (sea level pressure): Global sea-level pressure
  data were gathered between 1948 and 2010, with a spatial
  resolution of
  $2.5^\circ~\textnormal{latitude} \times
  2.5^\circ~\textnormal{longitude}$, and temporal resolution
  of five days. Here, data of 500 random sites over a span
  of 400 time units are considered. The dimensions of data
  $\vect{Y}$ are $(I_0, I_N) = (500, 400)$.

\end{enumerate}
The graph topology is constructed by the $k$-nearest
neighbors method as in \cite{mao2018spatio}.  In particular,
the Euclidean distance $d_{ij}$ between any pair of sensors
$i,j\in \mathcal{V}$ is calculated based on their given
geographical locations. Each node $i\in \mathcal{V}$ is
connected to $k$ nearest neighbors
$\mathcal{N}_i\coloneqq \{i_1, i_2, \ldots, i_k\} \subset
\mathcal{V}$ in terms of the Euclidean distance Thus,
$(i,i_j)\in\mathcal{E}, \forall i_j \in \mathcal{N}_i$.
Note that $i\notin \mathcal{N}_i$.  Then, the $(i,j)$th
entry of the adjacency matrix $\vect{W}$ is given as
$w_{ij}=1/d^2_{ij}$ if $(i,j)\in\mathcal{E}$, and $w_{ij}=0$
otherwise.  Eventually, the graph Laplacian matrix
$\vect{L}$ is calculated as described in
\cref{sec:graph.context}.

\noindent\textbf{Sampling patterns (Step 1 of \cref{fig:pipeline}).} With sampling ratio
$r\in \Set{0.1, 0.2, 0.3, 0.4, 0.5}$, the following sampling
patterns are considered.
\begin{enumerate}

\item P1 (random sampling)~\cite{qiu2017time}: signals of
  $\ceil{I_0 \cdot r}$ randomly sampled nodes at each time
  instant $t$ are observed, where $\ceil{\cdot}$ denotes the
  ceiling function.

\item P2 (entire-snapshots
  sampling)~\cite{giraldo2022reconstruction}: all node
  signals are sampled at $\ceil{I_N \cdot r}$ randomly
  chosen time instants. In other words, $\ceil{I_N \cdot r}$
  snapshots are randomly sampled.

\end{enumerate}

\noindent\textbf{Navigator data formation (Step 2 of \cref{fig:pipeline}).} The navigator data
$\check{\vect{Y}}_{\textnormal{nav}}$ of size
$\nu \times N_{\text{nav}}$, for some
$\nu, N_{\text{nav}} \in \IntegerPP$, are defined by the
sampled-and-zero-padded data $\mathscr{S}_\Omega(\vect{Y})$
via the following configurations.

\begin{enumerate}

\item Nav1 (snapshots):
  $\check{\vect{Y}}_{\textnormal{nav}}$ are defined by the
  snapshots of $\mathscr{S}_\Omega(\vect{Y})$, represented
  by the columns of $\mathscr{S}_\Omega(\vect{Y})$.  For
  pattern P1,
  $\check{\vect{Y}}_{\textnormal{nav}} \coloneqq
  \mathscr{S}_\Omega(\vect{Y})$, with $\nu=I_0$ and
  $N_{\text{nav}}=I_N$; for pattern P2, missing snapshots
  are discarded, \ie $\nu=I_0$ and
  $N_{\text{nav}}=I_N\cdot r$.

\item Nav2 (time profiles of nodes):
  $\check{\vect{Y}}_{\textnormal{nav}}$ are defined by the
  node time profiles of $\mathscr{S}_\Omega(\vect{Y})$,
  represented by the rows of $\mathscr{S}_\Omega(\vect{Y})$,
  \ie,
  $\check{\vect{Y}}_{\textnormal{nav}} \coloneqq
  \mathscr{S}_\Omega(\vect{Y})^\intercal$, with $\nu=I_N$
  and $N_{\text{nav}}=I_0$.

\item Nav3 (patches):
% for each node $i$ at time instant $t$ s.t.\ $\delta t < t \leq I_N - \delta t$, a ``patch'' is created by taking the signals
%   at $k$ nearest neighbors $\mathcal{N}_i$ of node $i$, over the time interval $[t-\delta t, t+ \delta t]$, where
%   $\delta t$ is a hyperparameter, and $\mathcal{N}_i\coloneqq
%   \{i_1, i_2, \ldots i_k\}$
%     are taken from the $k$-nearest neighbor graph of the sensor
%     network.
  for the user-defined $\delta t \in \IntegerP$,
  $\delta t < I_N/2$, and for each $i\in \mathcal{V}$,
  $t\in \IntegerPP$, $\delta t+1\leq t \leq I_n-\delta t$, a
  ``patch''
  $\vect{P}_{it} =
  \mathscr{S}_\Omega(\vect{Y})\vert_{\mathcal{N}_i\times
    \mathcal{N}_t}$ of size $k\times (2\delta t + 1)$ is
  constructed as the submatrix of
  $\mathscr{S}_\Omega(\vect{Y})$ corresponding to the rows
  and columns indicated by the index sets $\mathcal{N}_i$
  and $\mathcal{N}_t$, respectively, where $\mathcal{N}_i$
  is the $k$-nearest neighbors of node $i$ retrieved upon
  building $\vect{W}$, and
  $\mathcal{N}_t\coloneqq \{ \tau \in \IntegerPP \given
  t-\delta t \leq \tau \leq t +\delta t \}$. Then, each
  column of the navigator data matrix
  $\check{\vect{Y}}_{\textnormal{nav}}$ is a vectorized
  $\vect{P}_{it}$. The size of
  $\check{\vect{Y}}_{\textnormal{nav}}$ is
  $k (2\delta t + 1) \times I_0 (I_N-2\delta t)$.
  As post-processing, columns with all zeros of
  $\check{\vect{Y}}_{\textnormal{nav}}$ are removed. This
  formation offers the densest navigator point-cloud, but
  raises computational complexity for landmark-point
  selection step.

\item Nav4 (windows):
% for each time instant $t$ s.t.\ $\delta t < t \leq I_N - \delta t$,
%   snapshots of the interval $[t- \delta t, t+ \delta t]$ are vectorized to be navigator
%   points. Thus, the navigator data $\check{\vect{Y}}_{\textnormal{nav}}$ are of size
%   $(I_0\cdot (2\delta t + 1)) \times (I_N - 2\delta t)$.
  by the same notations used in the description of Nav3, a
  ``window'' centered at time instant $t$ is the submatrix
  $\mathscr{S}_\Omega(\vect{Y})\vert_{\mathcal{V}\times
    \mathcal{N}_t}$.  Each column of the navigator data
  matrix is one vectorized window, so the size of
  $\check{\vect{Y}}_{\textnormal{nav}}$ is
  $\nu=I_0\cdot (2\delta t + 1)$ and
  $N_\text{nav}= I_N - 2\delta t$. Finally, drops all-zeros
  columns of $\check{\vect{Y}}_{\textnormal{nav}}$.

\end{enumerate}

\noindent\textbf{Landmark point selection (Step 3 of \cref{fig:pipeline}).}
Based on Euclidean distance between navigator data points,
the following landmark-point selection strategies are
examined:
\begin{enumerate}
\item L1: greedy max-min-distance
  (maxmin)~\cite{de2004sparse},
\item L2: $k$-means clustering, and
\item L3: fuzzy $c$-means clustering, whose centroids are
  chosen to be the landmark points~\cite{rashmi2019optimal}.
\end{enumerate}

\noindent\textbf{Kernel matrix construction (Step 4 of \cref{fig:pipeline}).} In the
single-kernel case ($M=1$), the kernel from
\cref{app:tab.kernels.list} (supplementary file) which
yields optimal performance is used. The intercept $c$ of the
polynomial kernel (\cref{app:RKHS}) is set to be the
entry-wise mean of the landmark points. After extensive
experiments, multiple kernels ($M>1$) did not score
considerable improvement over the previous single-kernel
strategy.

The next sub-sections dive into different scenarios, each of
which contains four components, closely related to the steps
of the pipeline. A scenario is denoted by a tuple
($\cdot, \cdot, \cdot, \cdot$), corresponding to data,
sampling pattern, navigator data formation, and
landmark-point selection strategy. If an entry of the tuple
is left empty, then all possible values of the component are
validated. For example, (D1, P1, Nav1,$\cdot$) considers sea
surface temperature, random sampling, snapshots, and
validates all strategies L1, L2, and L3.

\subsection{Numerical tests}\label{sec:tests.tvgs}

\subsubsection{The sea-surface-temperature dataset}

To examine the effect of different navigator data formations
and landmark-point selection strategies, tests are conducted
on the (D1, P1,$\cdot$,$\cdot$) scenario, which are detailed
in \cref{app:exp.TVGS.nav_lm}. It is observed that Nav1 has
the lowest average errors among all navigator data
formations. On the other hand, there is no significant
performance difference between landmark-point selection
strategies L1, L2, and L3.  Henceforth, consider scenario
($\cdot$, P1, Nav1, L1); consider scenarios ($\cdot$, P2,
Nav1, L1) if $\ceil{I_N \cdot r}\geq 100$, and ($\cdot$, P2,
Nav3, L1) if $\ceil{I_N \cdot r}< 100$.

\cref{tab:results.avg}, \cref{fig:plot.d1p1,fig:plot.d1p2}
show that low-rank methods outperform others under both
sampling patterns. Averaged over sampling ratios,
MultiL-KRIM has lower MAE values than other methods by
\num{1}\% to \num{46}\% for pattern P1 and \num{5}\% to
\num{18}\% for pattern P2. Among low-rank methods, the
kernel-based NBP and MultiL-KRIM have slight advantages over
LRDS and MMF, especially at the lowest sampling ratio of
\num{10}\% (\Cref{fig:plot.d1p1,fig:plot.d1p2}). The better
performance of KGL and KRG over GraphTRSS highlights the
usefulness of highly structured data modeling. For pattern
P2, the gaps between all methods become narrow; this can
also be seen from \cref{fig:plot.d1p2}. Moreover, at higher
sampling ratios of \num{40}\% and \num{50}\%, the
performance of factorization methods converges.

\subsubsection{The PM2.5-concentration dataset}

\Cref{fig:plot.d2p1,fig:plot.d2p2} show that unstructured
methods LRDS and GraphTRSS score the highest errors across
all sampling rates in both patterns.  Unlike in the previous
dataset, all structured methods outweighs LRDS under all
sampling patterns and ratios.  Furthermore, the merits of
structured methods are emphasized at the lowest sampling
ratio of 10\%, where they outperform LRDS and GraphTRSS by
noticeable margins.  Also taken from
\cref{fig:plot.d2p1,fig:plot.d2p2}, MultiL-KRIM, NBP, and
MMF consistently outperform KGL and KRG in both patterns,
suggesting the necessity of low-rank factorization.
Compared to pattern P1, variance between all methods is
smaller in P2, especially among factorization methods, which
can also be observed from \cref{tab:results.avg}.  In
addition, the average MAE value of MultiL-KRIM is lower by
\num{3}\% to \num{22}\% in pattern P1, but the range of
improvement is \num{3}\% to \num{12}\% in pattern P2.

\subsubsection{The sea-level-pressure dataset}

\begin{figure}[!htb]
    \centering
    \includegraphics[width=0.65\linewidth]{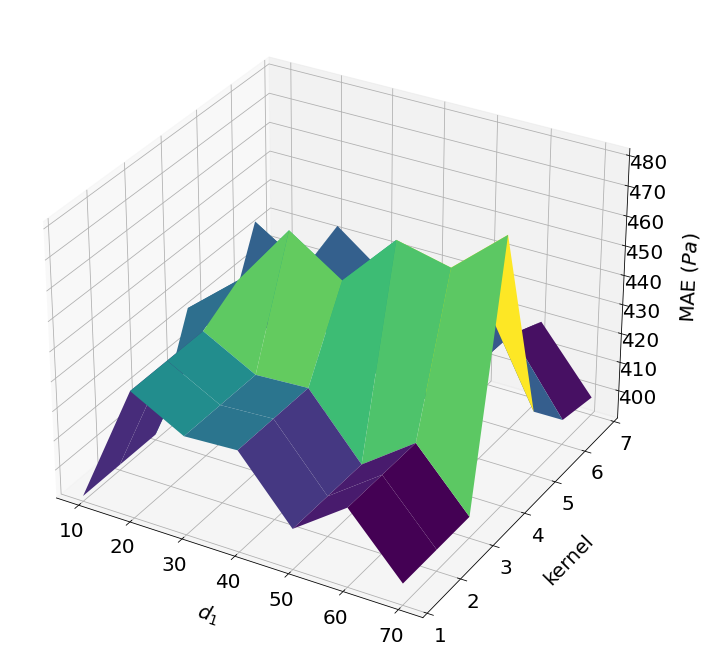}
    \caption{Sensitivity of MAE to variation of parameters
    $d_1$ and kernel choice in the context of (D3, P1, Nav1, L1)
    at 10\% sampling ratio.}
    \label{app:fig.P1.SLP10}
\end{figure}

\cref{app:fig.P1.SLP10} (supplementary file) depicts the
sensitivity of MAE values to variations of the inner
dimension $d_1 \in \Set{10, 20, 30, 40, 50, 60, 70}$ and the
choice of kernel from the dictionary in
\cref{app:tab.kernels.list} (supplementary file), under
scenario (D3, P1, Nav1, L1) at sampling ratio of
\num{10}\%. The lowest MAE is reached at $d_1=10$ and for
the Gaussian kernel, almost regardless of its parameter
$\sigma$. Notice that the linear kernel gives the highest
MAE across all different $d_1$.

\cref{tab:results.avg}, along with
\cref{fig:plot.d3p1,fig:plot.d3p2}, exhibits the continued
dominance of low-rank factorization methods MultiL- KRIM,
NBP, and MMF.  In terms of average MAE values, MultiL-KRIM
outperforms the all competing methods by \num{4}\% to
\num{21}\% for pattern P1, and by \num{1}\% to \num{10}\%
for pattern P2.  In the case of sampling pattern P1,
\cref{fig:plot.d3p1} illustrates a distinct gap between the
low-rank factorization methods MultiL-KRIM, NBP, and MMF,
compared to others. This suggests the need for both low-rank
and structured modeling.  From \cref{tab:results.avg}, the
average MAE of LRDS in the P1 case is 3\% and 4\% lower than
that of KGL and KRG, respectively.  This indicates a slight
advantage of low-rank modeling over kernel modeling in
pattern P1.  On the flip side, \cref{fig:plot.d3p2} shows
the outperformance of structured methods MultiL-KRIM, NBP,
MMF, KGL, and KRG, over the unstructured ones LRDS and
GraphTRSS under pattern P2, advocating the interpolation
strength of structured modeling.
% However, the marginal drop of MAE values
%   \cref{tab:results.avg} shows that,
%   in terms of average MAE values, MultiL-KRIM outperforms the
% all competing methods by \num{4}\% to \num{21}\% for pattern P1, but only \num{1}\% to
% \num{10}\% for pattern P2.

Computationally, MultiL-KRIM is more efficient than LRDS,
since LRDS requires SVD computations for nuclear-norm
minimization, which are expected to inflict computational
burdens if the dimensions of the dataset are large. Indeed,
for dataset D3 of dimensions \num{500}-by-\num{400}, LRDS
takes on average \num{1384} seconds for pattern P1 and
\num{1456} seconds for pattern P2, around three times the
run time of MultiL-KRIM (\num{444}s for P1 and \num{492}s
for P2). Details on computational times can be found in
\cref{app:tab.runTime.random_sampling} (supplementary file).

% \begin{figure}
%     \centering
%     \includegraphics[width=0.65\linewidth]{SLP10_param.png}
%     \caption{Sensitivity of MAE to variation of parameters
%     $d_1$ and kernel choice in the context of (D3, P1, Nav1, L1)
%     at 10\% sampling ratio.}
%     \label{fig:P1.SLP10}
% \end{figure}

%%%%%%%%%%%%%%%%%%%%%%%%%%%%%%%%%%%%%%%%%%%%%%%%%%%%%%%%%%%%%%%%%%%%%%%%%%%%%%%%%%%%%%%%%%%%%%%

\section{Application: Dynamic Magnetic Resonance Imaging}\label{sec:dmri}

\begin{figure*}[ht]
  \centering
  \subfloat[The (k,t)-space/domain \label{fig:dmri.ktspace}]
  {\includegraphics[height=100pt]{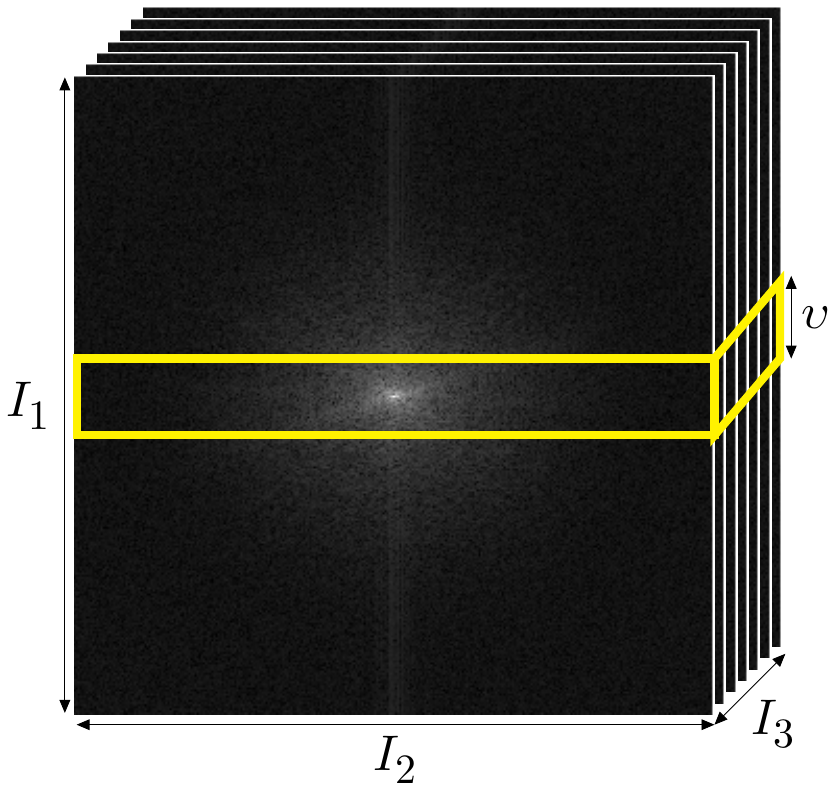}}\hspace{2pt}
  \subfloat[The image domain \label{fig:dmri.image}]
  {\includegraphics[height=100pt]{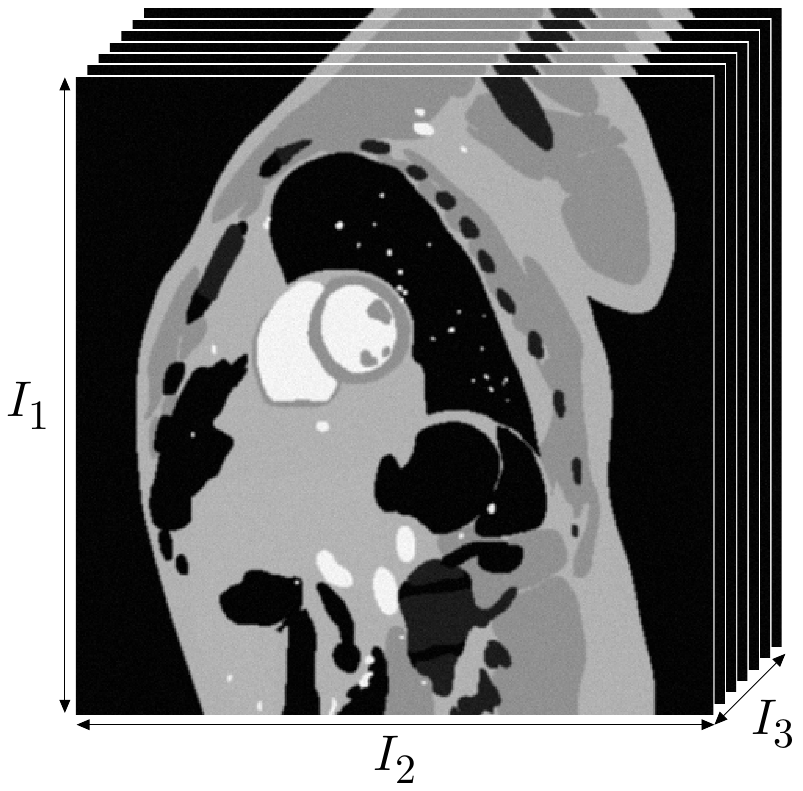}}\hspace{10pt}
  \subfloat[Cartesian sampling \label{fig:cartesian.sampling}]
  {\includegraphics[height=100pt]{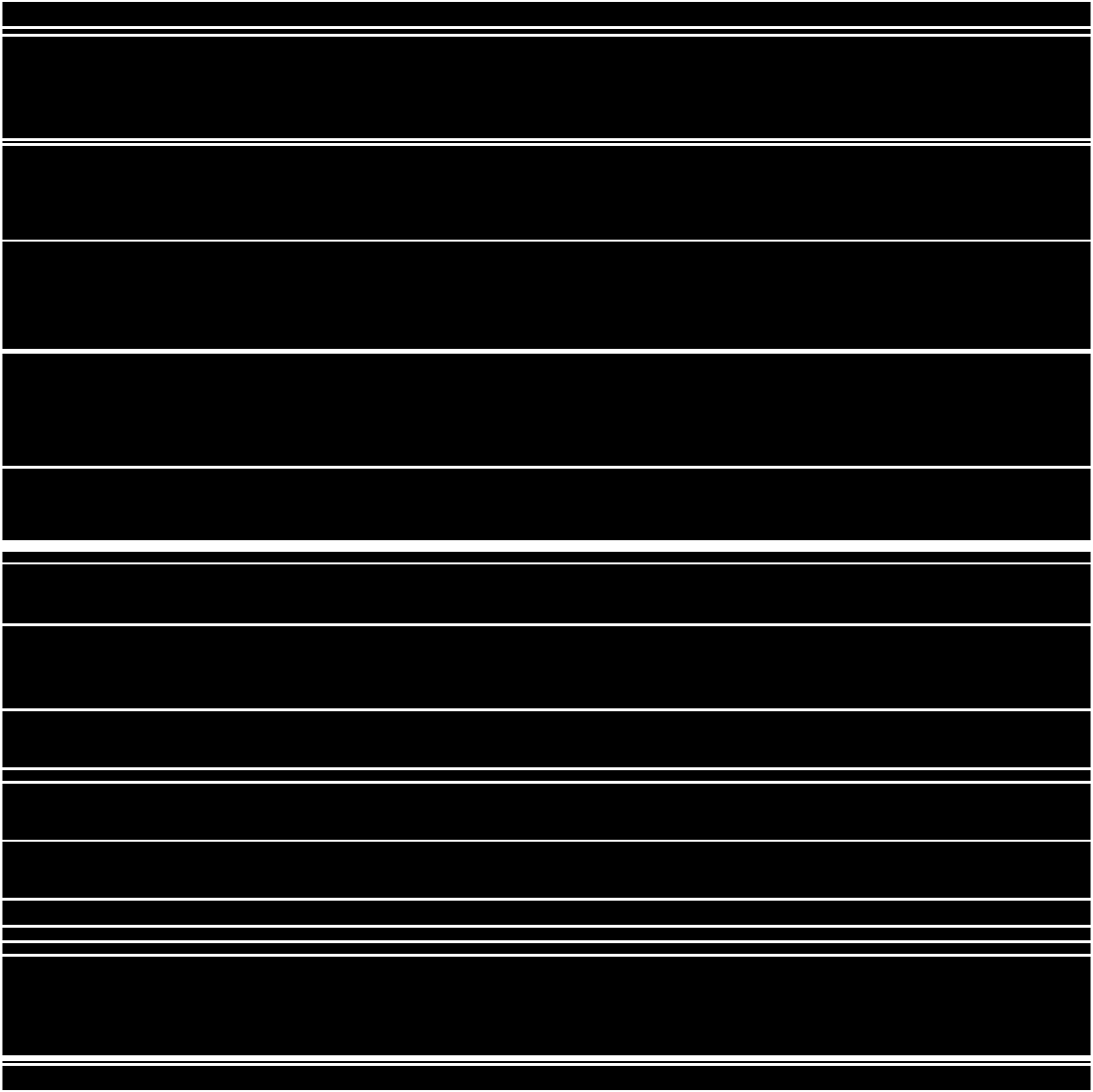}}\hspace{1pt}
  \subfloat[Radial sampling \label{fig:radial.sampling}]
  {\includegraphics[height=100pt]{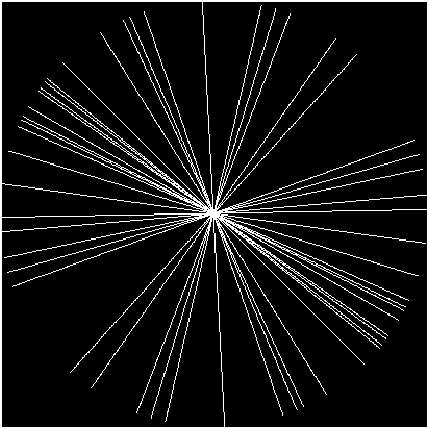}}\hspace{2pt}
  \caption{(a) Domain of size $I_1 \times I_2 \times I_3$
    where the dMRI data are observed and collected, with
    $I_3$ denoting the number of time frames. The marked
    $\upsilon \times I_2 \times I_3$ box shows the typical
    location of the faithful ``navigator/pilot'' data,
    corresponding, usually, to the ``low-frequency'' area of
    the domain. The ``white dots'' indicate the locations
    where data are collected, with the black-colored
    majority of the (k,t)-space to correspond to locations
    where there are missing data. After ``flattening'' the
    3D (k,t)-space into a 2D one (see
    \cref{sec:problem.formulation}), these ``white dots''
    define the index set $\Omega$, and thus the sampling
    operator $\mathscr{S}_{\Omega}$. (b) The
    $(I_1 \times I_2 \times I_3)$-sized image domain. The
    image-domain data are obtained by applying the two
    dimensional (2D) inverse DFT $\mathscr{F}^{-1}$ to the
    (k,t)-space data~\cite{zhi2000principles}. (c) 1D
    Cartesian and (d) radial sampling trajectories in
    k-space. The white-colored lines indicate locations
    where data are
    collected.} \label{fig:dmri.data.description}
\end{figure*}

This section focuses on dMRI, serving as a more detailed
exposition of~\cite{Thien:ICASSP24}. \cref{sec:dmri.context}
defines the problem of highly-accelerated dMRI recovery,
while MultiL-KRIM's inverse problem is stated in
\cref{sec:dmri.inv.prob}. The numerical tests in
\cref{sec:dmri.exp} demonstrate that MultiL-KRIM offers
significant speed-ups from its predecessor
KRIM~\cite{slavakis2022krim}, with no performance losses,
while outperforming state-of-the-art ``shallow'' learning
methods.

\subsection{DMRI-data recovery}\label{sec:dmri.context}

DMRI is a common non-invasive imaging technique for tracking
the motion of body organs, offering valuable insights in the
fields of cardiac and neurological
diagnosis~\cite{zhi2000principles}. All typical
data-analytic bottlenecks show up in dMRI: high
dimensionality of the image data, many missing data entries
due to severe under-sampling to achieve high acceleration in
data collection, and strong latent spatio-temporal
correlations within the data because of body-organ
structured movements; \eg, a beating
heart~\cite{liang1994efficient, zhi2000principles}.

Naturally, there is a rich literature of imputation methods
to address the highly under-sampled dMRI problem: compressed
sensing~\cite{lustig2007sparse, feng2016xd}, low-rank
modeling~\cite{lingala2011ktslr, zhao2012image,
  otazo2015low}, and learning methods based on
dictionaries~\cite{ravishankar2010mr, wang2014compressed,
  caballero2014dictionary},
manifolds~\cite{Usman.Manifold.15, Chen:MA:IEEETMI:17,
  poddar2016dynamic, nakarmi2017m, nakarmi2018mls,
  slavakis2022krim}, kernels~\cite{nakarmi2017kernel,
  nakarmi2018mls, poddar2019manifold, arif2019accelerated,
  slavakis2022krim}, tensors~\cite{liu2012tensor,
  Li_Ye_Xu_2017}, DeepL~\cite{knoll2020deep}, and
DIP~\cite{yoo2021time, qayyum2022untrained}.  Several of
those methods are selected for comparisons against
MultiL-KRIM in \cref{sec:dmri.exp}, including the
manifold-learning-based SToRM~\cite{poddar2016dynamic}, the
low-rank factorization with sparsity PS-Sparse
\cite{zhao2012image}, tensor factorization with total
variation LRTC-TV~\cite{Li_Ye_Xu_2017}, and the DIP-based
model TDDIP~\cite{yoo2021time}. SToRM follows a typical ManL
route and utilizes as a regularizer a quadratic loss
involving the graph Laplacian matrix. PS-Sparse uses a
``semi-blind'' factorization of two orthogonal low-rank
matrices, where one of those matrix factors is built from
navigator data.
% By the same token, LRTES \cite{he2016lrtes} leverages Tucker decomposition with ``explicit
% subspaces'', where all factors excluding the core tensor and the first mode matrix are
% orthogonal matrices constructed from the navigator data.
LRTC-TV instead enforces a ``blind'' decomposition of the data matrix, and imposes
low-rankness by nuclear-norm minimization. Meanwhile, TDDIP considers randomly generated points
on a user-defined manifold (a helix) as the navigator data.

The ``(k,t)-space/domain'' dMRI data are represented by a
3-way tensor
$\mathbfscr{Y} \in \Complex^{I_1 \times I_2 \times I_3}$
(\cref{fig:dmri.ktspace}), where $I_1$ and $I_2$ are the
number of phase and frequency encodings of the k-space,
respectively, while $I_3$ stands for the number of time
frames. Following the general data formation of
\cref{sec:problem.formulation}, $\vect{Y}$ is obtained by
vectorizing each ``slice/frame'' $\mathbfcal{Y}_t$ of
$\mathbfscr{Y}$ ($t$ denotes discrete time with
$t \in \{1, \ldots, I_3 \}$), or,
$\vect{Y} \coloneqq [ \vect{y}_1, \ldots, \vect{y}_{I_3} ]
\in \Complex^{ I_0 \times I_3}$, where
$I_0 \coloneqq I_1 I_2$ and
$\vect{y}_t \coloneqq \text{vec}(\mathbfcal{Y}_t)$. Usually,
the (k,t)-space data are severely under-sampled to promote
highly accelerated data
collection~\cite{liang1994efficient}, and to this end,
popular sampling strategies are 1-D Cartesian
(\cref{fig:cartesian.sampling}) and radial
(\cref{fig:radial.sampling}) sampling, which define the
index set $\Omega$ and the sampling operator
$\mathscr{S}_{\Omega}$ (\cref{sec:problem.formulation}). The
goal is to recover high fidelity images in the image domain,
despite the large number of missing entries in the
(k,t)-space. Since the end product of the dMRI recovery is
an image, the linear transform mapping $\mathscr{T}$,
introduced in \cref{sec:problem.formulation} to map data
from the image to the (k,t)-domain, becomes the 2D DFT
$\mathscr{F}$.

\subsection{The inverse problem}\label{sec:dmri.inv.prob}

With $\vect{X}$ denoting the image-domain data, the generic
MultiL-KRIM's inverse
problem~\eqref{eq:multil.manifold.right} takes here the
following special form:
\begin{subequations}\label{eq:dmri.task.general}%
  \begin{align}
    \min_{ (\vect{X}, \vect{Z}, \Set{\vectcal{D}_q}_{q=1}^Q, \vectcal{B}) }
    {}\ & {} \tfrac{1}{2} \norm{ \vect{X} - \vectcal{D}_1
          \vectcal{D}_2 \cdots
          \vectcal{D}_Q \vectcal{K} \vectcal{B}
          }^2_{\textnormal{F}} + \underbrace{ \lambda_1
          \norm{\vectcal{B}}_1 }_{\mathcal{R}_1(\vectcal{B})} \notag \\
        & + \underbrace{\tfrac{\lambda_2}{2} \norm{\vect{Z}
          - \mathscr{F}_{\text{t}}
          (\vect{X})}_{\textnormal{F}}^2 + \lambda_3
          \norm{\vect{Z}}_1}_{ \mathcal{R}_{21}
          (\vect{X}, \vect{Z}) } \notag \\
        & + \underbrace{\tfrac{\lambda_4}{2}
          \sum\nolimits_{q=1}^Q
          \norm{\vectcal{D}_q}_{\textnormal{F}}^2}_{
          \mathcal{R}_{22}( \Set{\vectcal{D}_q}_{q=1}^Q )
          } \label{eq:dmri.task.loss} \\
    \text{s.to} {}\
        & \mathscr{S}_{\Omega}(\vect{Y}) = \mathscr{S}_{\Omega}
          \mathscr{F}(\vect{X})\,, \label{dmri.task.consistency}
    \\
        & \vect{1}_{N_{\mathit{l}}}^{{\hermconj}} \vect{B}_m
          = \vect{1}_{I_N}^{{\hermconj}}\,,
          \forall m \in \{1, \ldots, M\}\,, \label{dmri.task.B}
  \end{align}%
\end{subequations}%
where the generic regularizer
in~\eqref{eq:multil.manifold.right} becomes
\begin{align*}
  \mathcal{R}( \vect{X}, \vect{Z},
  \Set{\vectcal{D}_q}_{q=1}^Q, \vectcal{B} )
  {} \coloneqq {} & \mathcal{R}_1( \vectcal{B} ) +
                    \mathcal{R}_{21}( \vect{X}, \vect{Z} )
  \\
  & + \mathcal{R}_{22}( \Set{\vectcal{D}_q}_{q=1}^Q ) \,.
\end{align*}
Regularizer $\mathcal{R}_{21}(\cdot, \cdot)$ is introduced
to capitalize on the prior knowledge that the dataset at
hand, to be used in \cref{sec:dmri.exp}, records the
periodic movement of a beating heart over a static
background. The mapping
$\mathscr{F}_{\textnormal{t}}(\vect{X})$ stands for the
temporal 1D DFT, which operates on rows of $\vect{X}$. Each
row of $\vect{X}$, of length $I_3$, is the time profile of a
single pixel in the image domain. The periodic or static
nature of each row of $\vect{X}$ suggests that its 1D DFT is
a sparse vector over the frequency domain. This observation
justifies the penalty term $\norm{\vect{Z}}_1$, well known
for promoting sparsity, with $\vect{Z}$ being an auxiliary
matrix-valued variable used to simplify computations when
solving the inverse problem. The solutions of the convex
sub-tasks of the inverse problem follow a similar routine to
that of \cref{sec:graph.inv}, and are thus deferred to
\cref{app:dMRI.solve} (supplementary file) due to space
limitations.

\subsection{Numerical tests}\label{sec:dmri.exp}

\begin{figure}[t!]
  \centering
  \subfloat[\label{fig:plot.cartesian}]
  {\includegraphics[width =
    .8\columnwidth]{{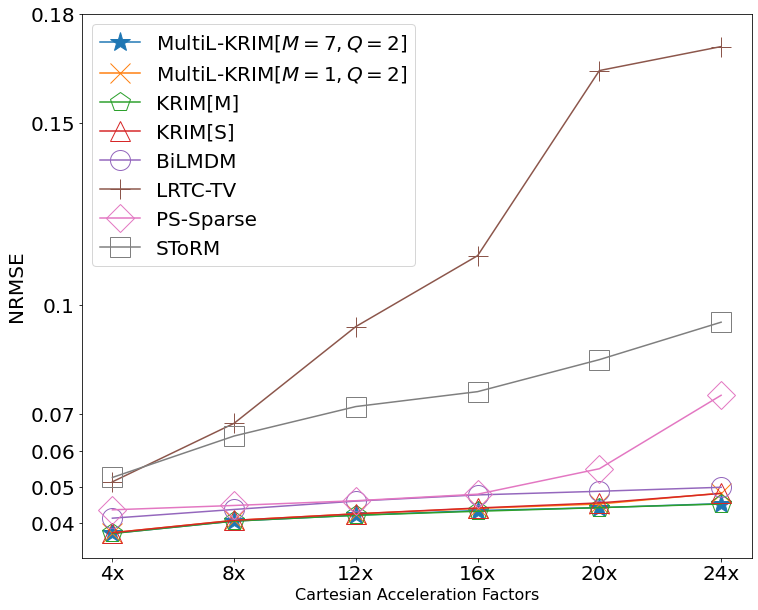}}} \\
  \subfloat[\label{fig:plot.radial}]
  {\includegraphics[width = .8\columnwidth]{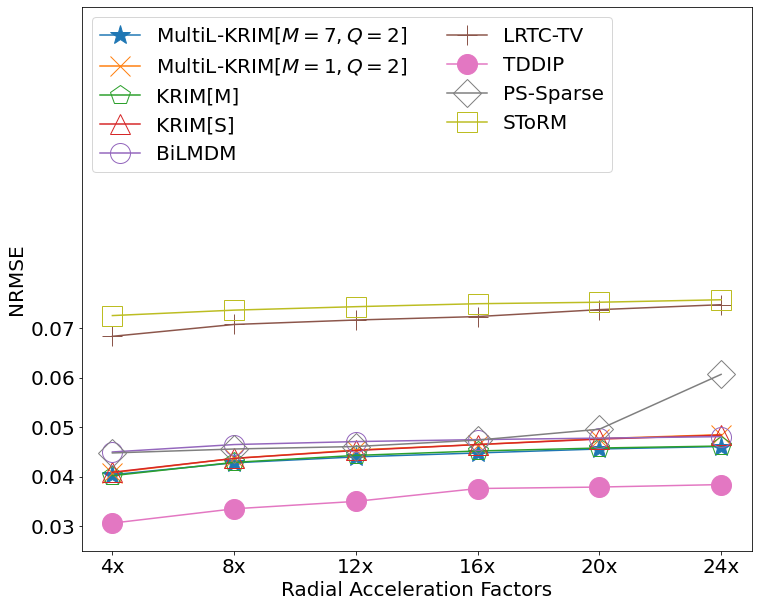}}
  \centering
  \caption{Normalized RMSE values (the lower the better)
    w.r.t.\ the ground truth vs.\ (sampling-time)
    acceleration factors. An acceleration factor
    ``$\beta$x'' means that only $1/\beta$ of the full-scan
    $I_1 \times I_2$ k-space is sampled (see
    \Cref{fig:cartesian.sampling,fig:radial.sampling}).} \label{fig:nrmse.v.acceleration}
\end{figure}

Following~\cite{zhao2012image, nakarmi2017m, nakarmi2018mls,
  shetty2020bilmdm, slavakis2022krim}, the proposed
framework is validated on the magnetic resonance extended
cardiac-torso (MRXCAT) cine phantom
dataset~\cite{wissmann2014mrxcat}, of size
$(I_1, I_2, I_3) = (408, 408, 360)$, under both radial and
Cartesian sampling. Similarly to~\cite{shetty2020bilmdm,
  slavakis2022krim}, the navigator data are formed by
vectorizing the heavily sampled region of the
(k,t)-space. In other words, the $\upsilon \times I_2$ box
of each frame in \cref{fig:dmri.ktspace}, with
$\upsilon = 4$, generates a number $I_3$ of
$(\upsilon I_2) \times 1$ navigator vectors. Then, the
maxmin strategy~\cite{de2004sparse} is adopted to select
$N_{\mathit{l}} \coloneqq 70$ landmark points. Parameter
$Q \in \Set{2, 3}$, and for each $Q$ the inner matrix
dimensions of $\Set{ \vectcal{D}_q}_{q=1}^Q$ are set as
follows:
\begin{enumerate*}[label=\textbf{(\roman*)}]
\item if $Q=2$, then $d_1 = 8$; and
\item if $Q=3$, then $(d_1, d_2) = (2, 8)$.
\end{enumerate*}
\cref{app:tab.N.Q} (supplementary file) shows the number of parameter unknowns~\eqref{eq:N.Q>1} for different settings
% \cref{tab:N.Q} shows the number of parameter unknowns~\eqref{eq:N.Q>1} for different settings
of $Q$ to emphasize the efficiency of MultiL-KRIM's
factorization. The inner-dimension parameter of the
predecessor KRIM~\cite{slavakis2022krim} is $d \coloneqq
8$. Parameter $M \in \Set{1, 7}$ for both MultiL-KRIM and
KRIM, with choices of kernels as in~\cite{slavakis2022krim}
(see also \cref{app:tab.kernels.list} (supplementary file)).

Besides the predecessors KRIM~\cite{slavakis2022krim} and
BiLMDM~\cite{shetty2020bilmdm}, MultiL-KRIM[$M, Q$] competes
also against the low-rank tensor factorization with total
variation (LRTC-TV)~\cite{Li_Ye_Xu_2017}, the DIP-based
TDDIP~\cite{yoo2021time} (designed especially for radial
sampling), the popular PS-Sparse~\cite{zhao2012image} and
ManL-based SToRM~\cite{poddar2016dynamic}. Tags KRIM[S] and
KRIM[M] refer to \cite{slavakis2022krim} for the case of a
single ($M=1$) and multiple ($M>1$) kernels,
respectively. Performance of several other state-of-the-art
methods and their comparison against KRIM and BiLMDM on the
same dataset can be found in~\cite{shetty2020bilmdm,
  slavakis2022krim}. Additionally, \cite{slavakis2022krim,
  shetty2020bilmdm} include also the validation of KRIM and
BiLMDM on several other dMRI datasets. The software code for
(MultiL-)KRIM and BiLMDM was written in
Julia~\cite{bezanson2017julia}, while the software codes of
the other methods were found available online. All tests
were run on an 8-core Intel(R) i7-11700 2.50GHz CPU with
32GB RAM. As mentioned after~\eqref{eq:NBP.approx}, this CPU
was not able to carry through NBP
computations~\cite{bazerque2013nonparametric}, since the
size $I_0 \times I_0$ ($>10^{10}$ entries) of
$\vect{K}_{\mathcal{Z}}$ seemed to be too large for the CPU
at hand.
% Examples of results in GIF format can be downloaded from
% \url{http://www.slavakislab.ict.e.titech.ac.jp/MRXCAT_GIFs.zip} .

\noindent\textbf{Performance metrics.} The main
evaluation metric is the normalized root mean square error
$\text{NRMSE} \coloneqq \norm{ \vect{X} - \hat{\vect{X}}
}_{\textnormal{F}} / \norm{ \vect{X} }_{\textnormal{F}}$,
where $\vect{X}$ denotes the image data obtained from fully
sampled (k,t)-space data, and $\hat{\vect{X}}$ represents
the estimate of $\vect{X}$. Additionally, reconstruction
quality of high-frequency regions are quantified by the
high-frequency error norm (HFEN)~\cite{ravishankar2010mr},
and two sharpness measures M1 (intensity-variance based) and
M2 (energy of the image gradient)~\cite[(43) and
(46)]{subbarao1993focusing}. Lastly, the structural
similarity measure (SSIM)~\cite{wang2004image} captures
local similarities between the reconstructed and the
ground-truth images.  Desirable outcomes are low values in
NRMSE and HFEN, and high values in SSIM, M1, and M2.  All
methods were finely tuned to achieve the lowest NRMSE. All
reported metric values are mean values of 10 independent
runs.

% \begin{table}
%   \centering
%   \caption{Number of parameter unknowns~\eqref{eq:N.Q>1}}\label{tab:N.Q}
%   \begin{tabular}{|c|c|c|c|} \hline 
%     $Q$ & $\Set{d_q}_{q=1}^{Q-1}$ & $N_Q[M=1]$ & $N_Q[M=7]$ \\ \hline 
%     1 & \_ & \num{11677680} & \num{81743760} \\ \hline 
%     2 & $\Set{8}$ & \num{1357472} & \num{9502304} \\ \hline 
%     3 & $\Set{2, 8}$ & \num{358704} & \num{2510928} \\ \hline
%   \end{tabular}
% \end{table}

It can be seen in \cref{fig:nrmse.v.acceleration}, and more
precisely in \Cref{tab:mrxcat.cartesian,tab:mrxcat.radial},
that MultiL-KRIM matches the performance of
KRIM. Nevertheless, MultiL-KRIM shows remarkable reduction
in computational time by up to \num{65}\% for multiple
kernels (red-colored) and \num{30}\% for a single kernel
(blue-colored). To showcase that the time reduction is
because of the novel factorization approach, the reported
times of KRIM[M] and KRIM[S] \textit{do not}\/ include the
time of KRIM's dimensionality-reduction
pre-step. Furthermore, looking within MultiL-KRIM[$M, Q$],
experiments verify that compared to when $Q=2$, there is
notable speed-up when $Q = 3$ and the inner dimensions
$\Set{ d_q }_{q=1}^{Q-1}$ are chosen appropriately. Due to
intensive fine-tuning, there is only slight difference
between the single-kernel case ($M=1$) versus the
multiple-kernel case ($M=7$). The tests also demonstrate the
better performance of MultiL-KRIM over LRTC-TV, which
deteriorates quickly with high acceleration rates under
Cartesian sampling, and has high computational footprint due
to nuclear-norm minimization. PS-Sparse is a contender with
a trade-off between NRMSE and runtime, but it declines
noticeably at high acceleration rates, while MultiL-KRIM
does not. Additionally, structural quality measures such as
SSIM and HFEN accentuate the difference between MultiL-KRIM
and PS-Sparse. Although MultiL-KRIM does not reach the NRMSE
and SSIM of TDDIP, it produces lower HFEN and higher
sharpness measure M1, \ie, sharper images. More importantly,
MultiL-KRIM offers a less computationally expensive and more
data-driven approach than TDDIP, which requires a carefully
handcrafted prior (manifold structure, here a helix) that
does not explicitly utilize the observed data. Extra
experiments on TDDIP, described in \cref{app:exp.dMRI}
(supplementary file), show that when the complexity of TDDIP
decreases, its NRMSE and SSIM approach those of MultiL-KRIM.
Albeit, increasing the number of unknowns of MultiL-KRIM,
either with more kernels $M$ or larger inner dimensions
$\Set{d_q}_{q=1}^{Q-1}$, does not improve its performance.
These observations intrigue with a ``deep'' architecture for
MultiL-KRIM, such as adding non-linear activation function
layers into the factorization.  This idea is based on the
results that deep matrix factorization outperforms the
standard MMF in matrix completion tasks \cite{fan2018matrix,
  de2021survey}.

\begin{table}[ht!]
  \renewcommand*{\arraystretch}{1.3}
  \caption{Performance on Cartesian sampling (acceleration rate: 20x)}
  \centering \resizebox{\columnwidth}{!}{%
    {\begin{tabular}{|c|l|l|l|l|l|l|} \hline Methods $\setminus$ Metrics & \multicolumn{1}{c|}{\textbf{NRMSE}}
       & \multicolumn{1}{c|}{\textbf{SSIM}} & \multicolumn{1}{c|}{\textbf{HFEN}}
       & \multicolumn{1}{c|}{\textbf{M1}} & \multicolumn{1}{c|}{\textbf{M2}}
       & \multicolumn{1}{c|}{\textbf{Time}} \\  \hline
       \rowcolor{tableMulti}
       \textbf{MultiL-KRIM[$M=7, Q=2$]} & ${\mathbf{0.0443}}$ & $\mathbf{0.8698}$ & ${0.1147}$
       & $37.26$ & ${1.4 \times 10^6}$ & 1.2hrs \\ \hline
       \rowcolor{tableSingle}
       \textbf{MultiL-KRIM[$M=1, Q=2$]} & ${0.0453}$ & ${0.8686}$ & ${0.1269}$ & $37.24$
                                          & ${1.3 \times 10^6}$ & 52min
       \\  \hline
       \rowcolor{tableMulti}
       \textbf{MultiL-KRIM[$M=7, Q=3$]} & ${{0.0444}}$ & ${0.8695}$ & ${0.1149}$ & $37.26$
                                          & ${1.4 \times 10^6}$ & 1.1hrs
       \\ \hline
       \rowcolor{tableSingle}
       \textbf{MultiL-KRIM[$M=1, Q=3$]} & ${0.0454}$ & ${0.8684}$ & ${0.1270}$ & $37.24$
                                          & ${1.3 \times 10^6}$ & 48min
       \\  \hline
       \rowcolor{tableMulti}
       \textbf{KRIM[M]}~\cite{slavakis2022krim} & $\mathbf{0.0443}$ & ${0.8696}$ & $\mathbf{0.1136}$ & $\mathbf{37.34}$
                                          & ${1.4 \times 10^6}$ & 3.3hrs \\ \hline
        \rowcolor{tableSingle}
       \textbf{KRIM[S]}~\cite{slavakis2022krim} & $0.0450$ & $0.8670$ & $0.1149$
       & $37.25$ & $1.3 \times 10^6$ & 55min \\
          \hline \textbf{BiLMDM}~\cite{shetty2020bilmdm} & $0.0488$ & $0.8589$ & $0.1423$ & $37.08$ & $1.3 \times 10^6$
& 40min \\
          \hline \textbf{LRTC-TV}~\cite{Li_Ye_Xu_2017} & $0.1645$ & $0.6943$ & $0.4330$ & $31.52$
& $\mathbf{1.5 \times 10^6}$ & 8.3hrs \\
        \hline
 % \textbf{FCTN}~\cite{zheng2021fctn}& $0.1995$& $0.6308$& $0.4114$& $31.96$&$9.5\times 10^5$ & 1.4hrs\\ \hline
          \textbf{SToRM}~\cite{poddar2016dynamic} & 0.0850
 & $0.8110$ & $0.2504$ & $37.02$ &
          $1.2 \times 10^6$ & 58min \\
          \hline \textbf{PS-Sparse}~\cite{zhao2012image} & $0.0550$ & $0.8198$ & $0.1483$ &
          $37.18$ & $1.2 \times 10^6$ & \textbf{15min} \\ \hline
          % \textbf{L+S} & $0.0466$ & $0.8532$ & $0.1416$ & $37.17$ & $\mathbf{1.4 \times
          % 10^6}$ & 3.9hrs \\ \hline
    \end{tabular}}
    }\label{tab:mrxcat.cartesian}

  \end{table}
  % \FloatBarrier

  % \FloatBarrier
\begin{table}[ht!]
    \renewcommand*{\arraystretch}{1.3}
    \caption{Performance on radial sampling (acceleration rate: 16x)}
    \centering \resizebox{\columnwidth}{!}{%
      {\begin{tabular}{|c|l|l|l|l|l|l|} \hline Methods $\setminus$ Metrics & \multicolumn{1}{c|}{\textbf{NRMSE}}
         & \multicolumn{1}{c|}{\textbf{SSIM}} & \multicolumn{1}{c|}{\textbf{HFEN}}
         & \multicolumn{1}{c|}{\textbf{M1}} & \multicolumn{1}{c|}{\textbf{M2}}
         & \multicolumn{1}{c|}{\textbf{Time}} \\  \hline
         \rowcolor{tableMulti}
         \textbf{MultiL-KRIM[$M=7, Q=2$]} & ${{0.0448}}$ & ${0.8680}$
                                              &${\mathbf{0.1023}}$ & ${37.35}$ & ${1.4 \times 10^6}$ & 1.2hrs
         \\ \hline
         \rowcolor{tableSingle}
         \textbf{MultiL-KRIM[$M=1, Q=2$]} & ${0.0465}$ & ${0.8618}$ & ${0.1305}$ & $37.26$
                                            &${1.4 \times 10^6}$ & 44min
         \\  \hline
         \rowcolor{tableMulti}
         \textbf{MultiL-KRIM[$M=7, Q=3$]} & $0.0448$ & ${0.8675}$ & ${0.1030}$ & ${37.35}$
                                            & ${1.4 \times 10^6}$ & 1hr
         \\ \hline
         \rowcolor{tableSingle}
         \textbf{MultiL-KRIM[$M=1, Q=3$]} & ${0.0465}$ & ${0.8618}$ & ${0.1305}$ & $37.26$
                                            & ${1.4 \times 10^6}$ & 41min
         \\  \hline
         \rowcolor{tableMulti}
         \textbf{KRIM[M]}~\cite{slavakis2022krim} & ${0.0450}$ & ${0.8670}$ & ${0.1136}$ & $37.34$
                                            & ${1.4 \times 10^6}$ & 3hrs \\ \hline
         \rowcolor{tableSingle}
         \textbf{KRIM[S]}~\cite{slavakis2022krim} & $0.0465$ & $0.8618$ & $0.1301$
         & $37.26$ & $1.4 \times 10^6$ & 58min \\
          \hline \textbf{BiLMDM}~\cite{shetty2020bilmdm} & $0.0475$ & $0.8560$ & $0.1491$ & $37.30$ & $1.4 \times 10^6$ & 1.4hrs \\
           \hline \textbf{TDDIP}~\cite{yoo2021time} & $\mathbf{0.0376}$ & $\mathbf{0.8896}$ & $0.1452$ & ${36.24}$ & $1.3 \times 10^6$ & 3.9hrs \\
            % \hline \textbf{TDDIP$^2$}~\cite{yoo2021time} & ${0.0388}$ & ${0.8855}$ & $0.1464$ & ${35.14}$ & $1.3 \times 10^6$ & 3.5hrs \\
            % \hline \textbf{TDDIP$^3$}~\cite{yoo2021time} & ${0.0415}$ & ${0.8691}$ & $0.1642$ & ${35.87}$ & $1.4 \times 10^6$ & 2hrs \\
            % \hline \textbf{TDDIP$^4$}~\cite{yoo2021time} & ${0.0422}$ & ${0.8680}$ & $0.1668$ & ${35.68}$ & $1.4 \times 10^6$ & 1.2hrs \\
           \hline \textbf{LRTC-TV}~\cite{Li_Ye_Xu_2017} & $0.0738$ & $0.8063$ & $0.3725$ & $37.06$ & $\mathbf{1.7 \times 10^6}$ & 8.6hrs \\
          \hline \textbf{SToRM}~\cite{poddar2016dynamic} & $0.0753$ & $0.8319$ & $0.3694$ & $\mathbf{37.38}$ &
          $1.6 \times 10^6$ & 30min \\
          \hline \textbf{PS-Sparse}~\cite{zhao2012image} & $0.0496$ & $0.7908$ & $0.1733$ &
          $37.31$ & $1.4 \times 10^6$ & \textbf{15min} \\ \hline
    \end{tabular}}
    }\label{tab:mrxcat.radial}
\end{table}

%%%%%%%%%%%%%%%%%%%%%%%%%%%%%%%%%%%%%%%%%%%%%%%%%%%%%%%%%%%%%%%%%%%%%%%%%%%%%%%%%%%%%%%%%%%%%%%

\section{Conclusions}\label{sec:conclusion}

This paper introduced MultiL-KRIM, a nonparametric
(kernel-based) data-imputation-by-regression framework. Two
important application domains were considered for
validation: time-varying-graph-signal (TVGS) recovery, and
reconstruction of highly accelerated
dynamic-magnetic-resonance-imaging (dMRI) data. Numerical
tests on TVGS recovery demonstrated MultiL-KRIM's
outperformance over state-of-the-art ``shallow''
methods. Furthermore, via tests on dMRI data, MultiL-KRIM
was shown to match the performance of its predecessor
KRIM~\cite{slavakis2022krim} with significant reduction in
computational time. MultiL-KRIM outperformed
state-of-the-art ``shallow'' methods also on dMRI data, and
although it did not reach the accuracy of the
deep-image-prior-based TDDIP~\cite{yoo2021time}, it offered
a more explainable and computationally tractable
path. Prospectively, to further enhance MultiL-KRIM's
performance, current research effort includes also its
``deep'' variations.

\bibliography{ref.bib}

%%%%%%%%%%%%%%%%%%%%%%%%%%%%%%%%%%%%%%%%%%%%%%%%%%%%%%%%%%%%%%%%%%%%%%%%%%%%%%%%%%%%%%%%%%%%%%%

\clearpage
\newpage

\setcounter{page}{1}

\begin{center}
  \Large\textbf{Supplementary File}
\end{center}

\appendices
\crefalias{figure}{appfig}
\crefalias{table}{apptab}
\crefalias{equation}{appeq}

\section{Reproducing Kernel Hilbert Spaces}\label[appendix]{app:RKHS}

A feature mapping $\varphi: \Complex^{\nu} \to \mathscr{H}$
maps a vector $\mathbfit{l}\in \Complex^{\nu}$ to
$\varphi(\mathbfit{l})$ in a feature space $\mathscr{H}$.
% , the essence of the
% modeling approach is that $\{ \varphi(\mathbfit{l}_k) \}_{k=1}^{N_{\mathit{l}}}$ lie into or
% close to an unknown-to-the-user smooth manifold $\mathscr{M}$~\cite{RobbinSalamon:22} embedded
% in $\mathscr{H}$; see \cref{fig:manifold.kernel.space}. 
To provide structured solutions, it is assumed that
$\mathscr{H}$ is a complex-valued reproducing kernel Hilbert
space (RKHS), associated with a reproducing kernel
$\kappa(\cdot, \cdot): \Complex^{\nu}\times \Complex^{\nu}
\to \Complex$, with rich attributes in approximation
theory~\cite{aronszajn1950theory, scholkopf2002learning,
  steinwart2006explicit, bouboulis2010extension,
  Slavakis:OL:2014}. In the RKHS setting, the feature
mapping becomes
$\varphi(\mathbfit{l}) \coloneqq \kappa(\mathbfit{l}, \cdot
) \in \mathscr{H}$,
$\forall \mathbfit{l}\in \Complex^{\nu}$. Most well-known
kernels for RKHSs are:
\begin{enumerate*}[label = \textbf{(\roman*)}]

\item The linear kernel
  $\kappa_{\text{L}} (\mathbfit{l} , \mathbfit{l}^{\prime})
  \coloneqq \mathbfit{l}^{\hermconj} \mathbfit{l}^{\prime}$,
  where $\hermconj$ denotes complex conjugate vector/matrix
  transposition;

\item the Gaussian kernel
  $\kappa_{\text{G}; \gamma} (\mathbfit{l},
  \mathbfit{l}^{\prime}) \coloneqq \text{exp} [ -\gamma
  (\mathbfit{l} - \overline{\mathbfit{l}}^{\prime}
  )^{\intercal} ( \mathbfit{l} -
  \overline{\mathbfit{l}}^{\prime} )
  ]$~\cite{steinwart2006explicit, bouboulis2010extension},
  where $\gamma \in \Real_{>0}$, superscript $\intercal$
  stands for vector/matrix transposition, and the overline
  symbol denotes entry-wise complex conju- gation of a
  scalar/vector/matrix; and

\item the polynomial kernel
  $\kappa_{\text{P}; (c,r)} ( \mathbfit{l},
  \mathbfit{l}^{\prime} ) \coloneqq (
  \mathbfit{l}^{{\hermconj}} \mathbfit{l}^{\prime} + c)^r$,
  where $r \in \IntegerPP$, $c \in \Complex$.

\end{enumerate*}
The list of kernels used for numerical tests is shown in
\cref{app:tab.kernels.list}.

\begin{table}[!ht]
    \centering
    \caption{Dictionary of kernels. }
    \begin{tabular}{|c|c|c|} \hline 
         Kernel index $m$  &  Kernel function & Parameters\\ \hline 
         1&  Gaussian& $\sigma=0.2$ \\ \hline 
         2&  Gaussian& $\sigma=0.4$ \\ \hline 
         3&  Gaussian& $\sigma=0.8$ \\ \hline 
         4&  Polynomial& $r=1$ \\ \hline 
         5&  Polynomial& $r=2$ \\ \hline 
         6&  Polynomial& $r=3$ \\ \hline 
         7&  Polynomial& $r=4$ \\ \hline
    \end{tabular}
    \label{app:tab.kernels.list}
\end{table}

\section{Derivation of a compact MultiL-KRIM data 
modeling}\label[appendix]{app:modeling.compact}

\begin{figure*}[!t]
  \begin{tikzpicture}
    \node[text width=\textwidth] (compact) {
      \begin{align}
        \vect{X} \approx
        \underbrace{
        \left[\begin{smallmatrix}
          \vect{D}_1^{(1)} & \vect{D}_2^{(1)} & \ldots & \vect{D}_M^{(1)}
        \end{smallmatrix}\right]
        }_{\vectcal{D}_1}
        \underbrace{\left[\begin{smallmatrix}
          \vect{D}_1^{(2)} & 0 & \cdots & 0 \\
          0 & \vect{D}_2^{(2)} & \cdots & 0 \\
          \vdots & \vdots & \ddots & \vdots \\
          0 & 0 & \cdots & \vect{D}_M^{(2)} \\
        \end{smallmatrix}\right]}_{\vectcal{D}_2}
        \ldots
        \underbrace{\left[\begin{smallmatrix}
          \vect{D}_1^{(Q)} & 0 & \cdots & 0 \\
          0 & \vect{D}_2^{(Q)} & \cdots & 0 \\
          \vdots & \vdots & \ddots & \vdots \\
          0 & 0 & \cdots & \vect{D}_M^{(Q)} \\
        \end{smallmatrix}\right]}_{\vectcal{D}_Q}
        \underbrace{\left[\begin{smallmatrix}
          \vect{K}_1 & 0 & \cdots & 0 \\
          0 & \vect{K}_2 & \cdots & 0 \\
          \vdots & \vdots & \ddots & \vdots \\
          0 & 0 & \cdots & \vect{K}_M \\
        \end{smallmatrix}\right]}_{\vectcal{K}}
        \underbrace{\left[\begin{smallmatrix}
          \vect{B}_1 \\
          \vect{B}_2 \\
          \vdots \\
          \vect{B}_M
        \end{smallmatrix}\right]}_{\vectcal{B}} \,.
        \label{app:eq.fig.multil.compact}
      \end{align}
    };
  \end{tikzpicture}
\end{figure*}

By defining the ``supermatrices''
\begin{alignat}{2}
  \vectcal{D}_1
  & {} \coloneqq {} [\vect{D}_1^{(1)}, \vect{D}_2^{(1)},
    \ldots, \vect{D}_M^{(1)}]
  && \in \Complex^{I_0 \times d_1M} \,, \notag\\
  \vectcal{D}_q
  & \coloneqq \bdiag{(\vect{D}_1^{(q)}, \vect{D}_2^{(q)},
    \ldots, \vect{D}_M^{(q)})}
  && \in \Complex^{d_{q-1}M \times d_{q}M} \,, \notag\\
  & \hphantom{{} \coloneqq {}} \forall q\in \{ 2, \ldots,
    Q\}\,, && \notag\\
  \vectcal{K}
  & \coloneqq \bdiag{(\vect{K}_1, \vect{K}_2, \ldots, \vect{K}_M)}
  && \in \Complex^{MN_{\mathit{l}} \times MN_{\mathit{l}}}
     \,, \label{app:eq.kernel.supermatrix}\\
  \vectcal{B}
  & \coloneqq [\vect{B}_1^{\hermconj},
    \vect{B}_2^{\hermconj}, \ldots,
    \vect{B}_{M}^{\hermconj}]^{\hermconj}
  && \in \Complex^{MN_{\mathit{l}} \times I_N} \,, \notag
\end{alignat}
data modeling \eqref{eq:multi.kernel.nodimred} of
MultiL-KRIM, incorporated in the generic inverse problem
\eqref{eq:multil.manifold.right}, takes the form in
\cref{app:eq.fig.multil.compact}.

\section{Solving the TVGS inverse problem}\label[appendix]{app:TVGS.solve}

This section details the solutions to the TVGS
sub-tasks~\eqref{eq:graph.min.X} and
\eqref{eq:graph.min.D}. Sub-task~\eqref{eq:graph.min.B}
involves a composite convex loss function with affine
constraints which can be efficiently solved
by~\cite{slavakis2018fejer}.

\subsection{Solving~\eqref{eq:graph.min.X}}\label[appendix]{app:TVGS.solve.X}

Given the index set $\Omega$, define the affine constraint
$\mathcal{A}_{\Omega} \coloneqq \{\vect{X}\in \Complex^{I_0
  \times I_N} \given \mathscr{S}_{\Omega} (\vect{Y}) =
\mathscr{S}_{\Omega} (\vect{X})\}$, and notice
that~\eqref{eq:graph.min.X} takes the form:
\begin{align}
  \hat{\vect{X}}^{(n+1/2)} = \arg\min\nolimits_{\vect{X} \in \mathcal{A}_{\Omega}}
  \mathcal{L}_X (\vect{X}) \,, \label{tvgs.subtask.x}
\end{align}
where the differentiable strongly convex loss
\begin{alignat*}{2}
  \mathcal{L}_X (\vect{X})
  & {} \coloneqq {}
  && \tfrac{1}{2} \norm{{\vect{X}} - \hat{\vect{M}}^{(n)}
     }_{\textnormal{F}}^2
      + \tfrac{\lambda_L}{2} \tr(\vect{X}^\intercal
      \vect{L}_{\epsilon}^{\beta} \vect{X} \mathbf{\Delta}
     \mathbf{\Delta}^\intercal) \\
  &&& + \tfrac{\tau_X}{2} \norm{ \vect{X} -
      \hat{\vect{X}}^{(n) }}_{\textnormal{F}}^2 \,,
\end{alignat*}
and
$\hat{\vect{M}}^{(n)} \coloneqq \hat{\vectcal{D}}_1^{(n)}
\cdots \hat{\vectcal{D}}_Q^{(n)} \vectcal{K}
\hat{\vectcal{B}}^{(n)}$. For convenience in notations, let
$\vect{X}_* \coloneqq \hat{\vect{X}}^{(n + 1/2) }$.

Problem~\eqref{tvgs.subtask.x} is a convex minimization task
with affine constraints, so it can be solved iteratively by
the efficient~\cite{slavakis2018fejer}. Nevertheless,
\eqref{tvgs.subtask.x} possesses also a closed-form
solution. To formulate that closed-form solution, define the
sampling matrix
$\vect{S}_{\Omega} \in \Complex^{I_0 \times I_N}$ as
follows: $[\vect{S}_{\Omega}]_{ij} \coloneqq 1$, if
$(i,j) \in \Omega$, while
$[\vect{S}_{\Omega}]_{ij} \coloneqq 0$, if
$(i,j) \notin \Omega$. Then,
$\mathscr{S}_{\Omega} (\vect{X}) = \vect{S}_{\Omega} \odot
\vect{X}$, where $\odot$ stands for the Hadamard
product. Furthermore, define the complement set
$\Omega^{\complement} \coloneqq \Set{ (i,j) \in \Set{1,
    \ldots, I_0} \times \Set{1, \ldots, I_N} \given (i,j)
  \notin \Omega }$. It can be verified that any
$\vect{X} \in \Complex^{I_0 \times I_N}$ can be written as
$\vect{X} = \mathscr{S}_{\Omega} (\vect{X}) +
\mathscr{S}_{\Omega^{\complement}}(\vect{X}) =
\vect{S}_{\Omega} \odot \vect{X} +
\vect{S}_{\Omega^{\complement}} \odot \vect{X}$. Moreover,
let
$\tovec\colon \Complex^{I_0 \times I_N} \to
\Complex^{I_0I_N}$ denote the standard vectorization
operator which turns matrices into vectors, and let its
inverse mapping be denoted by
$\tovec^{-1} \colon \Complex^{I_0I_N} \to \Complex^{I_0
  \times I_N}$. Let also the ``vectorization'' of $\Omega$
as the one-dimensional index set
$\tovec \Omega \coloneqq \Set{ i\in \Set{1, \ldots, I_0I_N}
  \given [\tovec ( \vect{S}_{\Omega} )]_i = 1 }$. Also, for
any $I_0I_N \times 1$ vector $\vect{v}$, let
$\vect{v} \vert_{\tovec \Omega}$ be the
$\lvert \tovec \Omega \rvert \times 1$ sub-vector of
$\vect{v}$ formed by the entries of $\vect{v}$ at indices
$\tovec \Omega$.

According to~\cite[Prop.~26.13]{hb.plc.book}, the desired
solution satisfies
$\vect{X}_* \in \mathcal{A}_{\Omega} \Leftrightarrow
\mathscr{S}_{\Omega} (\vect{Y}) = \mathscr{S}_{\Omega}
(\vect{X}_*)$ and
$\nabla \mathcal{L}_X ( \vect{X}_* ) \in
(\mathcal{A}_{\Omega} - \mathscr{S}_{\Omega}
(\vect{Y}))^{\perp} = \Set{ \vect{X} \in \Complex^{I_0
    \times I_N} \given \vect{0} = \mathscr{S}_{\Omega}
  (\vect{X})}^{\perp} = \Set{ \vect{X} \in \Complex^{I_0
    \times I_N} \given \vect{0} =
  \mathscr{S}_{\Omega^{\complement}} (\vect{X})}$, where
$\perp$ stands for the orthogonal complement of a linear
subspace. Hence,
\begin{alignat*}{2}
  &&& \hspace{-20pt} \nabla\mathcal{L}_X(\vect{X}_*) \\
  & {} = {} && (1 + \tau_X) \vect{X}_* + \lambda_L
               \vect{L}_{\epsilon}^\beta \vect{X}_* \mathbf{\Delta}\mathbf{\Delta}^\intercal -
               \hat{\vect{M}}^{(n)} - \tau_X
               \hat{\vect{X}}^{(n)} \\
  & = && (1 + \tau_X) \vect{X}_* + \lambda_L
         \vect{L}_{\epsilon}^\beta [
         \mathscr{S}_{\Omega^{\complement}} (\vect{X}_*) +
         \mathscr{S}_{\Omega} (\vect{X}_*) ] \mathbf{\Delta}\mathbf{\Delta}^\intercal \\
  &&& - \hat{\vect{M}}^{(n)} - \tau_X \hat{\vect{X}}^{(n)}
  \\
  & = && (1 + \tau_X) \vect{X}_* + \lambda_L
         \vect{L}_{\epsilon}^\beta
         \mathscr{S}_{\Omega^{\complement}} (\vect{X}_*)
         \mathbf{\Delta}\mathbf{\Delta}^\intercal \\
  &&& + \lambda_L \vect{L}_{\epsilon}^\beta
      \mathscr{S}_{\Omega} (\vect{Y})
      \mathbf{\Delta}\mathbf{\Delta}^\intercal -
      \hat{\vect{M}}^{(n)} - \tau_X
      \hat{\vect{X}}^{(n)} \,.
\end{alignat*}
The requirement $\mathscr{S}_{\Omega^{\complement}} (\nabla \mathcal{L}_X ( \vect{X}_* )) =
\vect{0}$ yields
\begin{align*}
  & (1 + \tau_X) \vect{S}_{\Omega^{\complement}} \odot
    \mathscr{S}_{\Omega^{\complement}}
    (\vect{X}_*) + \lambda_L \vect{S}_{\Omega^{\complement}}
    \odot \vect{L}_{\epsilon}^\beta
    \mathscr{S}_{\Omega^{\complement}} (\vect{X}_*) \mathbf{\Delta}\mathbf{\Delta}^\intercal \\
  & = - \lambda_L \vect{S}_{\Omega^{\complement}} \odot
    \vect{L}_{\epsilon}^\beta
    \mathscr{S}_{\Omega} (\vect{Y}) \mathbf{\Delta}\mathbf{\Delta}^\intercal +
    \vect{S}_{\Omega^{\complement}} \odot
    (\hat{\vect{M}}^{(n)} + \tau_X \hat{\vect{X}}^{(n)}) \,,
\end{align*}
and after vectorizing by $\tovec(\cdot)$,
\begin{align*}
  & (1 + \tau_X) \diag (\tovec
    \vect{S}_{\Omega^{\complement}}) \cdot \tovec \left(
    \mathscr{S}_{\Omega^{\complement}} (\vect{X}_*) \right)  \\
  & + \lambda_L \diag (\tovec
    \vect{S}_{\Omega^{\complement}}) \cdot (
    \mathbf{\Delta}\mathbf{\Delta}^\intercal \otimes
    \vect{L}_{\epsilon}^\beta) \tovec \left(
    \mathscr{S}_{\Omega^{\complement}} (\vect{X}_*) \right) \\
  & {} = {} - \lambda_L \diag (\tovec
    \vect{S}_{\Omega^{\complement}}) \tovec(
    \vect{L}_{\epsilon}^\beta \mathscr{S}_{\Omega} (\vect{Y})
    \mathbf{\Delta}\mathbf{\Delta}^\intercal ) \\
  & \hphantom{ {} = {} } + \diag (\tovec
    \vect{S}_{\Omega^{\complement}}) \tovec(
    \hat{\vect{M}}^{(n)} + \tau_X \hat{\vect{X}}^{(n)} ) \,,
\end{align*}
where $\otimes$ denotes the Kronecker product. Consequently,
by keeping only the entries at indices
$\tovec (\Omega^{\complement})$,
\begin{align*}
  & (1 + \tau_X) \tovec \left( \mathscr{S}_{\Omega^{\complement}} (\vect{X}_*) \right)
    \vert_{\tovec (\Omega^{\complement})} \\
  & + \lambda_L ( \mathbf{\Delta}\mathbf{\Delta}^\intercal \otimes
    \vect{L}_{\epsilon}^\beta)\vert_{ \tovec
    (\Omega^{\complement}) \times
    \tovec (\Omega^{\complement}) } \cdot \tovec \left(
    \mathscr{S}_{\Omega^{\complement}} (\vect{X}_*) \right)
    \vert_{\tovec
    (\Omega^{\complement})} \\
  & {} = {} - \lambda_L \tovec(
    \vect{L}_{\epsilon}^\beta \mathscr{S}_{\Omega} (\vect{Y})
    \mathbf{\Delta}\mathbf{\Delta}^\intercal )\vert_{\tovec
    (\Omega^{\complement})} \\
  & \hphantom{ {} = {} } + \tovec( \hat{\vect{M}}^{(n)} +
    \tau_X \hat{\vect{X}}^{(n)} )\vert_{\tovec
    (\Omega^{\complement})} \,,
\end{align*}
where
$( \mathbf{\Delta}\mathbf{\Delta}^\intercal \otimes
\vect{L}_{\epsilon}^\beta )\vert_{ \tovec
  (\Omega^{\complement}) \times \tovec
  (\Omega^{\complement})}$ is the submatrix extracted from
$\mathbf{\Delta}\mathbf{\Delta}^\intercal \otimes
\vect{L}_{\epsilon}^\beta$ by keeping only the rows and
columns indicated by $\tovec (\Omega^{\complement})$. It can
be readily verified that
\begin{alignat}{3}
  &&&&& \hspace{-30pt} \tovec \bigl( \mathscr{S}_{\Omega^{\complement}}
        (\vect{X}_*) \bigr) \vert_{\tovec
        (\Omega^{\complement})} \notag \\
  & {} = {} && \Bigl( (1 && + \tau_X)\vect{I}_{\vert \tovec
                            (\Omega^{\complement}) \rvert} +
                            \lambda_L (
                            \mathbf{\Delta}\mathbf{\Delta}^\intercal
                            \otimes
                            \vect{L}_{\epsilon}^\beta)\vert_{
                            \tovec (\Omega^{\complement})
                            \times \tovec
                            (\Omega^{\complement})}
                            \Bigr)^{-1} \notag\\
  &&& \cdot \Bigl( && - \lambda_L \tovec(
                      \vect{L}_{\epsilon}^\beta
                      \mathscr{S}_{\Omega}
      (\vect{Y}) \mathbf{\Delta}\mathbf{\Delta}^\intercal
                      )\vert_{\tovec (\Omega^{\complement})}
                      \notag\\
  &&& && + \tovec( \hat{\vect{M}}^{(n)} + \tau_X
         \hat{\vect{X}}^{(n)} )\vert_{\tovec
         (\Omega^{\complement})} \Bigr)
         \,. \label{Xstar.partial}
\end{alignat}
Recovering $\mathscr{S}_{\Omega^{\complement}} (\vect{X}_*)$
from~\eqref{Xstar.partial} is a straightforward task: take
an all-zero $I_0I_N \times 1$ vector, replace its zero
entries at positions $\tovec (\Omega^{\complement})$ by the
corresponding entries of~\eqref{Xstar.partial}, and then
apply $\tovec^{-1}(\cdot)$ to recover
$\mathscr{S}_{\Omega^{\complement}} (\vect{X}_*)$. Finally,
\begin{align*}
  \hat{\vect{X}}^{(n + 1/2) } = \vect{X}_*
  & = \mathscr{S}_{\Omega} (\vect{X}_*) +
    \mathscr{S}_{\Omega^{\complement}}(\vect{X}_*) \\
  & = \mathscr{S}_{\Omega} (\vect{Y}) +
    \mathscr{S}_{\Omega^{\complement}}(\vect{X}_*) \,.
\end{align*}

\subsection{Solving~\eqref{eq:graph.min.D}}\label[appendix]{app:TVGS.solve.D}

For compact notations, let
$\hat{\vectcal{D}}_{Q+1}^{(n)} \coloneqq \vectcal{K}$ and
$\hat{\vectcal{D}}_{Q+2}^{(n)} \coloneqq
\hat{\vectcal{B}}^{(n)}$. For any $q\in \Set{2, \ldots, Q}$,
define
\begin{align*}
  \hat{\vect{L}}_q^{(n)} & \coloneqq \prod\nolimits_{q^{\prime}=1}^{q-1}
                           \hat{\vectcal{D}}_{q^{\prime}}^{(n)}
                           \,, \\
  \hat{\vect{R}}_q^{(n)} & \coloneqq
                           \prod\nolimits_{q^{\prime} =
                           q+1}^{Q+2}
                           \hat{\vectcal{D}}_{q^{\prime}}^{(n)}
                           \,,
\end{align*}
while $\hat{\vect{L}}_1^{(n)} \coloneqq \vect{I}_{I_0}$ and
$\hat{\vect{R}}_1^{(n)} \coloneqq \prod_{q^{\prime} =
  2}^{Q+2} \hat{\vectcal{D}}_{q^{\prime}}^{(n)}$, so that
the gradient of the loss in~\eqref{eq:graph.min.D} takes the
form
\begin{alignat*}{2}
  \nabla \mathcal{L}_D(\vectcal{D}_q)
  & {} = {} && \hat{\vect{L}}_q^{(n)} \vectcal{D}_q
               \hat{\vect{R}}_q^{(n)} - \hat{\vect{X}}^{(n)}
  \\
  &&& + \lambda_2 \vectcal{D}_q + \tau_D (\vectcal{D}_q - \hat{\vectcal{D}}_q^{(n)}) \,.
\end{alignat*}
The solution $\hat{\vectcal{D}}_q^{(n + 1/2) }$
to~\eqref{eq:graph.min.D} takes a closed form, and the way
to obtain it follows the steps of \cref{app:TVGS.solve.X},
where the affine constraint to amount for the block-diagonal
form of $\vectcal{D}_q$, for $q\in \Set{2, \ldots, Q}$,
becomes
$\mathcal{A}_{\Omega} \coloneqq \{ \vectcal{D}_q \in
\Complex^{d_{q-1}M \times d_{q}M} \given \vect{0} =
\mathscr{S}_{\Omega} ( \vectcal{D}_q )\}$, with the index
set $\Omega$ comprising all pairs of indices positioned off
the diagonal blocks. In the case of $\vectcal{D}_1$, no
affine constraint (no block-diagonal structure) is necessary
to solve~\eqref{eq:graph.min.D}. Solving
$\nabla \mathcal{L}_D( \vectcal{D}_1 ) = \vect{0}$ for
$\vectcal{D}_1$ will do the job.

\section{Solving the dMRI inverse problem}\label[appendix]{app:dMRI.solve}

This section expands the details of the dMRI inverse problem
\eqref{eq:dmri.task.general}.  Similarly to the discussion
in \cref{sec:graph.inv}, and upon defining the following
tuple of estimates
$\forall n\in\IntegerP, \forall k\in\Set{0,1}$,
\begin{align}
  \hat{\mathbfcal{O}}^{(n+k/2)} \coloneqq (
  {} & {} \hat{\vect{X}}^{(n+k/2)}, \hat{\vect{Z}}^{(n+k/2)}, \hat{\vectcal{D}}_1^{(n+k/2)},
       \notag \\
     & \ldots, \hat{\vectcal{D}}_Q^{(n+k/2)}, \hat{\vectcal{B}}^{(n+k/2)}
       )\,, \label{Oh.tuple.dmri}
\end{align}
\cref{alg:multil.general}
solves~\eqref{eq:dmri.task.general} via computing solutions
of the following convex sub-tasks:
\begin{subequations}\label{eq:dmri.subtasks}%
  \begin{alignat}{3}%
    && \hat{\vect{X}}^{(n + 1/2) }
    && \in \arg\min_{ {\vect{X}} } {}
    & {}\ \tfrac{1}{2} \norm{{\vect{X}} -
      \hat{\vectcal{D}}_1^{(n)} \cdots
      \hat{\vectcal{D}}_Q^{(n)} \vectcal{K}
      \hat{\vectcal{B}}^{(n)} }_{\textnormal{F}}^2 \notag \\
    &&&&& {}\ + \tfrac{\lambda_2}{2} \norm{ \hat{\vect{Z}}^{(n)} -
          \mathscr{F}_{\text{t}}( {\vect{X}} )
          }_{\textnormal{F}}^2 \notag\\
    &&&&& {}\ + \tfrac{\tau_X}{2} \norm{ \vect{X} -
          \hat{\vect{X}}^{(n) }}_{\textnormal{F}}^2 \notag \\
    &&&& \text{s.to} {} & {}\ \mathscr{S}_{\Omega}(\vect{Y})
                          = \mathscr{S}_{\Omega}
                          \mathscr{F}( {\vect{X}} )
                          \,, \label{eq:dmri.min.X} \\
    && \hat{\vect{Z}}^{(n + 1/2) }
    && \in \arg\min_{\vect{Z}} {}
    & {}\ \tfrac{\lambda_2}{2} \norm{ \vect{Z} -
      \mathscr{F}_{\text{t}}( \hat{\vect{X}}^{(n)})
      }^2_{\textnormal{F}} + \lambda_3 \norm{ \vect{Z} }_1 \notag \\
    &&&&& {}\ + \tfrac{\tau_Z}{2} \norm{ \vect{Z} -
          \hat{\vect{Z}}^{(n)
          }}_{\textnormal{F}}^2 \,, \label{eq:dmri.min.Z}\\
    && \hat{\vectcal{D}}_q^{(n + 1/2)} {}
    && {} \in \arg\min_{\vectcal{D}_q} {}
    & {}\ \tfrac{1}{2} \norm{ \hat{\vect{X}}^{(n)} -
      \hat{\vectcal{D}}_1^{(n)} \cdots
      \vectcal{D}_q \cdots \hat{\vectcal{D}}_Q^{(n)} \vectcal{K}
      \hat{\vectcal{B}}^{(n)} }_{\textnormal{F}}^2 \notag \\
    &&&& {} & {}\ + \tfrac{\lambda_4}{2} \norm{
              \vectcal{D}_q }_{\textnormal{F}}^2
              + \tfrac{\tau_D}{2} \norm{ \vectcal{D}_q -
              \hat{\vectcal{D}}_q^{(n)} }_{\textnormal{F}}^2
              \,, \notag \\
    &&&& \text{s.to} {} & {}\ \vectcal{D}_q\ \textnormal{is
                          block diagonal}\ \forall q\in
                          \Set{2, \ldots, Q}
                          \,, \label{eq:dmri.min.D} \\
    && \hat{\vectcal{B}}^{(n + 1/2)}
    && \in \arg \min_{\vectcal{B}} {}
    & {}\ \tfrac{1}{2} \norm{ \hat{\vect{X}}^{(n)}-
      \hat{\vectcal{D}}_1^{(n)} \cdots
      \hat{\vectcal{D}}_Q^{(n)} \vectcal{K} \vectcal{B}
      }_{\textnormal{F}}^2 \notag \\
    &&&&& {}\ + \lambda_1 \norm{ \vectcal{B} }_1
          + \tfrac{\tau_B}{2} \norm{ \vectcal{B} -
          \hat{\vectcal{B}}^{(n)} }_{\textnormal{F}}^2 \notag \\
    &&&& \text{s.to} {}
    & {}\ \vect{1}_{N_{\mathit{l}}}^{{\hermconj}} \vect{B}_m
      = \vect{1}_{I_N}^{{\hermconj}}\,,
          \forall m \in \{1, \ldots, M\}
      \,. \label{eq:dmri.min.B}
  \end{alignat}%
\end{subequations}%
% Sub-task~\eqref{eq:dmri.min.B} is a composite convex minimization task under affine
% constraints, and thus can be solved by~\cite{slavakis2018fejer}, while~\eqref{eq:dmri.min.X},
% \eqref{eq:dmri.min.Z} and~\eqref{eq:dmri.min.D} have closed-form
% solutions~\cite{slavakis2022krim}. In particular, the unique solution to~\eqref{eq:dmri.min.Z}
% is given by the well-known soft-thresholding
% operator~\cite{slavakis2022krim}.
Sub-task \eqref{eq:dmri.min.B} is composite convex with
affine constraints and can be solved
by~\cite{slavakis2018fejer}; details are provided
in~\cite{slavakis2022krim}. Sub-task \eqref{eq:dmri.min.D}
is similar to sub-task \eqref{eq:graph.min.D} of the TVGS
case, so that its solution follows
\cref{app:TVGS.solve.D}. The following are the solutions to
sub-tasks \eqref{eq:dmri.min.X} and \eqref{eq:dmri.min.Z}.

\subsection{Solving \eqref{eq:dmri.min.X}}\label[appendix]{app:dMRI.solve.X}

Following~\cite{slavakis2022krim}, the unique solution
$\hat{\vect{X}}^{(n+1/2)}$ to \eqref{eq:graph.min.X} is
\begin{align*}
  \hat{\vect{X}}^{(n+1/4)}
  & \coloneqq C_X \left( \hat{\vect{M}}^{(n)} + \lambda_2
    I_N \mathscr{F}_t^{-1}(\hat{\vect{Z}}^{(n)}) + \tau_X\hat{\vect{X}}^{(n)}
    \right) \,, \notag \\
  \hat{\vect{X}}^{(n+1/2)}
  & \coloneqq
    \mathscr{F}^{-1}\mathscr{S}_\Omega(\vect{Y})
    + \mathscr{F}^{-1}\mathscr{S}_{\Omega^{\complement}} \mathscr{F}(\hat{\vect{X}}^{(n+1/4)})
    \,,
\end{align*}
where $C_X \coloneqq 1/(1+\lambda_2 I_N + \tau_X)$.

\subsection{Solving \eqref{eq:dmri.min.Z}}\label[appendix]{app:dMRI.solve.Z}

Consider a convex function
$g(\cdot): \Complex^{m\times n}\to \Real \cup \Set{ +\infty
}$ and $\lambda \in \Real_{>0}$.  The proximal mapping
$\prox_{\lambda g}(\cdot): \Complex^{m\times n} \to
\Complex^{m\times n}$ is defined as
$\prox_{\lambda g}(\vect{A}) \coloneqq \arg\min_{\vect{B}
  \in \Complex^{m\times n}} \lambda g(\vect{B}) + (1/2)
\norm{\vect{A}-\vect{B}}_{\text{F}}^2$~\cite{hb.plc.book}. When
$g$ is the $\ell_1$-norm $\norm{\cdot}_1$, the $(i,j)$th
entry of $\prox_{\lambda\norm{\cdot}_1}(\vect{A})$ is
obtained by the soft-thresholding rule~\cite{hb.plc.book}:
\begin{align}
    [\prox_{\lambda\norm{\cdot}_1}(\vect{A})]_{ij} = [\vect{A}]_{ij} \left(1 -
    \tfrac{\lambda}{\max\{\lambda, \lvert[\vect{A}]_{ij}\rvert\}} \right)
    \,. \label{eq:soft.thresholding}
\end{align}
Instead of $\prox_{\lambda\norm{\cdot}_1}(\vect{A})$, it is
more common to use notation
$\textnormal{Soft}(\vect{A}, \lambda)$. As such, the
solution $\hat{\vect{Z}}^{(n+1/2)}$ of \eqref{eq:dmri.min.Z}
is provided by
\begin{align*}
    \hat{\vect{Z}}^{(n+1/2)} =
    \text{Soft}[\mathscr{F}_t(\hat{\vect{X}}^{(n)})
    +\tfrac{\tau_Z}{\lambda_2} \hat{\vect{Z}}^{(n)}, \tfrac{\lambda_3}{\lambda_2}]
    \,.
\end{align*}

\section{More results for TVGS
  recovery}\label[appendix]{app:exp.TVGS}

This section provides additional figures and tables
containing experimental results of TVGS recovery (\cf
\cref{sec:graph.exp}).

\begin{table}[!ht]
  \centering
  \caption{Categorization of TVGS recovery methods.}
  \begin{tabular}{|c|ccc|}
    \hline
    Method & Factorization/structured & Low-rank & Kernel-based \\
    \hline
    GraphTRSS~\cite{giraldo2022reconstruction} & \ding{55} & \ding{55} & \ding{55} \\
    KRG~\cite{venkitaraman2019predicting} & \ding{51} & \ding{55} & \ding{51} \\
    KGL~\cite{pu2021kernel} & \ding{51} & \ding{55} & \ding{51} \\
    LRDS~\cite{mao2018spatio} & \ding{55} & \ding{51} & \ding{55} \\
    MMF~\cite{cichocki2007multilayer} & \ding{51} & \ding{51} & \ding{55} \\
    NBP~\cite{bazerque2013nonparametric} & \ding{51} & \ding{51} & \ding{51} \\
    \hline
    MultiL-KRIM & \ding{51} & \ding{51} & \ding{51} \\
    \hline
  \end{tabular}
    \label{app:tab.TVGS.category}
\end{table}

\subsection{Performance metrics}\label[appendix]{app:exp.TVGS.metrics}

Recall that the original TVGS data is $\vect{Y}$ and the
reconstructed data is $\vect{X}$, which are both matrices of
size $I_0\times I_N$.  The error metrics are defined as
follows,
\begin{align*}
  \text{RMSE} & \coloneqq \frac{\norm{\vect{X}-\vect{Y}}_{\textnormal{F}}} {\sqrt{I_0\cdot
                I_N}} \,, \\
  \text{MAE} & \coloneqq \frac{\norm{\vect{X} - \vect{Y}}_1}{I_0\cdot I_N} \,, \\
  \text{MAPE} & \coloneqq \frac{1}{I_0\cdot I_N} \sum \nolimits_{i=1}^{I_0}
                \sum\nolimits_{j=1}^{I_N} \left\lvert \frac{x_{ij}-y_{ij}}{y_{ij}} \right\rvert
                \,.
\end{align*}

\subsection{Navigator- and landmark-point analysis}\label[appendix]{app:exp.TVGS.nav_lm}

To examine the effect of different navigator-data formation
and landmark-point selection strategies, tests are conducted
on the (D1, P1,$\cdot$,$\cdot$) scenario.
\cref{app:tab.navigator.analysis} shows the error metrics of
MultiL-KRIM under scenario (D1, P1,$\cdot$, L1) when using
different navigator-data formations. Averaged over all
sampling ratios, navigator data created from snapshots
(Nav1) have advantage over other formations, while those
formed by nodes signals (Nav2) produce highest errors.  On
the other hand, \cref{app:tab.landmark.analysis} depicts
variants of landmark-point selection strategies under
scenario (D1, P1, Nav1,$\cdot$). Overall, there is no
significant difference when using maxmin or clustering
methods for landmark-point selection.

% \begin{table}[ht!]
% \centering
%     \resizebox{.7\columnwidth}{!}{\begin{tabular}{|c|cccc|}
%          \hline
%           Metric & Nav1 & Nav2 & Nav3 & Nav4 \\  
%          \hline 
%         MAE & \best{0.1460}& 0.1635& 0.1573& 0.1533\\
%         RMSE & \best{0.2315}& 0.2546& 0.2532& 0.2344\\
%         MAPE & \best{0.0148}& 0.0160& 0.0153& 0.0153\\
        
%          \hline
%     \end{tabular}
%     }
%       \caption{Average performance of MultiL-KRIM with different navigators formation strategies under scenario (D1, P1, \_, L1).}
%       \vspace{0.3cm}
%     \label{tab:navigator.analysis.avg}
% \end{table}

% \begin{table}[ht!]
% \centering
%     \resizebox{.6\columnwidth}{!}{\begin{tabular}{|c|ccc|}
%          \hline
%           Metric & L1 & L2 & L3 \\  
%          \hline 
%         MAE& \best{0.1460}& 0.1499& 0.1523\\
%         RMSE& 0.2315& \best{0.2303}& 0.2396\\
%         MAPE& 0.0148& \best{0.0147}& 0.0151\\
%          \hline
%     \end{tabular}
%     }
%       \caption{Average performance of MultiL-KRIM with different landmark points selection strategies under scenario (D1, P1, Nav1, \_).}
%       \vspace{0.3cm}
%     \label{tab:landmark.analysis.avg}
% \end{table}

\begin{table}[ht!]
\centering
      \caption{Performance on different navigators formation strategies under scenario (D1, P1, \_, L1).}
    \resizebox{\columnwidth}{!}{\begin{tabular}{c|c|cccc}
         \hline
          Sampling ratio & Metric & Nav1 & Nav2 & Nav3 & Nav4 \\  
         \hline 
         & MAE & \best{0.3004}& 0.3562& 0.3350& 0.3130\\
        0.1 & RMSE & \best{0.4211}& 0.5065& 0.5055& 0.4219\\
        & MAPE & 0.0296& 0.0299& 0.0283& \best{0.0271}\\
        \cline{1-6} 
        & MAE &  \best{0.1732}& 0.1794& 0.1883& 0.1811\\
        0.2 & RMSE & \best{0.2741}& 0.2899& 0.3046& 0.2872\\
        & MAPE &  \best{0.0175}& 0.0195& 0.0200& 0.0194\\
        \cline{1-6} 
        & MAE & \best{0.1176}& 0.1313& 0.1213& 0.1303\\
        0.3 & RMSE & \best{0.1861}& 0.2074& 0.1916& 0.2045\\
        & MAPE & \best{0.0137}& 0.0157& 0.0150& 0.0158\\
        \cline{1-6} 
        & MAE & 0.0824& 0.0865& 0.0834& \best{0.0812}\\
        0.4 & RMSE & \best{0.1607}& 0.1699& 0.1643& 0.1635\\
        & MAPE & \best{0.0082}& 0.0089& \best{0.0082}& 0.0084\\
        \cline{1-6} 
        & MAE & \best{0.0562}& 0.0638& 0.0585& 0.0608\\
        0.5 & RMSE & 0.1155& 0.0994& 0.0998& \best{0.0949}\\
        & MAPE & \best{0.0050}& 0.0059& \best{0.0050}& 0.0058\\
         \cline{1-6} 
        & MAE & \best{0.1460}& 0.1635& 0.1573& 0.1533\\
        Average & RMSE & \best{0.2315}& 0.2546& 0.2532& 0.2344\\
        & MAPE & \best{0.0148}& 0.0160& 0.0153& 0.0153\\
        
         \hline
    \end{tabular}
    }
    \label{app:tab.navigator.analysis}
\end{table}

\begin{table}[ht!]
\centering
      \caption{Performance on different landmark points selection strategies under scenario (D1, P1, Nav1, \_).}
    \resizebox{0.8\columnwidth}{!}{\begin{tabular}{c|c|ccc}
         \hline
          Sampling ratio & Metric & L1 & L2 & L3 \\  
         \hline 
         &MAE& \best{0.3004}& 0.3134& 0.3290\\
        0.1 &RMSE& 0.4211& \best{0.4180}& 0.4587\\
        &MAPE& 0.0296& \best{0.0285}& 0.0291\\
        \cline{1-5} 
        &MAE& \best{0.1732}& 0.1785& 0.1796\\
        0.2 &RMSE& \best{0.2741}& 0.2847& 0.2954\\
        &MAPE& \best{0.0175}& 0.0179& 0.0190\\
        \cline{1-5} 
        &MAE& 0.1176& 0.1193& \best{0.1166}\\
        0.3 &RMSE& 0.1861& 0.1885& \best{0.1848}\\
        &MAPE& \best{0.0137}& 0.0144& 0.0144\\
        \cline{1-5} 
        &MAE& 0.0824& 0.0819& \best{0.0811}\\
        0.4 &RMSE& 0.1607& 0.1605& \best{0.1596}\\
        &MAPE& \best{0.0082}& 0.0083& \best{0.0082}\\
        \cline{1-5} 
        &MAE& 0.0562& 0.0565&  \best{0.0551}\\
        0.5 &RMSE& 0.1155& 0.0998&  \best{0.0996}\\
        &MAPE& 0.0050& \best{0.0048}&  \best{0.0048}\\
        \cline{1-5} 
        &MAE& \best{0.1460}& 0.1499& 0.1523\\
        Average &RMSE& 0.2315& \best{0.2303}& 0.2396\\
        &MAPE& 0.0148& \best{0.0147}& 0.0151\\
         
         \hline
    \end{tabular}
    }
    \label{app:tab.landmark.analysis}
\end{table}

\begin{table}[hbt!]
\centering
    \caption{Runtime comparison (in seconds) between LRDS and MultiL-KRIM in dataset D3.}
    \resizebox{\columnwidth}{!}{
    \begin{tabular}{c||c|cc|c}
         \hline
         Sampling pattern & Sampling rate & LRDS & MultiL-KRIM & Speedup \\  
         \hline 
         \multirow{5}{*}{P1}
         &0.1& 1,393& \best{413} & 3.4x\\
         &0.2& 1,468& \best{490} & 3x\\
         &0.3& 1,358& \best{719} & 1.9x\\
         &0.4& 1,356& \best{330} & 4.1x\\
         &0.5& 1,346& \best{270}& 5x\\
         \hline \hline

         \multirow{5}{*}{P2}
         &0.1& 1,433 & \best{545} & 2.6x \\
         &0.2& 1,430 & \best{465} & 3.1x \\
         &0.3& 1,463 & \best{461} & 3.2x \\
         &0.4& 1,487 & \best{428} & 3.5x \\
         &0.5& 1,470 & \best{562} & 2.6x \\
         \hline \hline

    \end{tabular}
    }
    \label{app:tab.runTime.random_sampling}
\end{table}

\begin{table}[!ht]
  \centering
  \caption{Number of parameter unknowns~\eqref{eq:N.Q>1}}\label{app:tab.N.Q}
  \begin{tabular}{|c|c|c|c|} \hline 
    $Q$ & $\Set{d_q}_{q=1}^{Q-1}$ & $N_Q[M=1]$ & $N_Q[M=7]$ \\ \hline 
    1 & \_ & \num{11677680} & \num{81743760} \\ \hline 
    2 & $\Set{8}$ & \num{1357472} & \num{9502304} \\ \hline 
    3 & $\Set{2, 8}$ & \num{358704} & \num{2510928} \\ \hline
  \end{tabular}
\end{table}

\begin{table}[!htb]
  \centering
  \caption{Number of parameter unknowns of MultiL-KRIM and TDDIP}
  \label{app:tab.N.Q.methods}
  \begin{tabular}{|c|c|} \hline
    Method & Number of parameter unknowns \\ \hline 
    MultiL-KRIM[$M=7, Q=2$] & \num{9502304} \\ \hline 
    MultiL-KRIM[$M=1, Q=2$] & \num{1357472} \\ \hline 
    MultiL-KRIM[$M=7, Q=3$] & \num{2510928} \\ \hline
    MultiL-KRIM[$M=1, Q=3$] & \num{358704} \\ \hline
    TDDIP$^1$ & \num{3162025} \\ \hline
    TDDIP$^2$ & \num{2492713} \\ \hline
    TDDIP$^3$ & \num{394985} \\ \hline
    TDDIP$^4$ & \num{255417} \\ \hline
  \end{tabular}
\end{table}

\begin{table}[ht!]
    \renewcommand*{\arraystretch}{1.3}
    \caption{Performance of MultiL-KRIM and TDDIP on radial sampling (acceleration rate: 16x)}
    \centering \resizebox{\columnwidth}{!}{%
      {\begin{tabular}{|c|l|l|l|l|l|l|} \hline Methods $\setminus$ Metrics & \multicolumn{1}{c|}{\textbf{NRMSE}}
         & \multicolumn{1}{c|}{\textbf{SSIM}} & \multicolumn{1}{c|}{\textbf{HFEN}}
         & \multicolumn{1}{c|}{\textbf{M1}} & \multicolumn{1}{c|}{\textbf{M2}}
         & \multicolumn{1}{c|}{\textbf{Time}} \\  \hline
         \rowcolor{tableMulti}
         \textbf{MultiL-KRIM[$M=7, Q=2$]} & ${{0.0448}}$ & ${0.8680}$
                                              &${\mathbf{0.1023}}$ & $\mathbf{37.35}$ & $\mathbf{1.4 \times 10^6}$ & 1.2hrs
         \\ \hline
         \rowcolor{tableSingle}
         \textbf{MultiL-KRIM[$M=1, Q=2$]} & ${0.0465}$ & ${0.8618}$ & ${0.1305}$ & $37.26$
                                            &$\mathbf{1.4 \times 10^6}$ & 44min
         \\  \hline
         \rowcolor{tableMulti}
         \textbf{MultiL-KRIM[$M=7, Q=3$]} & $0.0448$ & ${0.8675}$ & ${0.1030}$ & $\mathbf{37.35}$
                                            & $\mathbf{1.4 \times 10^6}$ & 1hr
         \\ \hline
         \rowcolor{tableSingle}
         \textbf{MultiL-KRIM[$M=1, Q=3$]} & ${0.0465}$ & ${0.8618}$ & ${0.1305}$ & $37.26$
                                            & $\mathbf{1.4 \times 10^6}$ & \textbf{41min}
         \\  \hline
         % \rowcolor{tableMulti}
         % \textbf{KRIM[M]}~\cite{slavakis2022krim} & ${0.0450}$ & ${0.8670}$ & ${0.1136}$ & $37.34$
         %                                    & ${1.4 \times 10^6}$ & 3hrs \\ \hline
         % \rowcolor{tableSingle}
         % \textbf{KRIM[S]}~\cite{slavakis2022krim} & $0.0465$ & $0.8618$ & $0.1301$
         % & $37.26$ & $1.4 \times 10^6$ & 58min \\
         %  \hline \textbf{BiLMDM}~\cite{shetty2020bilmdm} & $0.0475$ & $0.8560$ & $0.1491$ & $37.30$ & $1.4 \times 10^6$ & 1.4hrs \\ \hline
            \textbf{TDDIP$^1$}~\cite{yoo2021time} & $\mathbf{0.0376}$ & $\mathbf{0.8896}$ & $0.1452$ & ${36.24}$ & $1.3 \times 10^6$ & 3.9hrs \\
            \hline \textbf{TDDIP$^2$}~\cite{yoo2021time} & ${0.0388}$ & ${0.8855}$ & $0.1464$ & ${35.14}$ & $1.3 \times 10^6$ & 3.5hrs \\
            \hline \textbf{TDDIP$^3$}~\cite{yoo2021time} & ${0.0415}$ & ${0.8691}$ & $0.1642$ & ${35.87}$ & $\mathbf{1.4 \times 10^6}$ & 2hrs \\
            \hline \textbf{TDDIP$^4$}~\cite{yoo2021time} & ${0.0422}$ & ${0.8680}$ & $0.1668$ & ${35.68}$ & $\mathbf{1.4 \times 10^6}$ & 1.2hrs \\
          %  \hline \textbf{LRTC-TV}~\cite{Li_Ye_Xu_2017} & $0.0738$ & $0.8063$ & $0.3725$ & $37.06$ & $\mathbf{1.7 \times 10^6}$ & 8.6hrs \\
          % \hline \textbf{SToRM}~\cite{poddar2016dynamic} & $0.0753$ & $0.8319$ & $0.3694$ & $\mathbf{37.38}$ &
          % $1.6 \times 10^6$ & 30min \\
          % \hline \textbf{PS-Sparse}~\cite{zhao2012image} & $0.0496$ & $0.7908$ & $0.1733$ &
          % $37.31$ & $1.4 \times 10^6$ & \textbf{15min} \\ 
          \hline
    \end{tabular}}
    }\label{app:tab.mrxcat.radial.tddip}
\end{table}

\begin{table*}[hbt!]
\centering
    \resizebox{\textwidth}{!}{\begin{tabular}{c||c|c|cccccc|c}
         \hline
          Dataset & Sampling ratio & Metric & LRDS & GraphTRSS & KRG & KGL &  MMF &NBP&MultiL-KRIM \\
         \hline 
         \multirow{3}{*}{Sea Temperature}&&MAE & 0.3450& 0.5822& 0.4487& 0.4438&  0.3338&0.3009&\best{0.3004}\\
         &0.1 &RMSE& 0.5615& 0.8645& 0.5906& 0.5824&  0.4773&\best{0.4018}&0.4211\\
         &&MAPE& 0.0327& 0.0507& 0.0430& 0.0405&  0.0277&\best{0.0269}&0.0296\\
         \cline{2-10} 
         &&MAE& 0.1817& 0.3200& 0.2478& 0.2352&  0.1758&\best{0.1726}&0.1732\\
         &0.2 &RMSE& 0.2814& 0.5168& 0.4156& 0.4049&  0.2806&0.2756&\best{0.2741}\\
         &&MAPE& 0.0186& 0.0295& 0.0222& 0.0211&  0.0185&0.0182&\best{0.0175}\\
        \cline{2-10} 
         &&MAE& 0.1185& 0.2093& 0.1669& 0.1649&  0.1199&0.1206&\best{0.1176}\\
         &0.3 &RMSE& 0.1867& 0.3659& 0.2880& 0.2832&  0.1901&0.1903&\best{0.1861}\\
         &&MAPE& \best{0.0131}& 0.0200& 0.0171& 0.0169&  0.0140&0.0142&0.0137\\
        \cline{2-10} 
         &&MAE& 0.0857& 0.1400& 0.1103& 0.1153&  0.0847&0.0840&\best{0.0824}\\
         &0.4 &RMSE& \best{0.1422}& 0.2584& 0.2328& 0.2438&  0.1662&0.1643&0.1607\\
         &&MAPE& 0.0098& 0.0145& 0.0092& 0.0093&  0.0083&\best{0.0079}&0.0082\\
        \cline{2-10} 
         &&MAE& 0.0621& 0.0982& 0.0749& 0.0782&  0.0602&0.0566&\best{0.0562}\\
         &0.5 &RMSE& 0.1079& 0.1966& 0.1236& 0.1292&  0.0941&\best{0.0889}&0.1155\\
         &&MAPE& 0.0066& 0.0101& 0.0061& 0.0063&  0.0052&0.0055&\best{0.0050}\\

        \hline \hline
         \multirow{3}{*}{PM2.5 Concentration}&&MAE& 2.586& 2.7698& 2.5111& 2.5214&  2.3684&2.2936&\best{2.2759}\\
         &0.1 &RMSE& 4.558& 4.7418& 4.8145& 4.8593&  4.3193&4.2110&\best{4.1863}\\
         &&MAPE& 0.4928& 0.4558& 0.4778& 0.4815&  0.3941&\best{0.3758}&0.4053\\
         \cline{2-10} 
         &&MAE& 2.0248& 2.1572& 1.7916& 1.8028&  1.7688&1.7041&\best{1.6893}\\
         &0.2 &RMSE& 3.7212& 3.6645& 3.3176& 3.3203&  3.2665&3.2666&\best{3.1932}\\
         &&MAPE& 0.3413& 0.3765& 0.3003& 0.2994&  0.3028&\best{0.2890}&0.2898\\
         \cline{2-10} 
         &&MAE& 1.6123& 1.7004& 1.5179& 1.4923&  1.4380&1.3917&\best{1.3255}\\
         &0.3 &RMSE& 3.10697& 3.0457& 3.1716& 3.1646&  3.0090&2.9376&\best{2.6244}\\
         &&MAPE& 0.2907& 0.292& 0.2440& 0.2380&  \best{0.2336}&0.2347&0.2354\\
         \cline{2-10} 
         &&MAE& 1.2519& 1.4574& 1.1746& 1.1553&  1.1251&1.0994&\best{1.0342}\\
         &0.4 &RMSE& 2.5083& 2.6089& 2.5277& 2.5188&  2.4154&2.3277&\best{2.2939}\\
         &&MAPE& 0.2504& 0.2554& 0.1947& 0.1912&  0.1933&0.1911&\best{0.1668}\\

         \hline \hline
         \multirow{3}{*}{Sea Level Pressure}&&MAE& 432.3553& 438.5799& 436.2410& 435.4914&  409.7814&408.8739&\best{393.5182}\\
         &0.1 &RMSE& 653.6381& 673.0891& 675.2066& 678.8780&  619.8458&618.0385&\best{577.7076}\\
         &&MAPE& 0.0044& 0.0043& 0.0041& 0.0041&  0.0040&0.0040&\best{0.0038}\\
         \cline{2-10} 
         &&MAE& 315.3191& 340.9664& 323.9920& 321.1525&  293.8725&290.9654&\best{277.9168}\\
         &0.2 &RMSE& 502.2517& 558.7858& 555.1179& 523.1133&  477.8935&481.9912&\best{463.2686}\\
         &&MAPE& 0.0031& 0.0034& 0.0032& 0.0029&  0.0028&0.0028&\best{0.0027}\\
         \cline{2-10} 
         &&MAE& 254.9302& 275.5929& 264.8433& 261.0184&  220.3090&220.6122&\best{208.6543}\\
         &0.3 &RMSE& 418.5478& 476.5155& 443.2093& 430.8483&  377.7021&380.5985&\best{363.3772}\\
         &&MAPE& 0.0025& 0.0027& 0.0025& 0.0023&  0.0023&0.0023&\best{0.0021}\\
         \cline{2-10} 
         &&MAE& 201.3499& 224.1063& 206.5154& 206.9803&  162.9623&163.7134&\best{158.0415}\\
         &0.4 &RMSE& 352.2565& 412.6796& 367.3505& 355.3074&  303.4321&311.2039&\best{296.1789}\\
         &&MAPE& 0.0021& 0.0022& 0.0019& 0.0018&  \best{0.0016}&\best{0.0016}&\best{0.0016}\\
         \cline{2-10} 
         &&MAE& 154.1453& 179.4160& 169.0522& 163.3182&  130.1137&123.4560&\best{119.3680}\\
         &0.5 &RMSE& 296.6426& 352.4861& 319.3949& 321.6328&  290.1288&\best{269.4014}&278.5416\\
         &&MAPE& 0.0015& 0.0018& 0.0016& 0.0015&  0.0013&\best{0.0012}&\best{0.0012}\\
         
        \hline
    \end{tabular}
    }
      \caption{Performance comparisons for Random Sampling (P1). Red-colored numbers are lowest errors.}
      \vspace{0.3cm}
    \label{app:tab.result_random_sampling}
\end{table*}

\begin{table*}[hbt!]
\centering
    \resizebox{\textwidth}{!}{\begin{tabular}{c||c|c|cccccc|c}
         \hline
          Dataset & Sampling ratio & Metric & LRDS & GraphTRSS & KRG & KGL  &MMF &NBP& MultiL-KRIM \\  
         \hline 
         \multirow{3}{*}{Sea Temperature}&&MAE & 1.1540& 1.2054& 1.1705& 1.1617&1.0946&1.0850& \best{0.9966}\\
         &0.1 &RMSE& 1.6223& 1.7232& 1.6684& 1.6596&1.4288&1.4154& \best{1.3542}\\
         &&MAPE& 0.1172& 0.1247& 0.1172& 0.1171&0.1175&0.1164& \best{0.1157}\\
         \cline{2-10} 
         &&MAE& 0.8621& 0.9962& 0.8935& 0.8786&0.8161&0.7927& \best{0.7762}\\
         &0.2 &RMSE& 1.3173& 1.5442& 1.2882& 1.2840&1.2215&1.2202& \best{1.1768}\\
         &&MAPE& 0.0898& 0.1040& 0.0930& 0.0922&0.0950&0.0892& \best{0.0865}\\
        \cline{2-10} 
         &&MAE& 0.6333& 0.6421& 0.6223& 0.6161&0.5879&0.5803& \best{0.5609}\\
         &0.3 &RMSE& 1.0650& 1.1240& 1.0641& 1.0697&1.0241&1.0148& \best{0.9876}\\
         &&MAPE& 0.0670& 0.0653& 0.0645& 0.0638&0.0575&0.0566& \best{0.0557}\\
         \cline{2-10} 
         &&MAE& 0.4536& 0.4520& 0.4023& 0.4018&0.3905&0.3876& \best{0.3770}\\
         &0.4 &RMSE& 0.8274& 0.8495& 0.7676& 0.7645&0.7366&0.7345& \best{0.7211}\\
         &&MAPE& 0.0509& 0.0467& 0.0419& 0.0419&0.0407&0.0403& \best{0.0385}\\
         \cline{2-10} 
         &&MAE& 0.2663& 0.3063& 0.2831& 0.2825&0.2605&0.2615&  \best{0.2560}\\
         &0.5 &RMSE& 0.5450& 0.5934& 0.5912& 0.5855&0.5580&0.5525& \best{0.5435}\\
         &&MAPE& \best{0.0263}& 0.0350& 0.0308& 0.0306&0.0306&0.0303& 0.0292\\

        \hline \hline
         \multirow{3}{*}{PM2.5 Concentration}&&MAE& 3.5170& 3.7224& 3.2365& 3.2133&3.2120&3.1685&\best{3.0717}\\
         &0.1 &RMSE& 5.7922& 6.5506& 5.4857& 5.4166&5.3981&5.3666& \best{5.2669}\\
         &&MAPE& 0.5125& 0.5987& 0.5828& 0.4831&\best{0.4811}&0.5309& 0.5278\\
         \cline{2-10} 
         &&MAE& 2.5809& 2.7509& 2.5363& 2.5412&2.4906&2.4782& \best{2.4226}\\
         &0.2 &RMSE& 4.4577& 4.8055& 4.5982& 4.6150&4.6008&4.5416& \best{4.2820}\\
         &&MAPE& \best{0.4588}& 0.5110& 0.4598& 0.4614&0.4834&0.4863& 0.4717\\
         \cline{2-10} 
         &&MAE& 2.0887& 2.1562& 2.0231& 2.0161&1.9740&1.9534& \best{1.9405}\\
         &0.3&RMSE& 4.1175& 4.0359& 3.9335& 3.9242&3.8544&3.8286& \best{3.8262}\\
         &&MAPE& \best{0.3420}& 0.3872& 0.3613& 0.3574&0.3583&0.3541& 0.3452\\
         \cline{2-10} 
         &&MAE& 1.7716& 1.7880& 1.5922& 1.5841&1.5715&1.5508& \best{1.5285}\\
         &0.4 &RMSE& 3.6206& 3.5443& 3.3331& 3.3171&3.3405&3.3078& \best{3.1535}\\
         &&MAPE& 0.2983& 0.3133& 0.2796& 0.2760&0.2809&0.2786& \best{0.2737}\\
         \cline{2-10} 
         &&MAE& 1.4244& 1.5043& 1.2903& 1.2876&1.2904&1.2867& \best{1.2645}\\
         &0.5 &RMSE& 2.9851& 3.1396& 2.9521& 2.9361&2.9533&2.9315& \best{2.8521}\\
         &&MAPE& 0.2506& \best{0.2260}& 0.2477& 0.2375&0.2303&0.2283& 0.2504\\

        \hline \hline
         \multirow{3}{*}{Sea Level Pressure}&&MAE& 452.3376& 469.8179& 460.6198& 459.5086&457.6376&452.2161& \best{444.3823}\\
         &0.1 &RMSE& \best{673.5309}& 704.6300& 722.8781& 715.2535&714.0710&707.0870& 696.8868\\
         &&MAPE& 0.0045& 0.0046& 0.0048& 0.0047&0.0047&0.0045& \best{0.0044}\\
         \cline{2-10} 
         &&MAE& 395.4735& 404.9930& 376.2283& 374.5962&365.2090&360.5889& \best{355.2410}\\
         &0.2 &RMSE& 602.4211& 609.5108& 641.7902& 606.5960&582.4397&581.1579& \best{578.3478}\\
         &&MAPE& 0.0039& 0.0040& 0.0038& 0.0038&0.0037&0.0036& \best{0.0035}\\
         \cline{2-10} 
         &&MAE& 319.3464& 328.3074& 309.2513& 308.1569&298.5832&296.9049& \best{293.2719}\\
         &0.3 &RMSE& 535.5085& 547.2752& 547.0275& 537.0053&525.3399&521.1263& \best{513.2530}\\
         &&MAPE& 0.0034& 0.0034& 0.0030& 0.0030&0.0030&0.0030& \best{0.0029}\\
         \cline{2-10} 
         &&MAE& 265.9806& 274.8237& 251.6239& 251.0470&249.6740&246.4856& \best{243.6812}\\
         &0.4 &RMSE& 483.8443& 476.1510& 479.6474& 480.5829&469.6758&461.5433& \best{456.8314}\\
         &&MAPE& 0.0028& 0.0027& 0.0026& 0.0026&\best{0.0025}&\best{0.0025}& \best{0.0025}\\
         \cline{2-10} 
         &&MAE& 222.3163& 224.0090& 206.3858& 205.5674&200.1395&198.1765& \best{195.2719}\\
         &0.5 &RMSE& 408.2798& 413.9014& 428.3704& 429.2268&411.1974&\best{370.1874}& 398.6022\\
         &&MAPE& 0.0022& 0.0022& 0.0021& 0.0021&\best{0.0020}&\best{0.0020}& \best{0.0020}\\

         \hline
    \end{tabular}
    }
      \caption{Performance comparisons for Entire Snapshots Sampling (P2). Red-colored numbers are lowest errors.}
      \vspace{0.3cm}
    \label{app:tab.result_entire_snapshots}
\end{table*}

\section{More results on dMRI reconstruction}\label[appendix]{app:exp.dMRI}

For demonstration, \cref{app:tab.N.Q.methods} examines
different sizes of TDDIP by varying the number of layers and
hidden dimensions, where TDDIP$^1$ is the setting scoring
the lowest NRMSE, TDDIP$^2$ and TDDIP$^3$ are close to
MultiL-KRIM[$M=7, Q=3$] and MultiL-KRIM[$M=1, Q=3$] in
number of parameters, respectively, while TDDIP$^4$ matches
the computational times of [$M=7$].  Specifically,
performance of these variations of TDDIP are reported in
\cref{app:tab.mrxcat.radial.tddip}. Note that if the
complexity of TDDIP decreases, its NRMSE and SSIM approach
those of MultiL-KRIM. Nevertheless, increasing the number of
unknowns of MultiL-KRIM, whether by increasing the number of
kernels $M$ or adopting larger inner dimensions, does not
significantly improve its performance.  These findings
suggest a research direction of developing a ``deep''
architecture for MultiL-KRIM, \eg, by adding non-linear
activation function layers into its factorization, inspired
by results showing that deep matrix factorization
outperforms the standard MMF in matrix completion
tasks~\cite{fan2018matrix, de2021survey}. More results on
this research front will be reported at other publication
venues.

\end{document}

%% file: preamble.tex
\IEEEoverridecommandlockouts
\usepackage{siunitx}
\usepackage{amsmath}
\usepackage{amsthm,amssymb,amsfonts,mathrsfs,mathtools,amsbsy}
\allowdisplaybreaks
\usepackage{bm}
\usepackage{algorithmic}
\usepackage{algorithm}
\usepackage{array}

\usepackage{subcaption}
\usepackage{textcomp}
\usepackage{multirow}
\usepackage{paralist, tabularx}
\usepackage{placeins}
\usepackage{url}
\usepackage{verbatim}
\usepackage{graphicx}
\usepackage{cite}
\usepackage[inline]{enumitem}
\usepackage{xspace}
\usepackage[table]{xcolor}
\usepackage{tcolorbox}
\usepackage[hidelinks]{hyperref}
\usepackage{pifont}% http://ctan.org/pkg/pifont

\usepackage{tikz}
\usetikzlibrary{shapes,arrows,positioning}

\usepackage[capitalize]{cleveref}

\crefname{thm}{Theorem}{Theorems}
\crefname{prop}{Proposition}{Propositions}
\crefname{assumption}{Assumption}{Assumptions}
\crefname{lemma}{Lemma}{Lemmata}
\crefname{definition}{Definition}{Definitions}
\crefname{example}{Example}{Examples}
\crefname{algo}{Algorithm}{Algorithms}
\crefname{fact}{Fact}{Facts}
\crefname{claim}{Claim}{Claims}
\crefname{appendix}{Appendix}{Appendices}
\crefname{coroll}{Corollary}{Corollaries}
\crefname{figure}{Figure}{Figures}
\crefname{section}{Section}{Sections}

\crefname{thmlisti}{Theorem}{Theorems}
\crefname{lemlisti}{Lemma}{Lemmata}
\crefname{proplisti}{Proposition}{Propositions}
\crefname{asslisti}{Assumption}{Assumptions}
\crefname{deflisti}{Definition}{Definitions}
\crefname{exlisti}{Example}{Examples}
\crefname{algolisti}{Algorithm}{Algorithms}
\crefname{factlisti}{Fact}{Facts}
\crefname{claimlisti}{Claim}{Claims}
\crefname{applisti}{Appendix}{Appendices}
\crefname{MyEnumSeci}{}{}
\crefname{MyEnumSubSeci}{}{}
\crefname{appfig}{Appendix Figure}{Appendix Figures}
\crefname{apptab}{Appendix Table}{Appendix Tables}
\crefname{appeq}{Appendix Equation}{Appendix Equations}

\colorlet{tableMulti}{red!20}
\colorlet{tableSingle}{cyan!30}

\newcommand\best[1]{\textcolor{red}{\textbf{#1}}}
\newcommand\nbest[1]{\textcolor{blue}{\textbf{#1}}}

\newcommand{\IntegerP}{\mathbb{N}}
\newcommand{\IntegerPP}{\mathbb{N}_*}
\newcommand{\Real}{\mathbb{R}}
\newcommand{\RealP}{\mathbb{R}_+}
\newcommand{\RealPP}{\mathbb{R}_{++}}
\newcommand{\Complex}{\mathbb{C}}
\newcommand\given{{\mathbin{}\mid\mathbin{}}}
\newcommand\vect[1]{\mathbf{#1}}
\newcommand\vectcal[1]{\mathbfcal{#1}}
\newcommand*{\hermconj}{\mathsf{H}}
\DeclarePairedDelimiter\norm{\lVert}{\rVert}
\DeclarePairedDelimiterX\innerp[2]{\langle}{\rangle}{#1
  \mathop{}\delimsize\vert\mathop{} #2}

\providecommand\given{} % so it exists
\newcommand\SetSymbol[1][]{
  \nonscript\,#1\vert \allowbreak \nonscript\,\mathopen{}}
\DeclarePairedDelimiterX\Set[1]{\lbrace}{\rbrace}%
{ \renewcommand\given{\SetSymbol[\delimsize]} #1 }

\DeclareMathAlphabet\mathbfcal{OMS}{cmsy}{b}{n}
\DeclareMathAlphabet\mathbfit{OML}{cmm}{b}{it}
\DeclareMathAlphabet\mathbfscr{OMS}{mdugm}{b}{n}

\DeclareMathOperator{\bdiag}{bdiag}
\DeclareMathOperator{\diag}{diag}
\DeclareMathOperator{\tr}{tr}
\DeclareMathOperator{\tovec}{vec}
\DeclareMathOperator{\Id}{Id}
\DeclareMathOperator{\prox}{Prox}

\DeclarePairedDelimiter{\ceil}{\lceil}{\rceil}

\makeatletter
\newcommand*{\ie}{%
  \@ifnextchar{,}%
  {i.e.}%
  {i.e.,\@\xspace}%
}
\newcommand*{\eg}{%
  \@ifnextchar{,}%
  {e.g.}%
  {e.g.,\@\xspace}%
}
\newcommand*{\cf}{%
  \@ifnextchar{,}%
  {cf.}%
  {cf.,\@\xspace}%
}
\makeatother

\usepackage{stackengine}
\stackMath
\newlength\matfield
\newlength\tmplength
\def\matscale{1.}
\newcommand\dimbox[4]{%
  \setlength\matfield{\matscale\baselineskip}%
  \setbox0=\hbox{\vphantom{X}\smash{#3}}%
  \setlength{\tmplength}{#1\matfield-\ht0-\dp0}%
  \fboxrule=1pt\fboxsep=-\fboxrule\relax%
  \fcolorbox{black}{#4}{\makebox[#2\matfield]{\addstackgap[.5\tmplength]{\box0}}}%
}
\newcommand\raiserows[2]{%
   \setlength\matfield{\matscale\baselineskip}%
   \raisebox{#1\matfield}{#2}%
}
\newcommand\matbox[6]{
  \stackunder{\dimbox{#1}{#2}{$#5$}{#6}}{\scriptstyle(#3\times #4)}%
}

\tikzstyle{block} = [rectangle, draw, fill=white, text width=10em, text centered, rounded corners, minimum height=5em, minimum width=5em]
\tikzstyle{line} = [draw, -latex']

\newcommand{\negphantom}[1]{\ifmmode\settowidth{\dimen0}{$#1$}\else\settowidth{\dimen0}{#1}\fi\hspace*{-\dimen0}}